\documentclass[reprint,
 superscriptaddress,
 amsmath,amssymb,amsthm,
 aps,prx,
]{revtex4-2}
\usepackage{graphicx}
\usepackage{dcolumn}
\usepackage{bm}
\usepackage{braket}
\usepackage[table]{xcolor}
\usepackage{xcolor}
\usepackage{amsthm}
\usepackage{upgreek}
\usepackage[normalem]{ulem}
\usepackage{times}
\usepackage[table]{xcolor}
\usepackage{graphicx} 
\usepackage{ulem}
\usepackage{bm, amsmath, amssymb, mathtools}
\usepackage{amsthm}
\usepackage{soul}

\usepackage{siunitx}
\newcommand{\aop}{{\bf a}}

\newcommand{\adag}{{\bf a^{\dag}}}
\newcommand{\bop}{{\bf b}}
\newcommand{\bdag}{{\bf b^{\dag}}}
\newcommand{\HH}{{\bf H}}
\newcommand{\LL}{{\bf L}}
\newcommand{\rhoo}{\boldsymbol{\normalrho}}

\usepackage{graphicx}
\usepackage{dcolumn}
\usepackage{bm}
\usepackage{braket}
\usepackage[table]{xcolor}
\usepackage{xcolor}
\usepackage{amsthm}
\usepackage{upgreek}
\usepackage[normalem]{ulem}

\newcommand{\varphiop}{\boldsymbol{\varphi}}

\let\normalrho\rho

\newcommand{\rop}{{\bf r}}

\usepackage{soul}
\usepackage{siunitx}

\usepackage{titletoc}

\usepackage[titles]{tocloft}
\makeatletter
\renewcommand{\@tocrmarg}{1.5em plus1fil}
\makeatother

\usepackage[multiuser,draft,inline,marginclue]{fixme}
\fxusetheme{colorsig}
\FXRegisterAuthor{las}{alas}{LA}
\FXRegisterAuthor{pci}{apci}{Ph}
\FXRegisterAuthor{pr}{apr}{PR}

\usepackage[colorlinks=false, linktocpage=true, hypertexnames=false]{hyperref}
\usepackage[all]{hypcap}

\usepackage[OT2,T1]{fontenc}
\DeclareSymbolFont{cyrletters}{OT2}{wncyr}{m}{n}
\DeclareMathSymbol{\Sha}{\mathalpha}{cyrletters}{"58}

\renewcommand{\dag}{\dagger}

\usepackage[capitalize]{cleveref}

\begin{document}

\preprint{APS/123-QED}

\title{Dissipating quartets of excitations in a superconducting circuit }

\author
{A. Vanselow,$^{1,2}$ B. Beauseigneur,$^{1,2}$ L. Lattier,$^{1,2}$ M. Villiers,$^{1,2}$ A. Denis,$^{1}$ P. Morfin,$^{1}$ Z. Leghtas,$^{1,2}$
P. Campagne-Ibarcq$^{1,2}$}
\affiliation{$^1$ Laboratoire de Physique de l’Ecole normale supérieure, ENS, Université PSL, CNRS, Sorbonne Université, Université de Paris, F-75005 Paris, France\\
$^2$ Inria, Mines Paris-PSL, Paris, France}

\date{\today}

\vspace{1cm}

\begin{abstract}

Over the past decade, autonomous stabilization of bosonic qubits has emerged as a promising approach for hardware-efficient protection of quantum information. However, applying these techniques to more complex encodings than the Schrödinger cat code requires exquisite control of high-order wave mixing processes. The challenge is to  enable specific multiphotonic dissipation channels  while avoiding unintended non-linear interactions. In this work, we leverage a genuine six-wave mixing process enabled by a near Kerr-free Josephson element to enforce  dissipation of quartets of excitations in a high-impedance  superconducting resonator. Owing to residual non-linearities stemming from stray inductances in our circuit, this dissipation channel is only effective when the resonator holds a specific number of photons. Applying  it to the fourth excited state of the resonator, we show an order of magnitude enhancement of the state decay rate while only marginally impacting the relaxation and coherence of lower energy states. Given that stray inductances could be strongly reduced through simple modifications in circuit design  and that our methods can be adapted to activate even higher-order dissipation channels, these results pave the way toward the dynamical stabilization of four-component Schrödinger cat qubits and even more complex bosonic qubits.

\end{abstract}

\maketitle

\section{Introduction}
\label{sec:introduction}

\begin{figure}[htbp]
		\centering
		\includegraphics[width=1\columnwidth]{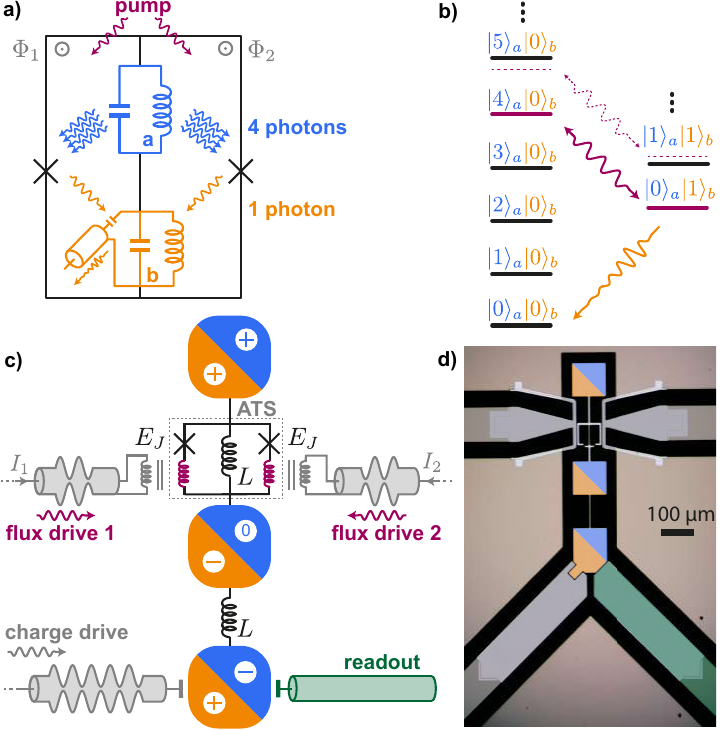}
		\caption{{\bf Dissipation engineering in a Josephson circuit. a)}. A  target oscillator (the \textit{memory}, blue) is coupled to an auxiliary dissipative mode (the \textit{buffer}, orange) by a  SQUID participating in both modes. When the circuit loops are biased with magnetic fluxes $\Phi_1$ and $\Phi_2$ following Eq.~\eqref{eq:fluxes} with $\epsilon(t)$  a pump oscillating at $4\omega_a-\omega_b$, the SQUID mediates a Kerr-free four-to-one photon exchange interaction, opening a four-photon dissipation channel acting on the memory. The mode impedances $Z_{a,b}$ set the strength of the interaction and  are of the order of $R_q/10=650~\Omega$ in our experiment.      {\bf b) } Energy level-diagram featuring the four-to-one photon  pump (purple wriggled arrow)  and the buffer single-photon dissipation (orange wriggled arrow). In presence of strong non-linearity, energy levels are unevenly spaced and the pump drives a single transition only (dashed purple arrow represents an off-resonantly driven transition). {\bf c)  } Schematic representation of the circuit used in our experiment. A three-pad superconducting structure hosts the memory and buffer modes with respective anti-symmetric and symmetric voltages across the pads. A  distributed $\lambda/2$ resonator capacitively coupled to the structure forms a third mode employed for readout. Foster synthesis of the linear circuit connected to the SQUID yields an effective circuit as in (a) but for the presence of a third mode and for series  inductances in the arms of the SQUID, responsible for spurious non-linearities (see Fig.~\ref{fig2}a and text). Magnetic biases are applied through two flux lines (top). A charge line (bottom) is capacitively connected to the circuit and enables linear drives and single-photon dissipation of the buffer. All lines are filtered by periodic loading to limit single-photon dissipation in the memory (see Sec.~\ref{sm:detailcrystal}). {\bf d)  } Optical microscope image of the sample, featuring all the structures pictured in (c) with matching false colors,  surrounded by a  ground plane (brown).  }
		\label{fig1}
	\end{figure}

Quantum Error Correction (QEC) relies on the measurement of error syndromes  to detect spurious evolutions of a  physical system embedding quantum information~\cite{nielsen2002quantum}. For stabilizer codes~\cite{gottesman1997stabilizer}, the regular approach consists in measuring a set of  stabilizers and decoding the measurement record to apply corrective feedback or feedforward controls.  Alternatively, QEC can be made autonomous by engineering  adequate dissipation channels~\cite{wolinsky1988quantum,zanardi1997noiseless,lidar1998decoherence,barnes2000automatic,sarovar2005continuous,ticozzi2008quantum,pastawski2011quantum,mirrahimi2014dynamically,cohen2014dissipation,albert2016geometry,kapit2016hardware,kapit2017upside, reiter2017dissipative,lihm2018implementation, royer2020stabilization,lebreuilly2021autonomous,de2022error,xu2023autonomous,rojkov2024stabilization}. Autonomous QEC is only practical in the most simple decoding cases, in particular when each feedback  control depends on the instantaneous value of a single stabilizer only.  This is the case for most single-mode bosonic codes~\cite{albert2025bosonic}, in which a qubit encoded in an oscillator is stabilized by a few operators only. This approach was pioneered in the setting of two-legged cat qubits~\cite{mirrahimi2014dynamically,leghtas2015confining,touzard2018coherent,lescanne2020exponential,berdou2023one,reglade2024quantum,marquet2024autoparametric,putterman2024hardware}. The two-legged cat code manifold is spanned by two coherent states $|\pm \alpha\rangle$ of a target oscillator $a$. It is stabilized by a Lindblad dynamics entailed by a single dissipation channel $\dot{\rhoo}_a=\gamma_2 \mathcal{D}[\aop^2-\alpha^2]\rhoo_a$, where $\gamma_2$ is the  dissipation rate and $\mathcal{D}[\LL]\rhoo= \LL \rhoo \LL^{\dagger}-\big(\LL^{\dagger} \LL \rhoo + \rhoo \LL^{\dagger}\LL \big)/2$.  The Lindblad operator $\aop^2-\alpha^2$  is of order two in the annihilation and creation operators $(\aop,\adag)$ and by extension, we characterize the stabilizing dissipation channel  as second order. It may be engineered  by activating a  two-to-one photon exchange interaction between the target oscillator  and an auxiliary mode $b$, combined with a resonant drive on $b$ to yield the Hamiltonian $\HH_2=g_2 (\aop^2 -\alpha^2 )\bdag +\mathrm{h.c.}$ If the auxiliary mode dissipates excitations into a cold environment at rate $\kappa_b $ much larger than the two-to-one photon exchange rate $g_2$, the multimode dynamics reduces to the desired single-mode Lindblad dynamics of the target mode with $\gamma_2=4g_2^2/\kappa_b$~\cite{lecturenotesmirrahimirouchon,robin2024convergence,note1}. Unless the two mode resonance frequencies are precisely targeted to reach $\omega_b=2\omega_a$~\cite{marquet2024autoparametric}, the interaction  needs to be parametrically modulated at $2\omega_a-\omega_b$ to become resonant. This can be realized with a variety of Josephson circuits such as the one pictured in Fig.~\ref{fig1}a,  all relying on the four-wave mixing capabilities of Josephson junctions. This strategy was successfully deployed to robustly suppress bit-flips of two-legged cat qubits. On the down side, this code does not offer any protection against phase-flips entailed by single-photon loss in the target mode, which is the dominant noise channel in most experimental platforms.\\

In order to generalize the approach to codes whose states have more complex structures in phase-space and offer some measure of protection against phase-flips, one typically needs higher order dissipation channels and thus higher order interactions. In particular, the fourth order channel corresponding to the Lindblad dissipator $\gamma_4 \mathcal{D}[\aop^4-\alpha^4]$ stabilizes a manifold of four coherent states, from which one can single out a sub-manifold of given parity to define a four-legged cat qubit~\cite{mirrahimi2014dynamically,rojkov2024stabilization}. This qubit is protected against bit-flips as its two-legged counterpart, and  single-photon loss events can be detected without perturbing the encoded information by monitoring the photon number parity~\cite{mirrahimi2014dynamically,sun2014tracking,ofek2016extending,hu2019quantum,ni2023beating}, thus suppressing phase-flips at first-order. Note that even higher order dissipation channels can correct parity flips autonomously~\cite{gertler2021protecting} or stabilize other rotation-symmetric codes~\cite{grimsmo2020quantum} and translation-symmetric codes~\cite{gottesman2001encoding,royer2020stabilization,de2022error}, offering a more robust protection of the encoded information. In earlier experimental works, high-order dissipation channels were obtained from low-order interactions. As discussed in Sec.~\ref{sec:discussion}, this approach  can lead to the opening of spurious dissipation channels directly corrupting the encoded information.\\

An alternative strategy circumventing this pitfall consists in directly engineering high-order dissipation channels from  high-order interactions. In this work, we take this approach and aim at engineering a four-photon loss channel---a key primitive to stabilize four-legged cat qubits---through genuine high-order interactions. Adapting the  methods developed to stabilize  two-legged cat qubits~\cite{lescanne2020exponential}, we proceed by   activating a resonant four-to-one photon exchange interaction between a long-lived superconducting resonator (coined the \textit{memory}) and an  auxiliary dissipative mode (coined the \textit{buffer}). In the idealized setting depicted in  Fig.~\ref{fig1}a, two $LC$ oscillators are connected in series with a balanced Superconducting Quantum Interference Device (SQUID). The junctions of the SQUID enable  mixing of four waves from the memory $a$, one wave from the buffer $b$, and one pump wave providing the energy needed for the interaction to be resonant. \\

In detail, the circuit features two superconducting loops which are biased with magnetic fields
\begin{equation}
\Phi_1(t)=\varphi_0\big(\pi +  \epsilon(t) \big) \qquad \qquad \Phi_2(t)= \varphi_0  \epsilon(t)
\label{eq:fluxes}
\end{equation}
where $\varphi_0=\hbar/(2e)$ is the reduced flux quantum and the AC signal $\epsilon$ is referred to as the pump. The SQUID couples to the sum of the modes reduced phase operators $\varphiop= \varphi^{ZPF}_a (\aop+ \aop^{\dagger} )+\varphi^{ZPF}_b (\bop+ \bop^{\dagger} )$ via
\begin{equation}
\begin{aligned}
\HH_J(t)&=- E_J \cos(\varphiop - \frac{\Phi_1}{\varphi_0}) - E_J \cos(\varphiop + \frac{\Phi_2}{\varphi_0})\\
&=2E_J~\sin\big(\epsilon(t)\big)\sin(\varphiop)
\end{aligned}
\label{eq:HATS}
\end{equation}
Expanding the \textit{sine} terms, we find that the SQUID  enables the mixing of an odd number of pump waves with an odd number of mode waves. In particular, by letting $\epsilon(t)=\xi e^{i(4 \omega_a - \omega_b)t} + \mathrm{c.c}.$, we activate a resonant four-to-one photon exchange interaction  between the memory and the buffer
\begin{equation}
\HH_4=\hbar g_4  \aop^4 \bop^{\dagger} ~+~ \mathrm{h.c.}
\label{eq:4to1}
\end{equation}
where the exchange rate $g_4\simeq 2 \frac{E_J}{\hbar 4!} \xi \varphi^{{ZPF}^4}_a \varphi^{ZPF}_b $ depends strongly on the vacuum phase fluctuations of the modes. For each mode $i$, these are set  by the mode impedance following $\varphi^{ZPF}_i=(\pi \frac{Z_i}{R_q})^{\frac{1}{2}}$ and $R_q\simeq 6.5~\mathrm{k}{\Omega}$ is the resistance quantum~\cite{nigg2012black,smith2016quantization,minev2021energy}. Therefore, we require the modes impedance to be large for the interaction rate to be significant (see Sec.~\ref{sec:discussion} for a discussion on the optimal parameter regime). In presence of strong single-photon dissipation at rate $\kappa_b\gg g_4$ in the buffer, the system dynamics reduces to that entailed by a four-photon dissipation channel $\frac{4g_4^2}{\kappa_b} \mathcal{D}[\aop^4]$ on the memory~\cite{lecturenotesmirrahimirouchon}. \\

Crucially, no other non-linear term in the expansion of $\HH_J(t)$ survives the Rotating Wave Approximation (RWA) when the mode frequencies are not commensurable. Therefore, the memory and buffer remain in principle perfectly linear. However, in our experiment, stray inductance of  the SQUID tracks (purple inductor in Fig.~\ref{fig1}c) induce a spurious non-linear mechanism that is not canceled at the system working point~\eqref{eq:fluxes}. As a result, the memory is sufficiently anharmonic 
that the four-to-one photon transitions $|n + 4\rangle_a|0\rangle_b \leftrightarrow |n \rangle_a|1\rangle_b$ (where $|n\rangle_i$ denotes a Fock state of mode $i$) cannot be indistinguishably driven with a single pump microwave (see Fig.~\ref{fig1}b). Nevertheless, driving the transition $|4\rangle_a |0\rangle_b \leftrightarrow |0\rangle_a |1\rangle_b$, we demonstrate that quartets of excitations can be dissipated at a rate an order of magnitude higher than the single-photon loss rate in our device, as detailed in the remaining of the paper. We review in Sec.~\ref{sec:discussion} simple design changes that should significantly suppress undesired non-linear mechanisms in a future iteration of the experiment, paving the way for the stabilization of the four-legged cat manifold.

\section{Circuit design and characterization}
\subsection{Design}

The 2D superconducting circuit we design to implement the idealized circuit  analyzed in the previous section is pictured in Fig.~\ref{fig1}c-d. At its core lie three superconducting islands (blue and orange squares). The top and central islands are bridged by an Asymmetrically Threaded SQUID (ATS)~\cite{lescanne2020exponential} whose shunt superinductor is made of a chain of 25 Josephson junctions for a total nominal inductance of $L=28.5$~nH.   A second identical inductor bridges the central and bottom islands. This symmetric structure supports the memory and buffer, with respective anti-symmetric and symmetric voltages across the pads. The SQUID participates strongly in both modes, resulting in strong phase-fluctuations as prescribed in the previous section.  A distributed $\lambda/2$ resonator, colored in green and coined the \textit{readout} mode, is capacitively coupled to the three-island structure with only a weak participation of the ATS. It is employed to probe the state of the memory as detailed in Sec.\ref{sec:prepmeas}. \\

\begin{figure}[htbp]
		\centering
		\includegraphics[width=1\columnwidth]{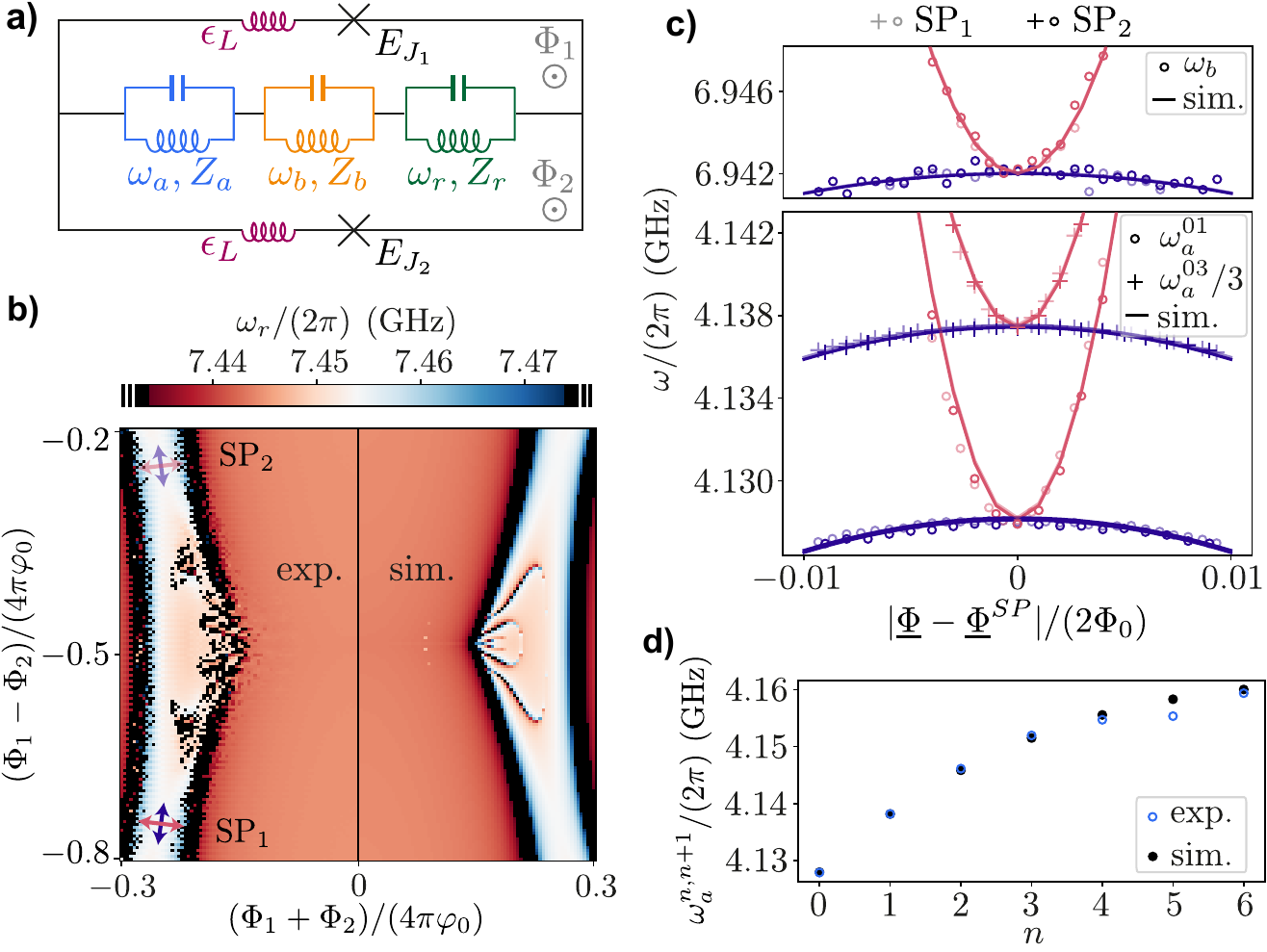}
		\caption{{\bf Circuit characterization. a)} Effective three-mode circuit obtained by Foster synthesis of the linear environment connected to the SQUID and used to reproduce data presented in (b-d). Fitted parameters are given in Table~\ref{tableparam}. {\bf b)} Measured and simulated resonance frequency of the readout mode as a function of applied DC magnetic biases. Black pixels indicate that the resonance frequency could not be extracted, either  because it lied out of the probed frequency range (encoded in color, simulation results clipped to the same range) or because the fit was  unreliable. Two saddle points, labeled $\mathrm{SP}_1$ and $\mathrm{SP}_2$, are identified by intersecting double-headed arrows. {\bf c)} Frequency of the memory and buffer low-energy transitions, along cuts of the frequency map intersecting $\mathrm{SP}_1$ and $\mathrm{SP}_2$ (respectively in bright and dim colors). Cut angles, indicated by double-headed arrows with matching colors in (b), either  maximize or minimize curvature. All  frequencies are near-identical around both saddle points, indicating that the ATS is near-perfectly balanced. {\bf d)} First seven transition frequencies of the memory measured at $\mathrm{SP}_1$ (empty blue circles). The  anharmonicity is positive and decreases with transition index. This spectrum is well captured by our model  accounting for stray inductance in series with the junctions of the ATS (filled black circles).  }
		\label{fig2}
	\end{figure}

To model and quantize our system, we perform  Foster synthesis of the linear circuit connected to the SQUID~\cite{foster1924reactance,nigg2012black,minev2021energy}, yielding the three system modes (labeled $a$, $b$, $r$) represented by a chain of parallel $LC$ oscillators with the hierarchy $Z_r\ll Z_a \sim Z_b \sim R_q/10$ (see Fig.~\ref{fig2}a). The ATS shunt inductor is accounted for in this synthesis, but \textit{not} the  SQUID itself, whose arms each feature a junction and a small series inductance. The SQUID is subsequently connected in parallel with the chain  to obtain the effective circuit pictured in Fig.~\ref{fig2}a. It is analogous to the ideal circuit pictured in Fig.~\ref{fig1}a but for the presence of a third mode and for the series inductances in the arms of the SQUID.  The fluxes~\eqref{eq:fluxes} biasing the loops of the ATS are delivered through two dedicated  feedlines, denoted as \textit{flux lines}~\cite{note2}.  A third feedline, denoted as the \textit{charge line}, is capacitively coupled to the three-pad structure and to the readout resonator. It allows direct linear drive of the modes and sets the single-photon decay rate of  the buffer and readout modes (see Table~\ref{tableparam}). Note that in principle, one could leverage the distinct symmetries of the memory and buffer modes to selectively couple a feedline to the buffer, thus preventing radiative decay of memory photons. This was not attempted in our experiment, in which the coupling of the feedline is of the same order for all modes. Instead, we filter all feedlines by sinusoidally  modulating  their geometry as detailed in Sec.~\ref{sm:detailcrystal}, thereby suppressing radiative decay at the memory frequency.\\

\subsection{Modes dispersion against bias fluxes}

In order to characterize the system, we record the modes resonance frequencies against  DC magnetic fluxes biasing the ATS loops  in absence of pump  ($\epsilon(t)=0$). We first probe the readout mode in single-tone spectroscopy through the charge line, yielding the frequency map represented in Fig.~\ref{fig2}b.  We identify the ATS working point corresponding to the DC biases introduced in Eq.~\eqref{eq:fluxes}  as a saddle point of the frequency map, along with analog points located at half-integer numbers of flux quanta away. This allows us to relate the direct currents applied through the  flux lines to the actual fluxes threading the ATS (Fig.~\ref{fig2}b uses calibrated flux coordinates~\cite{note3}, and two non-equivalent saddle points are identified by crossing arrows).  Next, we record the memory and buffer resonance frequencies---respectively in two-tone spectroscopy leveraging the memory-readout dispersive coupling and in single-tone spectroscopy---around the saddle points.  Fig.~\ref{fig2}c features each mode resonance frequency (empty circles) when sweeping applied fluxes through the saddle points of the frequency map, along directions of maximum and minimum curvatures (indicated by double-headed arrows with matching colors in Fig.~\ref{fig2}b).  Strikingly, the modes have near identical resonance frequencies at both saddle points (markers in bright colors corresponding to the saddle point $\mathrm{SP}_1$ are aligned with markers in dim colors corresponding to the  saddle point $\mathrm{SP}_2$), indicating that the Josephson junctions forming the SQUID  have a relative energy difference smaller than $0.01~\%$ (see Sec.~\ref{sec:maps}). We attribute this exceptionally low circuit asymmetry  to chance in fabrication and junction aging behavior. Indeed, the asymmetry was found to be two orders of magnitude higher in similar circuits that we have characterized, and  an order of magnitude higher during a previous cooldown of the same circuit.

\subsection{Stray inductances and memory non-linearity}
\label{sec:stray}
Unexpectedly,  when driving the memory on resonance  and subsequently probing the readout resonator, we detect a reflected signal that oscillates with the duration and amplitude of the memory drive pulse. Measuring such Rabi oscillations unambiguously reveals a significant non-linearity of the memory, whose transitions can be addressed individually. Driving the memory with sequences of charge drives  (to excite single-photon transitions $|n\rangle_a \leftrightarrow |n+1\rangle_a$) and/or flux drives (to excite three-photon transitions $|n\rangle_a \leftrightarrow |n+3\rangle_a$ leveraging the wave-mixing properties of the ATS), we  reconstruct the memory mode energy level structure at the saddle point (see Fig.~\ref{fig2}d). The anharmonicity   $\omega_a^{n+1,n+2}-\omega_a^{n,n+1}$ is positive for all $n$,  lies in the 1---10~MHz range and decreases sharply with $n$. As discussed in Sec.~\ref{sec:discussion}, we unambiguously relate this exotic pattern to a Cooper-pair pairing mechanism (renormalization of the Josephson potential with an added $\cos(2\varphi)$ contribution) owing to stray inductance in the tracks of the SQUID~\cite{smith2023spectral}  (inductive energy $\epsilon_L \gg E_J$, see Fig.~\ref{fig2}a). Including this mechanism in our model, we quantitatively reproduce all data presented in Fig.~\ref{fig2}.  The fitted parameters, presented in Table~\ref{tableparam}, are in good agreement with design values (see Sec.~\ref{sec:smNL} for details).

\begin{table}
\begin{tabular} {|c|c||c|c|}
\hline
\rowcolor{lightgray}
Parameter & Value &  Parameter & Value  \\
\hline
$\varphi_a^{ZPF}$ & 0.405 &$\omega_a$ & $2\pi \times 4.13$~GHz\\
  \hline 
  $\varphi_b^{ZPF}$ & 0.312&$\omega_b$ & $2\pi \times 6.94$~GHz\\
  \hline  
    $\varphi_r^{ZPF}$ & 0.044&$\omega_r$ & $2\pi \times 7.45$~GHz\\
    \hline  
$E_J$ & $h\times$17.7~GHz & $\delta E_J$ & $h\times$~0.35~MHz\\
  \hline  
$\kappa_b$ & $ 2\pi\times$2.05~MHz&$\kappa_r $ & $ 2\pi\times$1.6~MHz\\
\hline
$\kappa_{\phi}$ & $2\pi\times$32~kHz & $ \epsilon_L$ & $h\times1920$~GHz\\
  \hline  
\end{tabular}
	\caption{ {\bf Fitted circuit parameters.} $\delta E_J$ is the difference in Josephson energy between the two junctions of the SQUID, while $E_J$ labels the mean energy. Details of the fit procedure are presented in Sec.~\ref{sm:fits}. 
}
\label{tableparam}
\end{table}

\section{Activation and characterization of a high-order dissipation channel}
\label{sec:T4}

As detailed above, an adverse consequence of the strong non-linearity  of the memory is that the  memory dissipator $\mathcal{D}[\aop^4]$ cannot be activated with a single pump microwave. Combining pumps each addressing a specific transition $|n+4\rangle_a |0\rangle_b \leftrightarrow |n\rangle_a |1\rangle_b$ could in principle circumvent the problem. It is however technically challenging as the  amplitude and phase of each frequency component needs to be precisely adjusted, and was not attempted. We instead address a specific  transition, namely $|4\rangle_a  |0\rangle_b \leftrightarrow |0\rangle_a  |1\rangle_b $ and characterize the engineered dissipator $ \mathcal{D}[|0\rangle_a \langle 4|_a]$ so obtained.  \\

\subsection{Preparing a Fock state and measuring its occupation}
\label{sec:prepmeas}
In order to characterize the engineered dissipation channel, we need to  prepare the memory in each Fock state $|n\rangle_a$~\cite{note4} and to monitor the time decay of its population. Both the preparation and measurement are enabled by the same  non-linearity preventing us from addressing more than a single four-to-one photon transition. In principle, the former only requires to apply sequences of $\pi$-pulses on the $|k\rangle_a \leftrightarrow |k+1\rangle_a$ transitions for $k <n$, while the latter is enabled by the dispersive shift of the readout mode resonance frequency conditioned on $n$. However,   preparation and measurement of highly excited states are challenging due to the memory anharmonicity and the readout mode dispersive shift both tailing off with the level index $n$. We therefore limit ourselves to preparation and measurement of Fock states up to $n=4$. \\

In detail, the states are prepared with  pulse sequences combining $\pi$-pulses applied on the  single-photon transitions $|0\rangle_a \leftrightarrow |1\rangle_a$ and $|1\rangle_a \leftrightarrow |2\rangle_a$ (direct charge drive) and/or on the three-photon transitions $|0\rangle_a \leftrightarrow |3\rangle_a$ and $|1\rangle_a \leftrightarrow |4\rangle_a$ (flux drive of the ATS). The duration of these pulses being on par with the coherence time of the transitions, the preparation is imperfect. It is nonetheless sufficient to estimate the lifetime of each state, as detailed in the next section. As for the readout, the ground state can be  directly distinguished from excited states  through the pull exerted by excited states on the readout mode resonance frequency. However, their respective pulls saturate rapidly with increasing state index, making them hard to distinguish from one another. We  circumvent this shortcoming with the scheme depicted in Fig.~\ref{fig3}a. To estimate the occupation of 
state $|n\rangle_a$ ($n\geq1$), we  apply or not a  $\pi$-pulse on the $|n\rangle_a \leftrightarrow |n+3\rangle_a$ transition before probing the readout mode on resonance with a long pulse. Since the dynamics of the decay from $|n+3\rangle_a$ to the ground state is delayed compared to the decay from  $|n\rangle_a$, we detect a time-shifted signal in the reflected field $s_n^{ON}(t)$ detected after the application of the $\pi$-pulse with respect to the field $s_n^{OFF}(t)$ detected in absence of the pulse (see Sec.~\ref{sm:meas}). We integrate the detected fields with a time-envelope $\mathcal{E}(t)$ chosen to maximize the signal-to-noise ratio of the differential signal $S_n=\int_t \Re \Big( \big( s_n^{ON}(t)-s_n^{OFF}(t) \big) \mathcal{E}(t) \Big) \mathrm{d}t $ between these two situations. Since  the $\pi$-pulse  only modifies the memory state when $|n\rangle_a$ is occupied,  the integrated  signal $S_n$ is proportional to the occupation of state $|n\rangle_a$ at the instant when the pulse is delivered, even if population transfer is imperfect. Note that both for state preparation and detection, we favor  pulses applied to three-photon transitions to ones applied to single-photon transitions as the detuning between neighboring three-photon transitions is larger, allowing for stronger drives.\\

Before applying these preparation and measurement methods to characterize the memory population dynamics in presence of a pump, we report that some of the memory transitions---driven either during preparation or measurement, involving a single photon or three---were frequently and transiently observed  to shift or split (see Sec.~\ref{sec:TLS}). We attribute this effect to  hybridization of the memory to a bath of Two-Level-Systems (TLS). Such TLS baths, whose frequency landscape reconfigure on timescales of the order of a second, are frequently observed in superconducting circuits~\cite{oliver2013materials,muller2019towards}. While the exact coupling mechanism and density of TLS's was not characterized in our experiment,  at least one of the memory transitions was typically  affected by a TLS at any given time so that   waiting for the bath to reconfigure in a more favorable configuration (either passively or cycling the system to room temperature) was not a viable option. Instead, we use an off-resonant  flux drive of large amplitude  to shift the frequency of the memory transitions away from any resonant coupling with TLS's. The frequency of this drive was chosen to minimize decoherence induced on the memory, but still leads to  degradation of its energy relaxation  and dephasing times~\cite{note5}. The datasets presented in the following section were acquired in presence of such a drive.

\subsection{Characterization of the four-photon dissipation channel }

\begin{figure}[htbp]
		\centering
		\includegraphics[width=0.95\columnwidth]{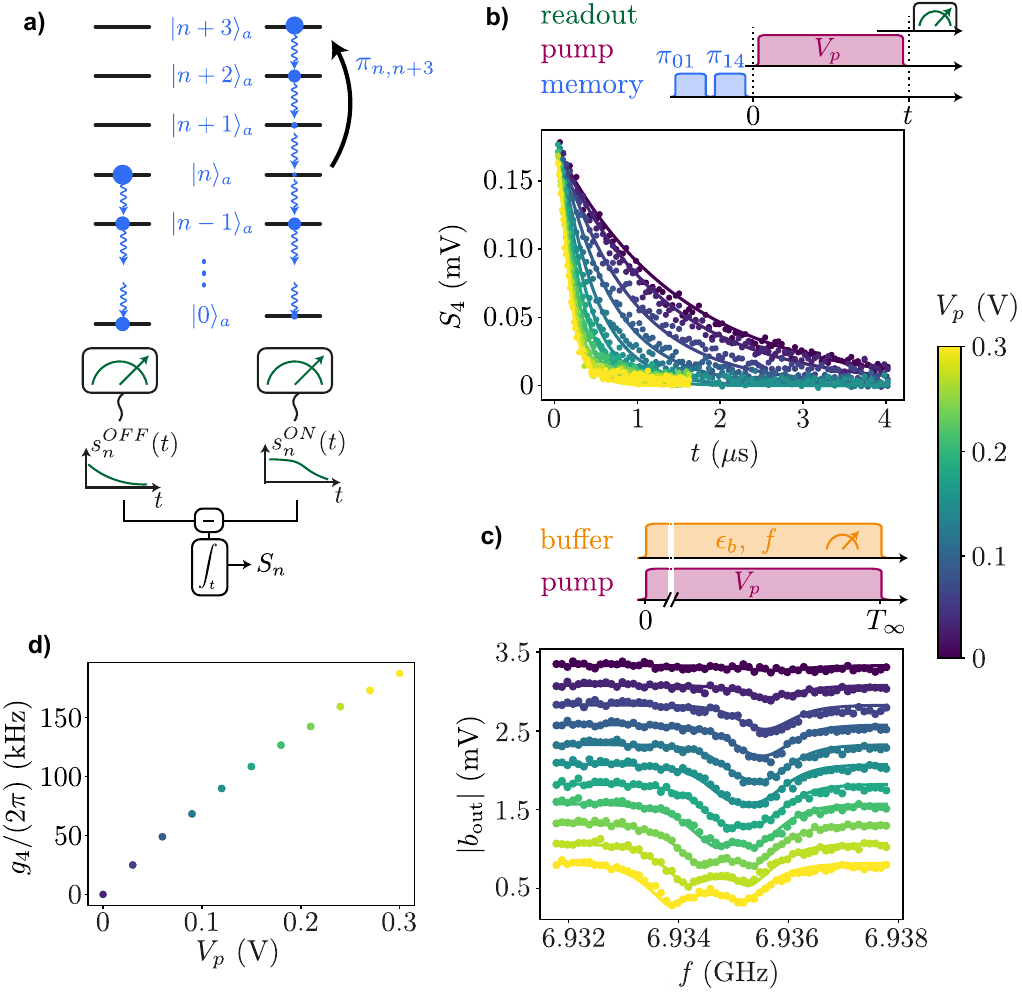}
		\caption{  {\bf Two-mode dynamics under the four-to-one photon pump a)} Principle of  the memory photon-number resolving measurement. Dispersive pull on the readout mode saturates for high energy states, preventing  their direct resolution. To estimate the occupation of  $|n\rangle_a$,   a $\pi$-pulse is applied on the $|n\rangle_a \leftrightarrow |n+3\rangle_a$ transition, modifying the dynamics of its relaxation to the ground state. Subsequently probing the readout mode on resonance, one obtains a  signal $s_n^{ON}$ which is time-shifted from the signal $s_n^{OFF}$ acquired in absence of pulse. Assuming that  $|n\rangle_a$ is the highest occupied state before the pulse is applied, the integrated differential signal is proportional to its occupation (see text). {\bf   b)} Time-decay of population in state $|4\rangle_a$ (measured up to an unknown proportionality factor) as a function of the pump amplitude (encoded in color and referenced by the voltage $V_p$ applied to the mixer generating the pump wave). The decay gets faster for increasing pump amplitude, and eventually deviates from an exponential law. {\bf c)} Buffer reflection spectrum acquired when simultaneously pumping the four-to-one photon transition (spectra  offset proportionally to the pump amplitude for readability). The buffer line shifts, broadens and eventually splits for increasing pump amplitudes, revealing a near-strong hybridization of the $|4\rangle_a|0\rangle_b$ and $|0\rangle_a|1\rangle_b$ states. Datasets (b) and (c) are reproduced by Lindblad master-equation simulations (lines), whose parameters are given in Table~\ref{tableparam}. For (c), we fit a displacement rate of $\epsilon_b=2\pi\times175$~kHz yielded by the resonant drive applied to the buffer. {\bf d)}  Four-to-one photon exchange rate against pump amplitude, extracted  from these simulations. It slightly deviates from the expected linear scaling.    }
		\label{fig3}
	\end{figure}

Armed with the preparation and measurement techniques described in the previous section, and after locating the frequency of the $|4\rangle_a  |0\rangle_b \leftrightarrow |0\rangle_a  |1\rangle_b $ transition  dressed by the pump  itself (see Sec.~\ref{sec:tuningPump}), we investigate  the time decay  of population prepared in  state $|4\rangle_a$ under pump (see Fig.~\ref{fig3}b). Clearly, the differential signal $S_4$---proportional to the occupation of $|4\rangle_a$---decays faster as the pump amplitude increases (here referenced by the voltage $V_p$ applied to the mixer generating the pump). For the lowest pump amplitudes ($V_p \leqslant 0.1~$V), the decay is exponential. However, for larger pump amplitudes,  we note a deviation  from an exponential law, indicating that reducing the joint dynamics of the memory and buffer modes to a single-mode Lindblad dynamics governed by the dissipator  $\mathcal{D}[|0\rangle_a \langle 4|_a]$ is no longer valid. We further investigate this regime with a complementary experiment in which, for the same range of pump amplitudes, we record the buffer reflection spectrum in continuous-wave  (see Fig.~\ref{fig3}c). As the pump amplitude increases, a dip in amplitude appears in the buffer line, which eventually splits into two almost separated lines. This behavior is expected at the onset of strong hybridization of the $|4\rangle_a|0\rangle_b$ and $|0\rangle_a|1\rangle_b$ levels, which occurs when $g_4 \sqrt{4!} \gtrsim \gamma$ with $\gamma$ the intrinsic width of the transition accounting both for homogeneous and inhomogeneous broadening. \\

These two datasets are quantitatively reproduced (lines in Fig.~\ref{fig3}b-c) by Lindblad master equation simulations modeling the memory and buffer interacting via the Hamiltonian~\eqref{eq:4to1}. The anharmonicity   and relaxation rate of each mode are independently calibrated and accounted for in these simulations. We leave as free fit parameters the value of $g_4$ for each pump amplitude and the strength of the buffer drive for the whole second dataset (see  Sec.~\ref{sec:fourtoonedynamics} for details of the fit procedures and assumptions behind this model). We also adjust in our model  a pure dephasing rate of $\kappa_{\phi}=2\pi\times 32$~kHz for the memory, close to the independently measured dephasing rate of the $|0\rangle_a \leftrightarrow |1\rangle_a$ transition. Finally, we fit a proportionality factor between the differential signal $S_4$ and the occupation of state $|4\rangle_a$. The fitted value of $g_4$ is reported in Fig.~\ref{fig3}d for each pump amplitude. It reaches $2\pi \times 185~$kHz for the strongest pump amplitude, corresponding to a normalized  amplitude  $|\epsilon(t)|=0.035$ with the convention used in the introduction. Note that the scaling of $g_4$ slightly deviates from the expected linear dependency against  pump amplitude at room temperature, which could be explained by the saturation of the mixer generating the pump pulses.\\

\begin{figure}[htbp]
		\centering
		\includegraphics[width=0.9\columnwidth]{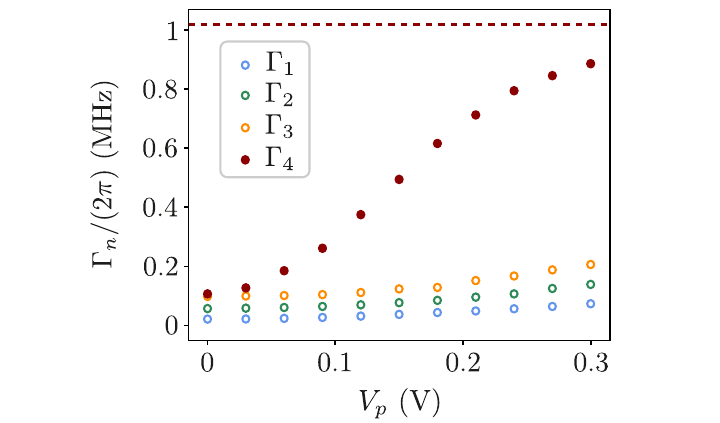}
		\caption{ {\bf Four-photon loss channel}. Decay rates $\Gamma_n$ of the first four excited states of the memory ($1\leq n\leq 4$) in presence of a pump of varying amplitude. $\Gamma_4$ is estimated through spectral analysis of the model Lindbladian used to reproduce the datasets presented in Fig.~\ref{fig3}b-c. Following an initial quadratic increase with respect to pump amplitude, it saturates near its maximum theoretical value at half the buffer damping rate (red dashed line). $\Gamma_n$ for $n\leq 3$ is directly obtained from an exponential fit of the time-decay of  population prepared in $|n\rangle_a$ (see Sec.~\ref{sec:gammaslow}). Their increase at large pump amplitude is not captured by our model.  }
		\label{fig4}
	\end{figure}

In order to quantitatively relate the largest fitted values of $g_4$ to four-photon dissipation rates, one needs to go beyond the simple model reduction allowed by adiabatic elimination of the buffer. We instead extract the lifetime of state $|4\rangle_a $ through spectral analysis of a reduced Liouvillian capturing the system dynamics, as detailed in Sec.~\ref{sec:fourtoonedynamics}. We report the extracted decay rate $\Gamma_4$ in Fig.~\ref{fig4} (filled red circles). Following an initial  quadratic increase of $\Gamma_4$ against pump amplitude---also predicted by the adiabatic elimination of the buffer---we observe a saturation of $\Gamma_4$ near its theoretical maximum value $\kappa_b/2$ set by the buffer damping rate~\cite{reglade2024quantum}. In this regime, engineered four-photon dissipation dominates the dynamics, inducing decay of population from state $|4\rangle_a$ at a rate that exceeds  that induced by native single-photon dissipation by a factor 10. As a sanity check, we also record, under the same conditions, the time-decay of populations prepared in states $|n\rangle_a$ for $n <4$ and report the corresponding decay rates $\Gamma_n$  in Fig.~\ref{fig4}. While these  should nominally not be affected by the pump (see Fig.~\ref{fig1}b), we observe a small but clear  increase  with pump amplitude. Such behavior was reported in similar parametrically pumped systems~\cite{leghtas2015confining,lescanne2020exponential}, and may stem from spurious near-resonant parametric processes~\cite{petrescu2020lifetime,venkatraman2024nonlinear,putterman2025preserving,carde2025flux}. In our case, we note that a two-to-one photon exchange process between the memory and the buffer, resonant at $2\omega_b -\omega_a$, is detuned by approximately 200~MHz from the target process. We estimate that this off-resonantly driven process results in an effective pump-dependent single-photon dissipation channel for the memory, contributing marginally to the observed degradation of the low-energy levels lifetime (dissipation rate in the $0-10$~kHz range for the considered pump amplitudes). In any case, the spurious increase of $\Gamma_n$ for $n<4$ remains an order of magnitude lower than the recorded increase of $\Gamma_4$.      \\

\section{Discussions}
\label{sec:discussion}

\subsection{High-order dissipators from high-order interactions}

In earlier experimental works realizing high-order dissipation channels, these were obtained from low-order interactions. Among other examples, autonomous parity recovery in a superconducting resonator  was achieved by combining dispersive interactions and a set of single and two-photon drives~\cite{gertler2021protecting},  autonomous stabilization of the GKP code  was achieved by combining Rabi interactions and sequences of single-photon drives~\cite{royer2020stabilization,de2022error,lachance2024autonomous}, and a four-to-two photon exchange interaction---which could be used to engineer a four-photon dissipation channel---was obtained by cascading off-resonant two-to-one photon exchange interactions in a Raman configuration~\cite{mundhada2019experimental,note6}. We argue that  this approach  can lead to the opening of spurious dissipation channels entailed by each low-order primitive involved in the generation of effective high-order interactions. For instance, in the Raman scheme, finite detuning of the pumps to two-to-one photon transitions can excite levels which should nominally be only virtually excited. Such excitations would directly lead to a logical flip of the encoded cat-qubit if the engineered dissipation channel were employed toward this goal. \\

In principle, low-order interactions can still be safely used leveraging  a biased-noise auxiliary mode~\cite{puri2018stabilized, siegele2023robust,ding2025quantum}, error-transparent interactions~\cite{vy2013error,rosenblum2018fault,reinhold2020error,ma2020path, putterman2024hardware} or carefully controlling the noise spectrum of the dissipative environment~\cite{putterman2022stabilizing}, but the experimental implementation of these ideas is a major challenge. On the other hand, in our approach,  the desired high-order dissipation is directly obtained from high-order interactions, acting on the memory mode via operators which commute with the logical operators of the encoded qubit (the interaction Hamiltonian ${\aop}^4 \bop^{\dagger} + \mathrm{h.c.}$ commutes with the four-legged cat code stabilizers and its Pauli operators). Therefore, even in presence of noise affecting the buffer, evolutions induced by these interactions do not affect the encoded information. Moreover, it is straightforward to adapt the circuit and method presented in this work to activate even higher-order dissipators.

\subsection{Parameter regime}

As discussed in this paper introduction, the memory and buffer phase fluctuations across the ATS should be large in order for the rate of the activated four-to-one photon exchange  to be significant. However, as detailed in Sec.~\ref{sec:targetparams}, the memory phase fluctuations cannot be chosen arbitrarily large when stabilizing four-legged cats. Indeed, for a pump $\epsilon(t)=\xi e^{i(4 \omega_a - \omega_b)t} + \mathrm{c.c}.$ resonant with the four-to-one photon transition, the RWA yields  corrections to the exchange interaction stemming from high-order terms in the ATS \textit{sine} Hamiltonian (see Eq.~\eqref{eq:HATS}):
\begin{equation}
\HH_4=\hbar g_4 :P_a(\adag \aop)P_b (\bdag \bop)~ \aop^4 \bdag: ~+~ \mathrm{h.c.}
\label{eq:4to1full}
\end{equation}
In this expression, $:\_:$ denotes the normal ordering of an operator, $g_4=2 \frac{E_J}{\hbar 4!} \xi e^{-(\varphi^{{ZPF}^2}_a+\varphi^{{ZPF}^2}_b)/2}\varphi^{{ZPF}^4}_a \varphi^{ZPF}_b $ and $P_a$ and $P_b$ are polynomials boiling down to unit factors if the SQUID  potential is truncated to leading order. These corrections yield a state-dependent four-to-one photon exchange rate which can result in instabilities when stabilizing a cat with size $\alpha \gtrsim 1/\varphi_a^{ZPF}$~\cite{rojkov2024stabilization}. We note that this effect also arises for two-legged cats, and is one of the limiting factors to their stabilization rate~\cite{note7}. On the upside, it was recently proposed to leverage the same effect to increase the number of attractors in phase-space---thus stabilizing higher-dimension cat codes---or to stabilize more complex rotation-symmetric codes with advantageous error-correcting capabilities~\cite{rojkov2024stabilization}.\\

In circuit design, we target the value of $\varphi_a^{ZPF}=0.45$, for which numerical simulations showed that cat states containing more than 10 photons could be stabilized with no visible impact from high-order corrections to the four-to-one photon interaction. Note that owing to the exponential scaling of the cat bit-flip time against the cat size, this mean photon number is typically sufficient for cats to reach macroscopic bit-flip times~\cite{reglade2024quantum}. For the buffer, we target $\varphi_b^{ZPF}=0.55$ (significantly larger than the measured value). This is lower than the unit value  maximizing the four-to-one photon exchange rate $g_4$. However, reaching this regime would have required to increase significantly the mode impedance for only a modest increase of the exchange rate.  As for the readout mode, its phase fluctuations across the ATS are an order of magnitude lower  owing to a weaker participation of the ATS. This mode was originally intended to provide a photon number resolving measurement by activating a term of the form $ \aop^{\dagger} \aop (\rop + \rop^{\dagger})$ (sometimes referred to as "longitudinal" coupling~\cite{touzard2019gated}) by pumping the ATS at the readout mode frequency~\cite{ reglade2024quantum}. In this context, setting $\varphi_r^{ZPF}\ll \varphi_a^{ZPF}$ ensures that unwanted cubic interactions of readout photons, which are also activated by the pump, are significantly weaker than the desired term. In practice, the memory and readout modes were found to be dispersively coupled due to circuit non-idealities  such that dispersive measurement was sufficient to characterize the  state of the memory (see Sec.~\ref{sec:prepmeas}).

\subsection{Spurious non-linearities and  mitigation measures}

In the introduction of this paper, we discussed how an idealized circuit comprising a SQUID placed in series with two resonators  enabled dissipation engineering while preserving the target mode linearity. Various non-idealities can break this property and induce anharmonicity. We here outline how the observed memory anharmonicity can be traced back to stray inductance in the SQUID and give possible mitigation measures for future experiments. We refer the reader to Sec.~\ref{sec:smNL} for a detailed analysis of the various non-linear mechanisms arising in our circuit. \\

The memory anharmonicity pattern presented in Fig.~\ref{fig2}d significantly differs from that of a Kerr oscillator, for which a constant anharmonicity is expected. It is characteristic of a strongly non-linear, \textit{i.e.} high-order, perturbation of weak amplitude. Such a perturbation  impacts low-energy states but is overcome by the quadratic part of the Hamiltonian at high energy. This disqualifies the chain of Josephson junctions forming the circuit superinductors---expected to induce a negative quartic non-linearity---as the dominant source of non-linearity. Imbalance of the SQUID junctions---often responsible for spurious nonlinearities in circuits comprising an ATS~\cite{lescanne2020exponential}---is here negligible and cannot explain the magnitude of the anharmonicity. On the other hand, the anharmonicity pattern is quantitatively reproduced when fitted with a small second harmonic contribution to the potential of each Josephson junction forming the SQUID. Including this contribution, the ATS Hamiltonian~\eqref{eq:HATS} becomes
\begin{equation}
\begin{aligned}
\HH_J(t)&=- E_J \cos\big(\varphiop - \frac{\Phi_1}{\varphi_0}\big) - E_J \cos\big(\varphiop + \frac{\Phi_2}{\varphi_0}\big)\\
&\qquad- E_{J_2} \cos\big(2(\varphiop - \frac{\Phi_1}{\varphi_0})\big) - E_{J_2} \cos\big(2(\varphiop + \frac{\Phi_2}{\varphi_0})\big)\\
&=2E_J~\sin\big(\epsilon(t)\big)\sin(\varphiop)\\
& \qquad - 2E_{J_2}~\cos\big(2\epsilon(t)\big)\cos(2\varphiop)
\end{aligned}
\end{equation}
We immediately see that the added contribution does not cancel at the ATS working point, even in absence of pump ($\epsilon=0$). Moreover, 
the vacuum phase fluctuations of each mode are effectively doubled and approach unity for this term, such that the \textit{cosine} cannot be approximated by a low-order polynomial. For the fitted value $E_{J_2}=E_J / 500$, we thus find a strongly non-linear perturbation of weak amplitude as sought. While this term has a large impact on the memory-mode anharmonicity, raising it by roughly two orders of magnitude, it has no significant effect on the four-to-one exchange mechanism we aim to activate.  \\

A known mechanism  renormalizing the Josephson energy with a second harmonic stems from small stray inductances in the arms of the SQUID~\cite{smith2023spectral,putterman2025preserving}. For series inductance with inductive energy $\epsilon_L \gg E_J$ (see Fig.~\ref{fig2}a), the amplitude of the  second harmonic is $E_{J_2}=-E_J^2/(4\epsilon_L)$. In our case, the observed amplitude corresponds to an inductance value of $\varphi_0^2/\epsilon_L=85$~pH  matching our estimate for the geometric inductance of the tracks of the SQUID (see Sec.~\ref{sec:smNL}). Note that an alternative mechanism yielding harmonics in the Josephson potential---stemming from pinholes in the Josephson junctions oxide barrier---was recently proposed to explain reported anomalies in transmon energy level structures~\cite{willsch2024observation}. The amplitude of the second harmonic found in our experiment  is smaller than the values reported in that work  and can be entirely attributed to stray circuit inductances. This different behavior may be explained 
by a thicker oxide barrier in our junctions.\\

In a future iteration of the experiment, stray inductances could be significantly reduced with simple design changes. Indeed, the rather large SQUID loop in the present experiment (dimensions $\sim50\times50~ \mathrm{\mu m}^2$) was chosen to accommodate at its center the long chain of junctions forming the shunt inductor of the ATS. However, physically placing this chain in parallel but \textit{outside} of the SQUID loop would yield an equivalent circuit---simply requiring to reallocate bias fluxes between loops---while allowing to reduce the SQUID perimeter down to $\sim10~\mu$m as routinely achieved in flux qubits~\cite{braumuller2020characterizing}. Combined with a reduction of the junctions energy to $E_J=h\times 5~$GHz,  we estimate that the anharmonicity of the memory mode---scaling quadratically with $E_J$---will become sufficiently low to be overcome by the four-photon dissipation---scaling linearly with $E_J$ for a well chosen value of $\kappa_b$. \\

Finally, we note that while our experiment is particularly sensitive to  non-linearity resulting from various circuit non-idealities (imbalance of the SQUID, non-linearity of the superinductors, harmonics in the Josephson potential), these mechanisms also exist in experiments aiming at dissipatively stabilizing other bosonic codes with similar circuits. While these experiments typically operate in a lower impedance regime in order to to suppress the anharmonicity inherited by the memory (dominant non-linear term scaling as ${\varphi_a^{ZPF}}^4\propto Z_a^2$), the rate of desired interactions  decreases in the same limit. Therefore, even when targeting lower-order dissipation channels scaling more favorably with the modes impedance, the optimal parameter choice results from a tradeoff between the rate of engineered dissipation and the strength of spurious non-linearities. Our platform is an ideal testbed to investigate this tradeoff.  Our experiment also highlights that the anharmonicity inherited from different mechanisms cannot be boiled down to the amplitude of a single Kerr-like term. As a result, compensating the effect of one mechanism with another as was recently realized~\cite{putterman2025preserving} can only lead to imperfect cancellation. An intriguing exception detailed in Sec.~\ref{sec:smNL} is the possibility to perfectly cancel, in the RWA, the anharmonicity induced by a weak but non-zero imbalance of the SQUID with a dynamical mechanism involving off-resonant pumps. In future experiments, this scheme should prove instrumental  in order not to rely on perfect SQUID symmetry as was achieved in the present experiment.

\section{Conclusion and outlook}
\label{sec:conclusion}

We  demonstrate the engineering of a  four-photon dissipation channel whose rate far exceeds the intrinsic single-photon loss rate of the target oscillator. Our approach relies on a near Kerr-free Josephson mixing element embedded in a high-impedance environment, mediating genuine high-order interactions while inducing only residual non-linearity on the involved modes. Even though  this non-linearity is larger than the engineered dissipation rate in our experiment---preventing us from dissipating quartets of photons from more than a specific oscillator state---we show that it dominantly originates from a renormalization of the SQUID inductive potential by stray inductances in  its tracks. We  thus estimate that simple design changes would decrease the circuit non-linearity down to a level where  it can be overcome by the four-photon dissipation, enabling the dissipative stabilization of  a four-legged cat manifold. \\

Looking forward, stabilization of the four-legged cat manifold could be alternated with parity checks~\cite{sun2014tracking,ofek2016extending,hu2019quantum,ni2023beating} or autonomous parity recovery operations~\cite{gertler2021protecting} to provide a first-order error-corrected qubit. However, tying up with the discussion given in the introduction, these techniques usually rely on low-order interactions to probe the infinite-order parity operator, with the risk of opening spurious dissipation channels corrupting the encoded qubit. Alternatively, one could increase further the memory impedance presented to the SQUID to enable a genuine parity detection compatible with the manifold stabilization~\cite{cohen2017degeneracy}. In a similar regime, one could consider stabilizing Schrödinger cat states with more than four components via  wave-mixing processes of even higher order than considered in this work, or combining several such processes to stabilize GKP grid states ~\cite{sellem2025dissipative,kolesnikow2024gottesman,nathan2024self}. These examples illustrate the potential of driven high-impedance circuits for autonomous and robust quantum error-correction.\\

\section*{Acknowledgments}
The authors thank S. Jezouin, R. Lescanne, A. Murani, and  W.C. Smith for fruitful discussions on circuit design, J. Palomo, A. Pierret, M. Rosticher and J.-L. Smirr  for assistance in microfabrication, and M. Mirrahimi, P. Rouchon and L.-A. Sellem for valuable discussions on dissipation engineering.  Our circuit was fabricated within the consortium Salle Blanche Paris Centre and at  the Collège de France. This work has been funded by the European Research Council (ERC) under the European Union’s Horizon 2020 research and innovation program (project Dancingfool, grant agreement No. 101042304 and project Q-feedback grant agreement No. 884762),  by the French ANR-22-PETQ-0003 grant and by the ANR-22-PETQ-0006 grant under the ‘France 2030' plan.

\section*{Data availability}
The experimental data and numerical simulations presented in this manuscript are available from the corresponding author upon request.

\section*{Author contributions}

A. V. designed and fabricated the sample, performed the experiment and analyzed the data under the supervision of P.C.-I. B.B. and L.L. performed complementary calibration measurements and supporting calculations. P.C.-I. and Z.L.  conceived the project. M.V., A.D. and P.M. designed the sample holder used in the experiment. All authors contributed in preparing the manuscript.

\section*{Correspondence}

Any correspondence should be addressed to philippe.campagne-ibarcq@inria.fr.

\appendix




\section{Circuit design and measurement setup}

\subsection{Target parameters and circuit design}
\label{sec:targetparams}
Our sample was designed to dynamically stabilize a four-legged cat qubit. The choice of circuit parameters was  motivated by this goal. Here, we give a brief summary of the target dynamics stabilizing a four-legged cat and of the target parameter regime.\\

In an ideal setting, dynamical stabilization of the four-legged cat manifold (spanned by the four coherent states $|+\alpha\rangle,~|+i\alpha\rangle,~|-\alpha\rangle,~|-i\alpha\rangle$) is achieved through the interaction Hamiltonian
\begin{equation}
\HH_{4}=\hbar g_4 (\aop^4-\alpha^4)\bdag+\mathrm{h.c.}
\label{eq:H4}
\end{equation}
with a lossy buffer $b$ (single-photon dissipation rate $\kappa_b$), here expressed in the rotating frame of both modes~\cite{mirrahimi2014dynamically}. When $\kappa_b \gg g_4$, this two-mode dynamics reduces to a single-mode Lindblad-type dynamics 
\begin{equation}
\dot{\rhoo} = \gamma_4 \mathcal{D}[\aop^4-\alpha^4]\rhoo
\label{eq:gamma4}
\end{equation}
with $\gamma_4=4g_4^2/\kappa_b$, and the coherent states spanning the cat manifold are attractors of the dynamics. Note that this property remains true in the regime $g_4 \gtrsim \kappa_b$, even though the two-mode dynamics is no longer accurately captured by \eqref{eq:gamma4}~\cite{robin2024convergence}. In Eq.\eqref{eq:H4}, the term $\alpha^4 \bdag +\mathrm{h.c.}$ is trivially implemented with a resonant drive on the buffer, so that we now focus on the engineering of the more challenging four-to-one photon exchange interaction. \\

In practice, we engineer this interaction using a pumped ATS participating in the memory and buffer (see Fig.~\ref{fig1}). After Foster decomposition of the linear circuit connected to the ATS~\cite{nigg2012black, lescanne2020exponential}, we reach the Hamiltonian:
\begin{equation}
\begin{split}
\HH(t)=&~\hbar \omega_a \adag \aop + \hbar \omega_b \bop \bdag \\
&~+ 2E_J\epsilon(t) \sin\big(\varphi^{ZPF}_a (\aop+ \adag )+\varphi^{ZPF}_b (\bop+ \bdag )\big)
\end{split}
\label{eq:simpleATS}
\end{equation}
where we have neglected the weakly participating readout mode and considered an ideal circuit  for simplicity (see Sec.~\ref{sec:smNL} for an analysis of non-idealities). Note that this expression is valid in the limit $|\epsilon(t)|\ll 1$ as it results from a linearization of $\sin(\epsilon(t))$~\cite{lescanne2020exponential}.\\

We now detail the constraints on the parameters of this Hamiltonian and motivate our choice of parameters.

\subsubsection*{Zero-point fluctuations of modes across the ATS}

We first move to the rotating frame of both modes ($\aop \rightarrow \aop e^{-i\omega_a t }$ and $\bop \rightarrow \bop e^{-i\omega_b t }$). Using Baker-Campbell-Hausdorff formula, we can normal order exponential functions following
\begin{widetext}
\begin{equation}
\begin{split}
    e^{i \varphi^{ZPF}_a(\aop e^{-i\omega_a t } + \adag e^{+i\omega_a t }) }&=e^{-\frac{{\varphi_a^{ZPF}}^2}{2}} \sum_{n,m}  \frac{(i{\varphi_a^{ZPF}})^{n+m}}{n! m!} e^{i(n-m)\omega t} \adag^n \aop^m\\
    e^{-i {\varphi_a^{ZPF}}(\aop e^{-i\omega_a t } + \adag e^{+i\omega_a t })}&=e^{-\frac{{\varphi_a^{ZPF}}^{2}}{2}} \sum_{n,m}  \frac{(-i{\varphi_a^{ZPF}})^{n+m}}{n! m!} e^{i(n-m)\omega t} \adag^n \aop^m
\end{split}
\end{equation}
\end{widetext}
with similar formula for $b$.\\

If the system is pumped at the  frequency $\epsilon_p(t)=\xi e^{i(4 \omega_a - \omega_b)t} + \mathrm{c.c}.$ and after selecting resonant terms assuming $\omega_a$, $\omega_b$ incommensurate, we get in the RWA the static Hamiltonian
\begin{equation}
\tilde{\HH}_4=\hbar g_4 :P_a(\adag \aop)P_b (\bdag \bop)~ \aop^4 \bdag: ~+~ \mathrm{h.c.}
\label{eq:Hwithhigher}
\end{equation}
where $:\_:$ denotes the normal ordering of an operator, $g_4=2\frac{E_J}{\hbar 4!} \xi {\varphi_a^{ZPF}}^{4} {\varphi_b^{ZPF}} e^{-\frac{{\varphi_a^{ZPF}}^{2}}{2}-\frac{{\varphi_b^{ZPF}}^{2}}{2}}$ and
\begin{equation}
\begin{split}
P_a(X)= 4!\sum_{n \geqslant 0} (-1)^{n} \frac{{\varphi_a^{ZPF}}^{2n}}{n! (n+4)!}  X^n   \\
P_b(X)= \sum_{n \geqslant 0} (-1)^{n} \frac{{\varphi_b^{ZPF}}^{2n}}{n! (n+1)!}  X^n  
\end{split}
\end{equation}
Compared to the target Hamiltonian \eqref{eq:H4}, higher-order terms appear (corresponding to terms indexed by $n\geqslant 1$ in $P_a$ and $P_b$). These terms renormalize the four-to-one photon exchange rate between states $|k+4\rangle_a |l\rangle_b$ and $|k\rangle_a |l+1\rangle_b$ for $(k,l)\neq (0,0)$. Note that in our experiment, the modes are non-linear so that  the pump is not resonant with any of these transitions. Therefore, the system dynamics is the same as that yielded by  \eqref{eq:H4} (see Sec.~\ref{sec:smNL} and Sec.~\ref{sec:fourtoonedynamics}).\\

The four-to-one photon exchange rate $g_4$ in \eqref{eq:Hwithhigher} depends strongly on ${\varphi_a^{ZPF}}$, which should thus be maximized. However, when stabilizing a four-legged cat in a linear system, high-order terms in the sine expansion can have a a significant impact on the dynamics when $\varphi^{ZPF}_a\alpha \gtrsim 1$ such that the sine potential cannot be truncated at leading order. In practice, we saw in numerical simulations that beyond this limit, the cat blobs are distorted and eventually become unstable (not shown). In particular, we found that one should limit ${\varphi_a^{ZPF}} \leq 0.5$ to stabilize a cat containing up to 10 photons on average. As for high-order terms in $P_b$, they should have no impact on the system dynamics when it is close to its steady state~\cite{sellem2025dissipative}. However, the prefactor $e^{-\frac{{\varphi_b^{ZPF}}^{2}}{2}}$ limits the four-to-one photon exchange rate which reaches a maximum for $\varphi^{ZPF}_b=1$. In practice, designing a circuit with modes whose vacuum fluctuations across the ATS approach 1 is challenging and one gains only marginally beyond $\varphi^{ZPF}_b\sim 0.5$, so that we targeted in design ${\varphi_a^{ZPF}}=0.45$ and ${\varphi_b^{ZPF}}=0.55$.\\

As for the readout mode which we have neglected so far, we targeted ${\varphi_r^{ZPF}}=0.05$ which was deemed sufficient to perform photon-number resolved measurements of the memory leveraging the non-linearity of the ATS~\cite{note8} while ensuring that the memory anharmonicity remained smaller than its linewidth at all the ATS DC bias points. This last property is convenient to characterize the system by monitoring the readout mode frequency as a function of bias fluxes (see Fig.~\ref{fig2}).

\subsubsection*{Modes resonance frequencies}

The mode resonance frequencies are not particularly constrained toward the goal of stabilizing a four-legged cat. However, the resonance frequencies should be sufficiently separated  that the feedlines can be filtered to obtain a strong single-photon dissipation on the buffer while retaining a large intrinsic quality factor for the memory. Moreover, we generate the pump microwave with standard microwave equipment (in the $1-20$~GHz range) and  should avoid frequency collisions with processes resonant at $n\omega_a+m\omega_b$ with $(n,m)\neq (4,-1)$. In practice, we targeted $\omega_a=2\pi \times 4.5~$GHz and $\omega_b=2\pi \times 8~$GHz. For the resonance frequency of the readout mode, we aimed at the center of the working range of our parametric amplifier (see Fig.~\ref{fig:wiring}).\\

We adjusted the design of the three-pad structure and shunt inductance values (see Fig.~\ref{fig1}) to reach these frequencies and phase fluctuations of each mode across the ATS in finite-element electromagnetic simulations (performed with \textit{Ansys HFSS} and using the method of energy participation ratios~\cite{minev2021energy}).

\subsubsection*{Josephson junction energy}

The four-to-one photon exchange rate $g_4$ scales linearly with $E_J$ so that naively, one would aim for large $E_J$ values to increase the engineered dissipation rate. However, spurious non-linearities stemming from the tunneling of pairs of Cooper pairs through the  SQUID  scales with $E_J^2$ (see Sec.~\ref{sec:smNL}). Through numerical simulations, we estimated that for the target values of ${\varphi_a^{ZPF}}$ and ${\varphi_b^{ZPF}}$ given above, the smallest Josephson junctions that we could fabricate while ensuring symmetry of the SQUID at the $\sim 1~\%$ level   would yield a sufficiently large four-to-one photon exchange rate  to enable  inflation of a coherent Schrödinger cat state from vacuum~\cite{mirrahimi2014dynamically}). We therefore targeted this minimum value of $E_J=h\times$5~GHz.\\

The target parameters listed above are summarized in Table~\ref{tableSM}. The value of the buffer single-photon dissipation rate was optimized in numerical simulations to minimize the inflation time of Schrödinger cat states (for $|\alpha|^2\sim 4-10$).

\begin{table}
\begin{tabular} {|c|c|c|}
\hline
\rowcolor{lightgray}
Parameter & Target value &  Measured value   \\
\hline
${\varphi_a^{ZPF}}$ (SQUID) & 0.43& 0.405\\
\hline
  $\varphi_b^{ZPF}$ (SQUID) & 0.51 & 0.312\\
  \hline
    $\varphi_r^{ZPF}$ (SQUID) & 0.05 & 0.044\\
    \hline
${\varphi_{a, \mathrm{series}}^{ZPF}}$ (series inductor) & 0.0012& 0.0037\\
\hline
  $\varphi_{b, \mathrm{series}}^{ZPF}$ (series inductor) & 0.0014 & 0.0028\\
  \hline
    $\varphi_{r,\mathrm{series}}^{ZPF}$ (series inductor) & 0.00014 & 0.0004\\
    \hline
$\omega_a$ & $2\pi \times 4.5$~GHz & $2\pi \times 4.13$~GHz\\    
\hline
$\omega_b$ & $2\pi \times 8$~GHz & $2\pi \times 6.94$~GHz\\  
\hline
 $\omega_r$ & $2\pi \times 7.5$~GHz & $2\pi \times 7.45$~GHz\\   
  \hline    
$E_J$ & $h\times$5.6~GHz & $h\times$17.7~GHz \\
  \hline  
$\kappa_b$ & $ 2\pi\times$10~MHz & $ 2\pi\times$2.05~MHz\\

  \hline  
\end{tabular}
	\caption{ {\bf Targeted and measured system parameters.} The phase vacuum fluctuations of each mode are given across the whole SQUID and across the inductors in series with the Josephson junctions (see  Fig.~\ref{fig2}a). These inductors account  for the inductance of the SQUID arms in our model. As detailed in Sec.~\ref{sec:smNL}, their inductance is either estimated from electromagnetic simulations (for the target values, estimated inductance 80~pH) or from a fit of the memory anharmonicity pattern (measured value, estimated inductance 85~pH).   Note that the estimated Josephson  energy in our experiment is three times larger than its target value, which explains that a larger fraction of the modes' phase fluctuations is allocated to the series inductors for the measured values than for the target values. This stronger participation of the series inductors entails a stronger  non-linearity of  all modes~ than anticipated (see Sec.~\ref{sec:smNL}). The achieved  dissipation rate for the buffer is also significantly smaller than its nominal values, which limits the  engineered four-photon dissipation rate (see Fig.~\ref{fig4}).
}
\label{tableSM}
\end{table}

\subsection{Fabrication process}
\label{sm:fab}

\textit{Wafer preparation}: The circuit is fabricated from a 430 $\mu$m-thick wafer of 0001-oriented, double-side epipolished Sapphire C. The sapphire wafer is initially cleaned through a stripping process in a reactive ion etching (RIE) machine, after which it is loaded into a sputtering system. After one night of pumping, we initiate an argon milling cleaning step, followed by the sputtering of 120 nm of niobium. Subsequently, we apply a protective layer of optical resist (S1805), dice the wafer and clean the small chips in solvents. \\

\textit{Circuit patterning}: We spin optical resist (S1805) and pattern the large features (ground plane, feedlines and readout resonator) using a laser writer. After development (MF319), we rinse in de-ionized water for 1 min, and etch the sample in SF6 with a 20 s over-etch. Finally, the sample is cleaned for 10 min by sonication in acetone at $50^{\circ}$ C.\\

\textit{Junction patterning}: Next, we apply a bilayer of methacrylic acid/methyl methacrylate [MAA EL13] and poly(methyl methacrylate) (PMMA E3), capped with a layer of conductive resist Electra 92 to evacuate charges during lithography. The entire circuit  is patterned in a single electron-beam lithography step. The Electra 92 is removed in deionized water. Subsequently, the development takes place in a 3:1 isopropyl alcohol (IPA)/water solution at $6^{\circ}$ C for 90 s, followed by 10 s in IPA.\\

\textit{Aluminum deposition}: The chip is then loaded in an e-beam evaporator. We start with a thorough argon ion milling for 2~min at $\pm30^{\circ}$ angles. We then evaporate 35 nm and 70 nm of aluminum, at $\pm30^{\circ}$ angles, separated by an oxidation step in 100~mbar of pure oxygen for 10 min. After evaporating the two layers, the sample is oxidized for 5~min in 100~mbar of oxygen to provide a controlled oxide cap.\\

\textit{Sample mounting}: The chip was subsequently glued with PMMA onto a PCB, wire-bonded and mounted into a sample holder. The device was then thermally anchored to the base plate of a Bluefors dilution refrigerator, surrounded by three concentric cans for magnetic and infrared shielding (outer: cryoperm, middle: aluminum, inner: copper, see Fig.~\ref{fig:wiring}). \\

We here note that the energy of each junction forming the SQUID is 3 times larger than its design value (see Table~\ref{tableSM}), indicating that something unexpected occurred in fabrication. While we cannot identify for certain the origin of the problem, the fact that the memory and buffer mode frequencies are off by only 10~\% (indicating that the superinductances have a value 20~\% higher than intended) whereas the inductance of the SQUID junctions are lower than intended by 60 \% points towards a mechanism that affects more strongly junctions of small area than larger ones. We thus suspect that the electron-beam was poorly focused during patterning.

\label{sec:fab}
\subsection{Circuit layout details and measurement setup}
\label{sm:detailcrystal}

\begin{figure}[htbp]
		\centering
		\includegraphics[width=1\columnwidth]{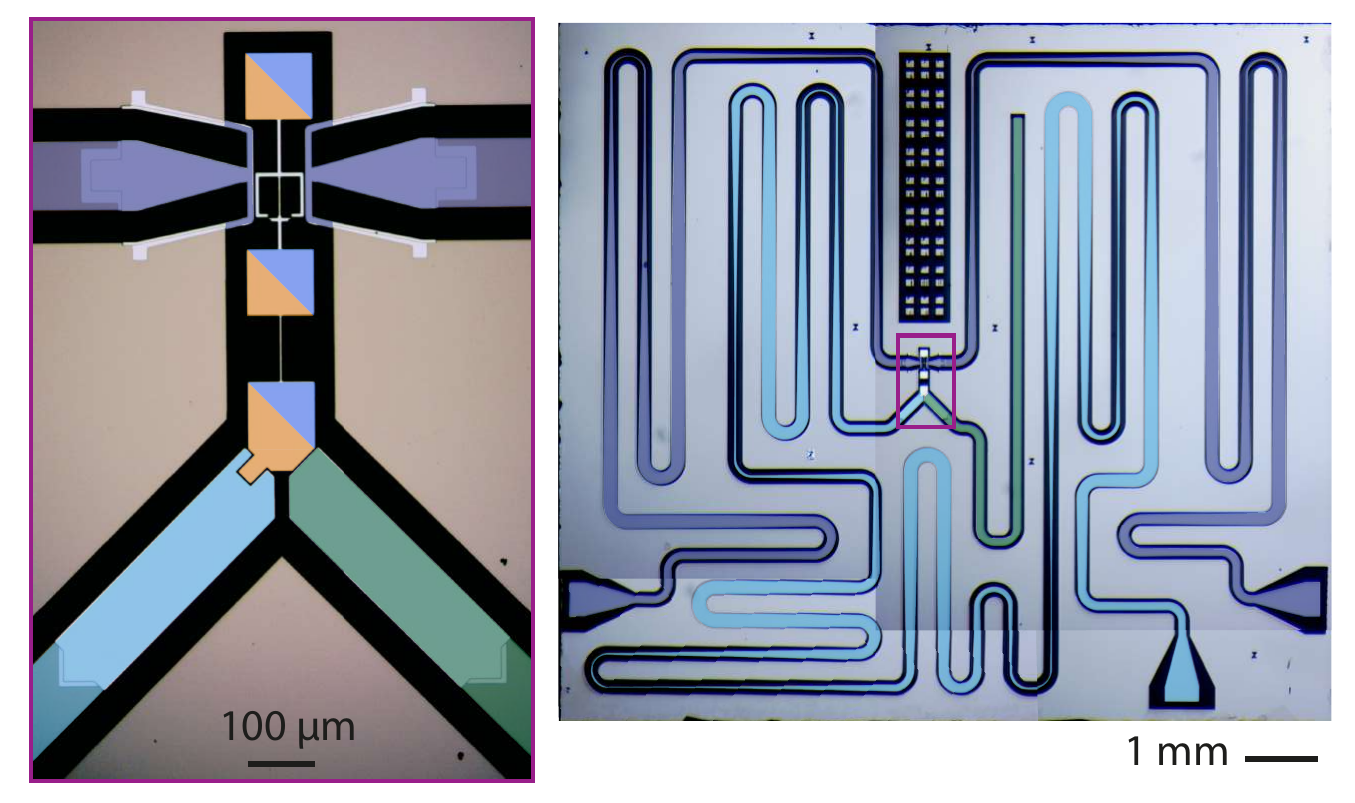}
		\caption{{\bf Optical microscope pictures of the circuit (false colors)}. The three-pad structure supporting the memory and buffer modes is visible in blue and orange in the zoomed-in picture on the left (corresponding to the purple rectangle in the zoomed-out picture on the right). The readout resonator is colored in green. The width of the center track of the  charge line (light blue) and that of the flux lines (mauve) is sinusoidally modulated  to prevent radiative decay of the memory (see Fig.~\ref{fig:crystals}). The lengths of the flux lines are not exactly matched.}
		\label{fig:circuit}
	\end{figure}

\subsubsection*{Circuit layout and filtering of the feedlines}

An optical microscope image of our circuit is shown in Fig.~\ref{fig:circuit}. The core part of the circuit (zoomed-in picture), comprising the three-pad structure hosting the memory and buffer, the ATS and the termination of the feedlines and readout resonator, is made out of evaporated aluminum. The rest of the circuit (zoomed-out picture), comprising the ground plane, the feedlines and the readout resonator is made out of sputtered niobium. Most of the chip is covered by long meandering coplanar waveguide (CPW) transmission lines. The width of the center track of the lines is periodically modulated to limit radiative decay of the memory, which we now detail.\\

 A 1D microwave transmission line is characterized by its inductance $l$ and capacitance $c$ per unit length. If these electrical properties  are periodically modulated with a spatial period $2\pi/\eta$, one obtains a microwave photonic crystal featuring a gap in its energy spectrum. More specifically, for a small modulation amplitude $\epsilon \ll 1$ of both the inductance per unit length $l(x)=l(1+\epsilon \cos(\eta x))$ and the capacitance per unit length $c(x)=c (1+\epsilon \cos(\eta x))^{-1}$---yielding a constant wave propagation velocity $v=1/\sqrt{lc}$ and a modulated   characteristic impedance $Z(x)=Z(1+\epsilon \cos(\eta x))$---the gap is centered at   $\omega_G= v \eta /2 $ and its width  increases with $\epsilon$~\cite{taguchi2015mode,note9}. A consequence of this gapped spectrum is that the line  behaves as a stopband filter for incoming microwave signals, whose in-gap isolation increases exponentially with the number of modulation periods. Note that  \emph{Stepped impedance} filters~\cite{pozar2011microwave}, which are the analog of  \emph{Bragg mirrors}  used in optics, were previously employed to this same end~\cite{bronn2015broadband}. They may be seen as  photonic crystals with multiple spatial frequencies $(2n+1)\eta$ such that stopbands appear in their transmission at $(2n+1)v\eta/2$. Photonic crystals were also employed in traveling-wave parametric amplifiers to enable phase-matching~\cite{planat2020photonic}.\\

\begin{figure}[htbp]
		\centering
		\includegraphics[width=0.7\columnwidth]{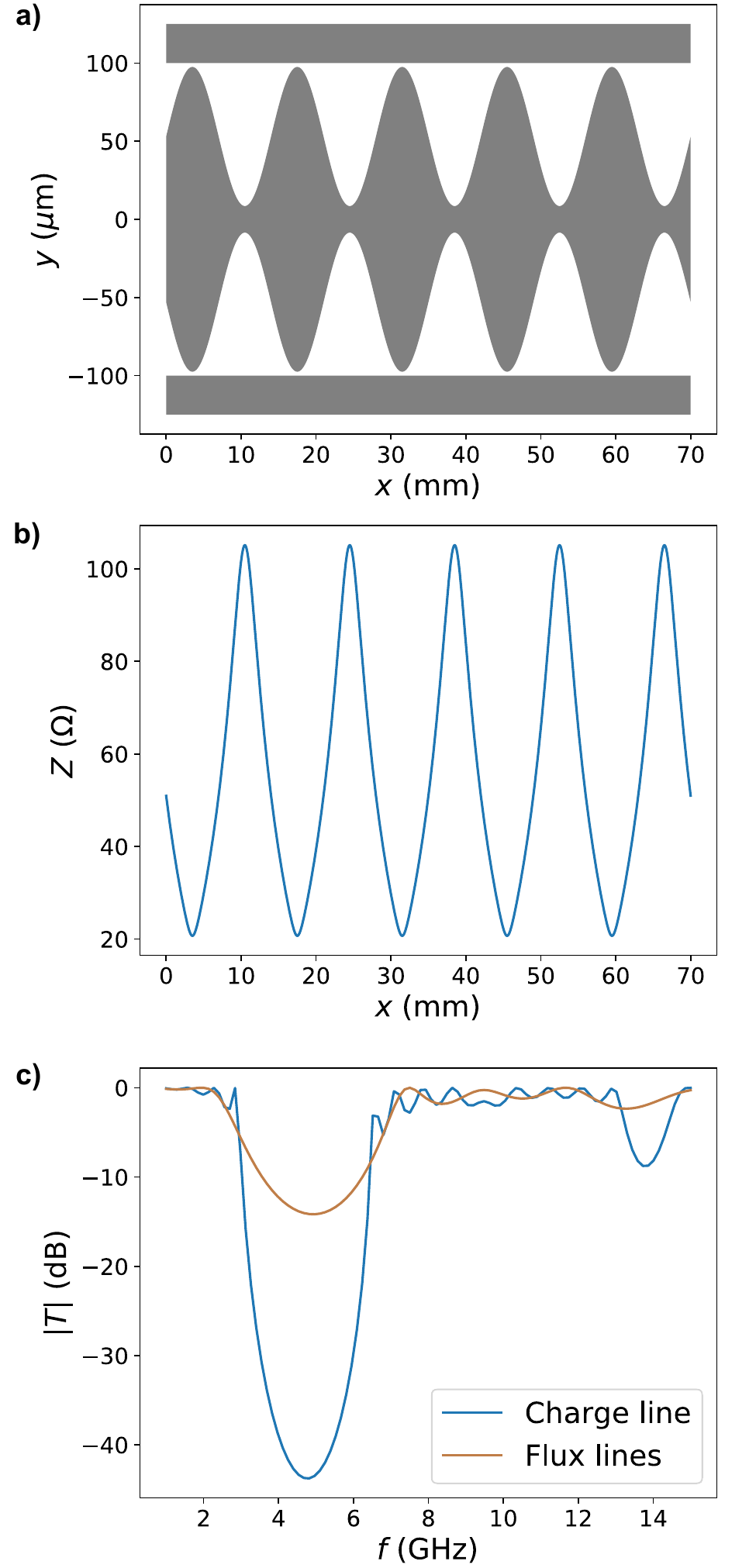}
		\caption{{\bf Photonic crystal filters. a)} Schematic representation of the charge line CPW geometry, featuring the width of its center track and gaps.  Metalized areas are colored in gray (the ground planes at the top and bottom extend beyond the graphic frame, and are connected together by wirebonds). The flux lines layout is identical, except for a reduced length (two periods of modulation instead of five). {\bf b)} Estimated characteristic impedance of the line, which is locally set by the center track width. {\bf c)} Numerically computed transmission coefficient of the charge line when connected to 50~$\Omega$ ports (blue). The stopband center frequency is set by the line modulation period, and its width by the amplitude of  modulation of its impedance. The flux lines, which are shorter, are not as strongly attenuated in-band.   }
		\label{fig:crystals}
	\end{figure}

In our sample, all three feedlines are filtered with these photonic crystals. As illustrated in Fig.~\ref{fig:crystals}a, the modulation period is $2\pi/\eta=14$~mm for a stopband centered at 4.5~GHz (expected memory frequency). The total CPW width is held constant at 200~$\mu$m, and the inner conductor width is sinusoidally varied from $17~\mu$m to $195~\mu$m. With these parameters, the line impedance is modulated in the range $20-100~\Omega$, and is $50~\Omega$ after an integer number of periods. The latter property ensures minimal reflections at the far end of the lines (connected to 50~$\Omega$ ports) for out-of-band microwaves. Note that the impedance variation is not small as in the simple picture presented above, and the modulation is not strictly sinusoidal. Owing to the properties of CPW lines, the phase velocity does not depend significantly on the line geometry. \\

We perform numerical simulations in order to quantitatively estimate the in-band attenuation of the feedlines. In a nutshell, we use a discrete-cell model for the line (cell electrical length much shorter than relevant wavelengths) and compute the transfer matrix for each cell. We then multiply these matrices to obtain the line transfer matrix. We finally  translate the result in a scattering matrix assuming that the ports at each end of the line have an impedance of $50~\Omega$~\cite{pozar2011microwave}. Intuitively, the transmission coefficient so obtained informs us on the extra protection provided by the modulated line as compared to the case where the system is directly connected to a 50~$\Omega$ port. We report the line transmission against probe frequency in Fig.~\ref{fig:crystals}c. The charge line, which is more strongly coupled to the circuit modes, is modulated over 5 periods, yielding more than 40~dB of attenuation. For lack of space on the chip, the flux lines are shorter, with only 2 periods of modulation each, yielding about 15~dB of attenuation. We estimate with electromagnetic simulations that  these attenuations are sufficient for  radiative decay through the lines  not to limit the lifetime of the memory. For all lines, the attenuation varies smoothly over the stopband, whose width exceeds 1~GHz. This feature is particularly appealing as filtering is robust against fabrication disorder, which can result in imprecise targeting of the modes frequency (see Table.~\ref{tableSM}). Note that the buffer, nominally out of band of the filters, is measured  to be close to the stopband. This may partly explain the unexpectedly low value of its single-photon dissipation rate (five times smaller than its nominal value).

\begin{figure*}[htbp]
		\centering
		\includegraphics[width=1.8\columnwidth]{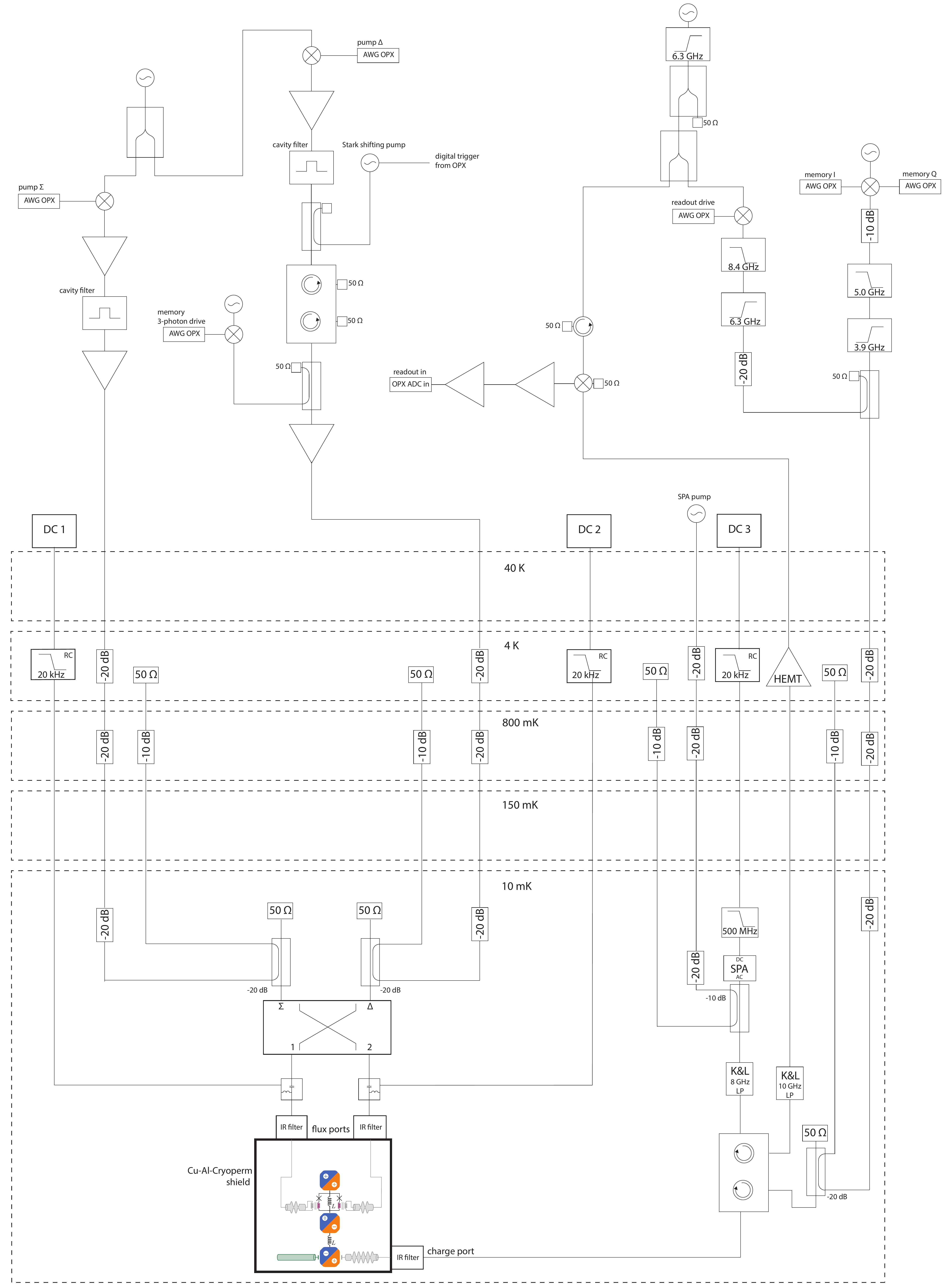}
		\caption{{\bf Wiring diagram} Microwave pulses are generated and detected at room temperature by mixing local oscillators (\textit{APUAS} from \textit{Anapico}) with waveforms in the $10-200$~MHz range generated by an \textit{OPX} from \textit{Quantum machines} (ADC and DAC ports labeled by their function). DC currents biasing the circuit and SNAIL-parametric amplifier~\cite{frattini2018optimizing} (SPA) are provided by  voltage sources from \textit{Yokogawa} and are thermalized with \textit{Therma uD} RC filters  from \textit{Aivon} anchored on the 4~K plate of the dilution cryostat. Microwave input lines are attenuated and filtered, in particular with infrared filters (IR) from \textit{Bluefors}. Output signals are amplified by the SPA and HEMT amplifiers from \textit{Low-noise factory} and other amplifiers at room temperature. }
		\label{fig:wiring}
	\end{figure*}

\section{Non-idealities and  non-linearity of the modes}
\label{sec:smNL}
\begin{figure*}[htbp]
		\centering
		\includegraphics[width=1.4\columnwidth]{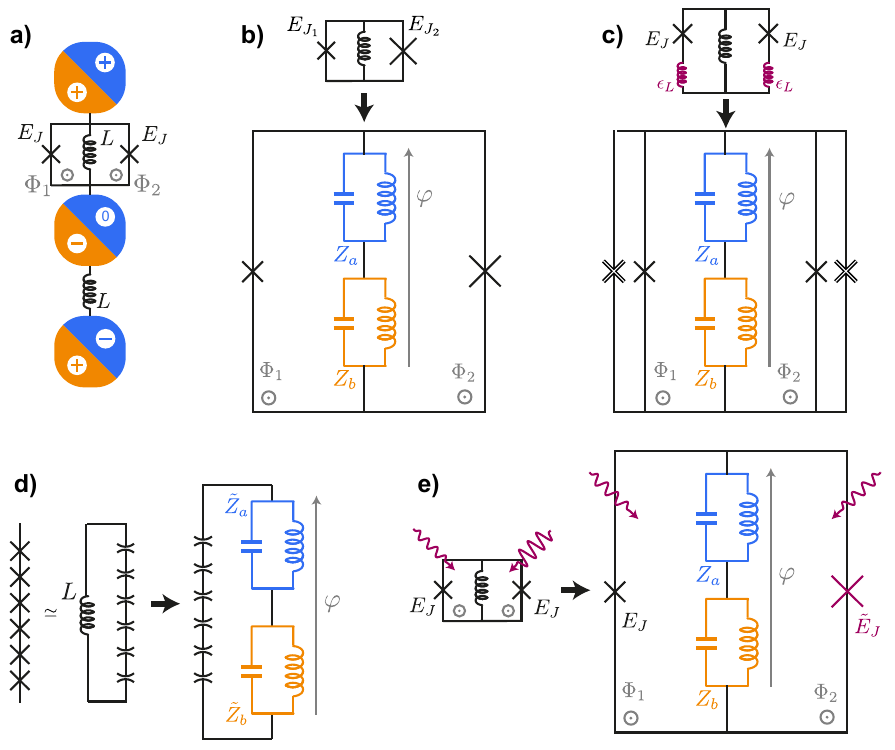}
		\caption{{\bf Sources of non-linearity. a)} Ideal circuit (no source of non-linearity) layout featuring the memory and buffer modes only. {\bf b) SQUID imbalance} directly appears in the effective circuit found by Foster synthesis of the environment seen by the SQUID. {\bf c) Series inductance} in the arms of the SQUID enable a  two-Cooper pair tunneling mechanism~\cite{smith2023spectral} (double-cross junctions) which is not suppressed at the circuit bias point. {\bf d) A superinductor} made of finite-length chains of Josephson junctions may be viewed as an inductor in parallel of a chain of inductive elements with quartic potential (spider symbols following the convention of Ref.~\cite{nigg2012black}). In general, the effective modes may present a different impedance $\tilde{Z}_{a,b}$ to this element from the impedance presented to the SQUID. However, given the circuit symmetry, we find $\tilde{Z}_{a,b}=Z_{a,b}$. {\bf e) Asymmetric pump waves} yield  non-linear contributions to the circuit Hamiltonian in the RWA. }  
		\label{fig:nonidealities}
	\end{figure*}
Following the simple analysis given in Sec.~\ref{sec:introduction}, the ATS~\cite{lescanne2020exponential} behaves, when properly biased, as a Kerr-free Josephson mixing element. In this section, we review mechanisms that break this property and estimate the  non-linearity induced on the circuit modes.

\subsection{Asymmetry of the SQUID junctions}

We consider the effective two-mode circuit pictured in Fig.~\ref{fig:nonidealities}b, which is obtained after Foster decomposition of the environment seen by the SQUID in the physical circuit (Fig.~\ref{fig:nonidealities}a), and neglecting all modes but the memory and buffer. We derive the SQUID inductive potential when the junctions forming its SQUID are imbalanced. Following the convention of Fig.~\ref{fig:nonidealities}b, we let $\varphi$ the reduced phase across the modes in series with the  SQUID,  $\Phi_1$, $\Phi_2$ the fluxes biasing each loop,  $E_{J_1}$, $E_{J_2}$ the energies of each junction. Defining
\begin{equation}
\begin{split}
\varphi_{\Sigma}&=\frac{\Phi_1+\Phi_2}{2\varphi_0}\\
\varphi_{\Delta}&=\frac{\Phi_1-\Phi_2}{2\varphi_0}\\
E_{\Sigma}&=E_{J_1}+E_{J_2}\\
E_{\Delta}&=E_{J_1}-E_{J_2}
\label{eq:cdfluxes}
\end{split}
\end{equation}
($\varphi_0=\hbar/(2e)$ is the reduced flux quantum) and using Kirchhoff's law, we find the SQUID potential to be
\begin{equation}
\begin{split}
V^{SQUID}_{\varphi_{\Sigma},\varphi_{\Delta}}(\varphi)=&-E_{\Sigma}\cos(\varphi_{\Sigma}) \cos(\varphi-\varphi_{\Delta}) \\
&-E_{\Delta}\sin(\varphi_{\Sigma}) \sin(\varphi-\varphi_{\Delta})
\end{split}
\end{equation}
Now focusing on the ATS operating point at $(\varphi_{\Sigma},\varphi_{\Delta})=(-\pi/2,- \pi/2)$ ($\mathrm{SP}_2$ in Fig.~\ref{fig2}b), the static circuit Hamiltonian (no pump applied here) reads
\begin{equation}
\HH_{asym}=\sum_i \hbar \omega_i \aop^{\dagger}_i \aop_i  -E_{\Delta} \cos(\varphiop)
\label{eq:NLdelta}
\end{equation}
where $i=a,b$ ($\aop_i$ denoting the annihilation operator of  mode $i$), $\varphiop=\sum_i \varphi^{ZPF}_i (\aop_i+\aop_i^{\dagger})$ and  $\varphi^{ZPF}_i =\big(\pi Z_i/R_q \big)^{1/2}$. The \textit{cosine term} in this Hamiltonian features non-linear, photon-number preserving terms that survive the RWA. At leading order (truncating the \textit{cosine} at fourth order), we find that each mode inherits a self-Kerr non-linearity of amplitude  $\chi_{aa}=\frac{1}{2} E_{\Delta}{\varphi_a^{ZPF}}^4$, $\chi_{bb}=\frac{1}{2} E_{\Delta}{\varphi_b^{ZPF}}^4$ and both modes are dispersively coupled through a cross-Kerr term $\chi_{ab}= E_{\Delta}{\varphi_a^{ZPF}}^2{\varphi_b^{ZPF}}^2$. Given the near perfect symmetry of the ATS  in our experiment, these terms are negligible ($\chi_{aa},\chi_{bb},\chi_{ab}<2\pi\times$10~kHz). \\

\subsection{Stray inductances and multi-Cooper pair tunneling}

In this subsection, we consider two models accounting for inductors (with identical inductive energy $\epsilon_L$) placed in series with the Josephson junctions forming the SQUID of the ATS (see Fig.~\ref{fig:nonidealities}c). These inductors stem from the geometric inductance of the long metallic tracks forming the SQUID (each arm is 120~$\mu$m long, see Fig.~\ref{fig:circuit}).\\

\begin{figure*}[htbp]
		\centering
		\includegraphics[width=1.5\columnwidth]{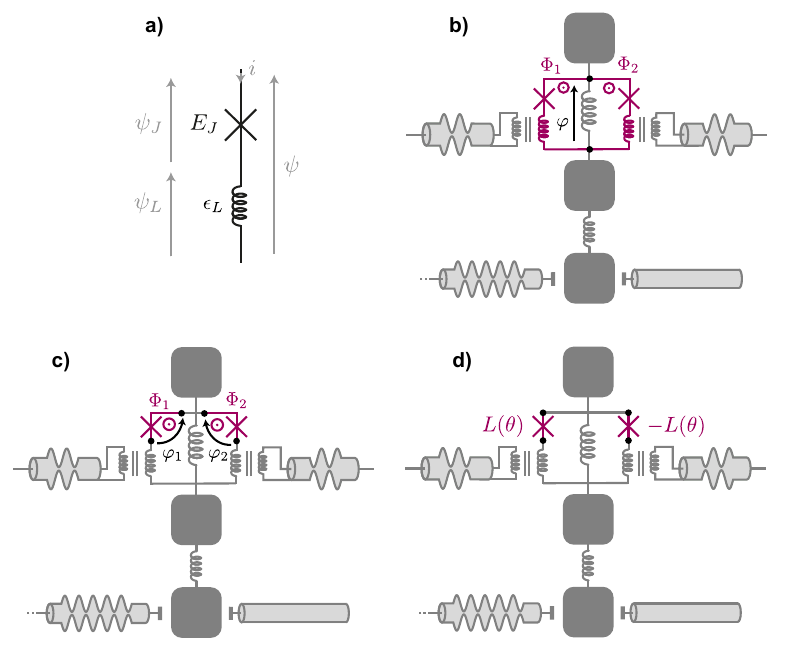}
		\caption{ {\bf a)} Inductive dipole made of a junction in series with an inductor. Using Kirchhoff's law, the dipole potential may be expressed as a function of its generalized phase $\psi$ only. {\bf b)} Circuit partition considered in the model 1 and throughout the paper. We perform Foster synthesis of the linear portion of the circuit colored in gray, yielding a set of normal modes $\{\aop_i\}$. This synthesized environment is connected through a single port with phase $\varphi$ (black circles) to a non linear dipole colored in purple (the SQUID, which includes the series inductors accounting for the inductance of tracks and the externally applied magnetic fluxes). The inductive potential of the dipole is expressed as a function of $\varphi$ and of the applied magnetic fluxes $\Phi_1$ and $\Phi_2$ in Eq.~\eqref{eq:potrenorm}. {\bf c)} Circuit partition considered in the model 2. We synthesize the linear portion of the circuit including the series inductors in the SQUID arms, yielding a set of normal modes $\{\bop_i\}$. This environment is connected to the two junctions via 2 ports with with phases $\varphi_1$ and $\varphi_2$. The potential of each junction is expressed as a function of each applied magnetic flux in Eq.~\eqref{eq:mode2pot}. Note that we choose a gauge in which the magnetic fluxes are allocated to the junctions and not to the linear circuit. {\bf d)} At a DC flux point $\varphi_{\Sigma},\varphi_{\Delta}=(\pi/2,\theta)$, the circuit in (c) is well approximated by a linearized version in which junctions have been replaced with inductances $\pm \frac{L_J}{\mathcal{B}\sin(\theta)}.$ This linear circuit is fed to an electromagnetic solver to estimate the inductance of the SQUID arms (see text). }  
		\label{fig:seriesSM}
	\end{figure*}

In the first model, each junction and its series inductor is considered as a non-linear inductive element (see Fig.~\ref{fig:seriesSM}a).  The circuit is analyzed  through Foster synthesis of the linear portion of the circuit connected to these elements  (see Fig.~\ref{fig:seriesSM}b). This is the point of view adopted throughout this paper, and it allows us to predict the non-linearity induced on the circuit modes as a function of $\epsilon_L$.\\

In the second model, we analyze the circuit through Foster synthesis of the linear circuit \textit{including} the series inductors and directly connected to the junctions (see Fig.~\ref{fig:seriesSM}c). This model allows us to extract the value of $\epsilon_L$ through finite-element electromagnetic simulation of the circuit.

\subsubsection{Model 1: dressed Josephson junction}

Following the derivation in Ref.~\cite{smith2023spectral}, a junction in series with a small inductor (inductive energies $E_J \gg \epsilon_L$) behaves as an inductive dipole with reduced phase $\psi$ (see Fig.~\ref{fig:seriesSM}a). Letting $i$ the current threading the dipole, $\psi_L$ and $\psi_J$ the reduced phases across the inductor and the junction, we find using Kirchhoff's law
\begin{equation}
\begin{split}
\psi_L+\psi_J=\psi\\
E_J \sin(\psi_J)=\epsilon_L \psi_L
\end{split}
\label{eq:cos2phi}
\end{equation}
The dipole energy reads $V=-E_J \cos(\psi_J)+\frac{\epsilon_L}{2}\psi_L^2$. Given the hierarchy $E_J \ll \epsilon_L$, we can solve the equation~\eqref{eq:cos2phi} to leading order in $E_J/\epsilon_L$ and inject the solution in the dipole potential to get
\begin{equation}
V(\psi)=-E_J\cos(\psi) +\frac{E_J^2}{4\epsilon_L}\cos(2\psi)+\mathcal{O}\big(\frac{E^3_J}{\epsilon^2_L}\big)
\end{equation}
The first-order correction to the Josephson potential corresponds to the coherent tunneling of pairs of Cooper pairs. Remaining perturbative terms  corresponding to tunneling of larger multiplets of Cooper pairs are neglected. \\

When replacing standard junctions with two such dipoles in the circuit  SQUID (see Fig.~\ref{fig:nonidealities}c), the inductive potential is renormalized as
\begin{equation}
\begin{split}
\tilde{V}^{SQUID}_{\varphi_{\Sigma},\varphi_{\Delta}}(\varphi)=&-E_{\Sigma}\cos(\varphi_{\Sigma}) \cos(\varphi-\varphi_{\Delta}) \\
&+ e_{\Sigma}\cos(2\varphi_{\Sigma}) \cos(2\varphi-2\varphi_{\Delta})
\end{split}
\label{eq:potrenorm}
\end{equation}
where $e_{\Sigma}=\frac{E^2_J}{2\epsilon_L}$ and we recall  the definition of the reduced magnetic fluxes $\varphi_{\Sigma/\Delta}=\frac{\Phi_1~ +/-~\Phi_2}{2\varphi_0}$. Note that  we have assumed the SQUID to be balanced for simplicity.  \\

In order to relate predictions from this model to those from the model described below, we note that at a DC flux bias of $(\varphi_{\Sigma},\varphi_{\Delta})=(\pi/2,\theta)$, the static circuit Hamiltonian (no pump applied here) reads
\begin{equation}
\HH_{\theta}=\sum_i \hbar \omega_i \aop^{\dagger}_i \aop_i  - e_{\Sigma} \cos(2\varphiop -2\theta)
\end{equation}
where we remind that $\varphiop=\sum_i \varphi^{ZPF}_i (\aop_i+\aop_i^{\dagger})$ and the $a_i$'s are modes found by through Foster synthesis of the linear circuit connected to the two dipoles each formed by a Josephson junction in series with an inductor (see Fig.~\ref{fig:seriesSM}b).\\

Given that $e_{\Sigma}\ll \omega_i$, we can estimate the frequency shift of each mode induced by the second term in the RWA. We first note that, neglecting rotating terms and for low-energy states, $e_{\Sigma}\cos(2\varphiop-2\theta)$ is well approximated by $-2e_{\Sigma} \mathcal{A}^4\cos(2\theta)\varphiop^2$, where $\mathcal{A}=e^{-\frac{1}{2}\sum_i {\varphi_i^{ZPF}}^2}$ and  $\mathcal{A}^4$ is a prefactor appearing when normal-ordering the operator $\cos(2\varphiop)$. Still in the RWA, this term provides each mode with an additional inductive energy that adds to the energy $E_{L_i}=\hbar \omega_i {\varphi_i^{ZPF}}^2/2$ determined by the Foster synthesis of the linear circuit. Compared to the case where the series inductors in the SQUID are neglected, we find that the modes' resonance frequencies to be dressed as 
\begin{equation}
\omega_i \rightarrow \omega_i\Big(1+\frac{\mathcal{A}^4 E_J^2}{\epsilon_L E_{L_i}}\cos(2\theta)\Big)
\label{eq:eLdispersion}
\end{equation}

\subsubsection{Model 2: dressed modes}

We now adopt a different viewpoint and synthesize the linear portion of the circuit \textit{including} the inductors in the SQUID arms   as depicted in Fig.~\ref{fig:seriesSM}c.  The synthesis yields a set of modes $\{\bop_i\}$ and allows us to express the phase of each mode across the two ports to which the Josephson junctions are  connected:
\begin{equation}
\varphiop_1=\sum_i \varphi^{ZPF}_{i,1}(\bop_i+\bop_i^{\dagger}) \qquad \qquad  \varphiop_2=\sum_i \varphi^{ZPF}_{i,2}(\bop_i+\bop_i^{\dagger})
\end{equation}
Given the circuit symmetry, each mode  either has symmetric phases across the two ports ($\varphi^{ZPF}_{i,2}=\varphi^{ZPF}_{i,1}$), or antisymmetric phases ($\varphi^{ZPF}_{i,2}=-\varphi^{ZPF}_{i,1}$). We now label with an index $k_{\Sigma}$ the modes with symmetric phases (with phase vacuum fluctuations $\varphi^{ZPF}_{k_{\Sigma}}$), and with an index $k_{\Delta}$ the modes with antisymmetric phases (with vacuum fluctuations $\varphi^{ZPF}_{k_{\Delta}}$). \\

In a similar spirit to the analysis made in this paper introduction, we can write the inductive potential of each junction as 
\begin{equation}
\begin{split}
&V_1(\varphi_1)=-E_J\cos(\varphi_1-\varphi_{\Sigma}-\varphi_{\Delta}) \\
 &V_2(\varphi_2)=-E_J\cos(\varphi_2+\varphi_{\Sigma}-\varphi_{\Delta})
\end{split}
\label{eq:mode2pot}
\end{equation}
where we have chosen a gauge in which the externally applied magnetic fluxes are allocated to the junctions. Expressing $\varphiop_1$ and $\varphiop_2$ in the normal mode basis, the circuit Hamiltonian at the DC flux point $(\varphi_{\Sigma},\varphi_{\Delta})=(\pi/2,\theta)$

\begin{widetext}
\begin{equation}
\tilde{\HH}_{\theta}=\sum_{k_\Sigma} \hbar \omega_{k_{\Sigma}} \bop^{\dagger}_{k_{\Sigma}} \bop_{k_{\Sigma}} + \sum_{k_\Delta} \hbar \omega_{k_{\Delta}} \bop^{\dagger}_{k_{\Delta}} \bop_{k_{\Delta}} - 2E_J\sin \Big(\sum_{k_{\Delta}}\varphi^{ZPF}_{k_{\Delta}}(\bop_{k_{\Delta}}+\bop^{\dagger}_{k_{\Delta}})\Big)\cos \Big(\sum_{k_{\Sigma}}\varphi^{ZPF}_{k_{\Sigma}}(\bop_{k_{\Sigma}}+\bop^{\dagger}_{k_{\Sigma}})-\theta\Big)
\end{equation}
As we are concerned with predicting the modes' frequency dispersion as a function of $\theta$ and in the low-energy regime, we now approximate the Josephson term to leading order and drop linear drive terms that do not impact the modes' frequencies
\begin{equation}
\tilde{\HH}_{\theta}\approx\sum_{k_\Sigma} \hbar \omega_{k_{\Sigma}} \bop^{\dagger}_{k_{\Sigma}} \bop_{k_{\Sigma}} + \sum_{k_\Delta} \hbar \omega_{k_{\Delta}} \bop^{\dagger}_{k_{\Delta}} \bop_{k_{\Delta}}  - 2\mathcal{B} E_J \sin(\theta)   \Big(\sum_{k_{\Delta}}\varphi^{ZPF}_{k_{\Delta}}(\bop_{k_{\Delta}}+\bop^{\dagger}_{k_{\Delta}})\Big) \Big(\sum_{k_{\Sigma}}\varphi^{ZPF}_{k_{\Sigma}}(\bop_{k_{\Sigma}}+\bop^{\dagger}_{k_{\Sigma}})\Big)
\label{eq:linearizedmod1}
\end{equation}
 where the prefactor $\mathcal{B}=e^{-\frac{1}{2}\sum_{k_{\Sigma}} {\varphi_{k_{\Sigma}}^{ZPF}}^2-\frac{1}{2}\sum_{k_{\Delta}} {\varphi_{k_{\Delta}}^{ZPF}}^2}$ appears when normal ordering the non-linear operators. Finally, we note that this Hamiltonian encodes the dynamics of the linear circuit depicted in Fig.~\ref{fig:seriesSM}d, in which the Josephson junctions are replaced with linear inductors with  inductances $ \frac{L_J}{\mathcal{B}\sin (\theta)}$  and $ -\frac{L_J}{\mathcal{B}\sin (\theta)}$.  \\
\end{widetext}
Before moving on to estimating the inductance of the tracks of the SQUID, we make two observations 
\begin{itemize}
    \item We kept our formalism general and indexed modes with labels $k_{\Sigma}$, $k_{\Delta}$. Indeed, in a realistic setting including the circuit feedlines terminated by a 50~$\Omega$ environment, the number of modes to include in the Foster synthesis is infinite. Nevertheless, among symmetric modes, the synthesis is dominated by three modes that can be identified to the memory, buffer and readout modes. Other modes are expected to resonate at much higher frequency and to participate only weakly in the junctions. On the anti-symmetric side, the modes have a low impedance or do not participate strongly in the ATS (only a mode corresponding to charge oscillations between the intrinsic junction capacitors is expected to participate strongly but  resonates at a much higher frequency than the modes considered so far and is of low impedance).  Overall we can approximate the Josephson energy reduction factor $\mathcal{B}$ with the factor $\mathcal{A}$ found in the model 1.
    \item In the model 1, the frequency dispersion of the circuit modes as a function of the applied differential flux $\theta$ is the consequence of an additional inductive energy stemming from the (unsuppressed) tunneling of pairs of Cooper pairs through the junction. In the  model 2, the frequency dispersion comes from the hybridization of the symmetric modes with the anti-symmetric environment. However, since choosing where to set the limit of the linear portion of the circuit to be synthesized is arbitrary, the two models should yield identical results and therefore these two mechanisms are equivalent. We note that in the model 2, the coupling factor of symmetric modes to anti-symmetric ones depends on $\theta$ as $g\propto \sin(\theta)$ such that in the weak hybridization limit, we expect a frequency dispersion $\propto \sin^2(\theta)$, compatible with the dispersion found in the model 1. 
\end{itemize}

\subsubsection{Non-linearity of the modes and estimates for the inductance of the SQUID arms.}

In principle, one can estimate the inductance $\varphi_0^2/\epsilon_L$ of the SQUID tracks by recording the modes's resonance frequency as a function of applied DC fluxes $(\varphi_{\Sigma},\varphi_{\Delta})=(\pi/2,\theta)$ and fitting the obtained data with Eq.~\eqref{eq:eLdispersion}. However, this data was not acquired in our experiment. In the following, we rather estimate $\epsilon_L$ through the memory mode anharmonicity pattern at the saddle point. Subsequently, we compare this estimate to the value found in electromagnetic simulations based on the model 2.\\

Now focusing on the ATS operating point at $(\varphi_{\Sigma},\varphi_{\Delta})=(-\pi/2,- \pi/2)$, the static  Hamiltonian derived in the model 1 reads
\begin{equation}
\HH_{2\varphi}=\sum_i \hbar \omega_i \aop^{\dagger}_i \aop_i  +e_{\Sigma} \cos(2\varphiop)
\end{equation}

We fit the memory energy spectrum recorded at the ATS saddle point (see Fig.~\ref{fig2}b) with this model,  thus estimating $e_{\Sigma}$ and finally $\epsilon_L=h\times 1920~$GHz. This inductive energy corresponds to a series inductance of 85~pH in each arm of the SQUID. Note that we include in our fit corrections from the Josephson junctions asymmetry (independently calibrated) and a non-linear contribution from the circuit superinductors (see following section).\\

As a sanity check, we estimate the geometric inductance of the tracks of the SQUID leveraging the equivalence between model 1 and  model 2. In electromagnetic simulations (with \textit{HFSS} by \textit{Ansys}), we replace the junctions in our circuit with linear lumped elements with inductance $\pm \frac{L_J}{\mathcal{B}\sin(\theta)}$ (see comment below Eq.~\eqref{eq:linearizedmod1} and Fig.~\ref{fig:seriesSM}d). For consistency with the model 1, we do not include intrinsic junction capacitance in these simulations.  For each value of $\theta$, the EM solver returns the dressed frequency of each mode.  We then fit the frequency dispersion of each of the three modes with  Eq.~\eqref{eq:eLdispersion}, yielding three estimates for $\epsilon_L$. The corresponding inductances are $79~$pH, $88~$pH and $74~$pH respectively, in good agreement with the value estimated from the memory mode anharmonicity. We also tried a more direct approach in which the inductance of the whole ATS (comprising the SQUID and the shunt inductor) is estimated through EM simulations before subtracting the contribution from the junctions and the shunt inductor. This approach yielded a slightly higher estimate  of 100~pH for the inductance of the SQUID arms. We consider this direct approach as less reliable as the ATS's inductance is estimated from the impedance seen between two particular  points of the circuit, which does not account for the finite size of the circuit tracks.\\

Here, we note that a recent paper~\cite{willsch2024observation} reported deviations from the standard $\cos(\varphi)$  potential for  junctions embedded in several transmon circuits, with significant higher-order harmonics attributed to pinholes in the junction oxide barrier.  In our case, the observed $\cos(2\varphi)$ term can be entirely attributed to geometric inductance in the SQUID arms. Moreover, the amplitude of this term is only 0.2~\%  of the amplitude of the standard $\cos(\varphi)$ contribution. This ratio is an order of magnitude lower than the values reported in~\cite{willsch2024observation}.  Since our Josephson junctions are fabricated with a long, high-pressure oxidation step (see Sec.~\ref{sec:fab}), their oxide barrier is probably thicker than for  junctions analyzed in that work. We can thus conclude, in line with the model presented in~\cite{willsch2024observation}, that junctions with a thick oxide barrier are less prone to the formation of pinholes and are  described by a nearly pure $\cos(\varphi)$ potential. \\

To conclude this section, we note that the  $\cos(2\varphiop)$ term is the dominant contributor to the modes non-linearity at the saddle point. In particular, it sets the memory anharmonicity   $\omega_a^{n+1,n+2}-\omega_a^{n,n+1}$, which is positive for all $n$,  lies in the 1---10~MHz range and decreases sharply with $n$ (see Fig.~\ref{fig2}b). This last property is a consequence of the large zero point fluctuations of the phase of the mode, which are effectively doubled by the Cooper pair pairing mechanism. Indeed, the $\cos(2\varphiop)$ term cannot   be truncated at fourth order, which would have yielded a Kerr non-linearity (constant anharmonicity). Beyond introducing non-linearity into the memory mode, which prevents stabilizing a full four-legged cat state, we do not expect the stray inductance of the SQUID arms to impact significantly the results and calculations presented in this work. In particular, while series inductors slightly dilute the participation of the Josephson junctions in the circuit modes, the impact on the four-to-one photon exchange rate is negligible (see Table~\ref{tableSM}: more than 99~\% of the modes's phase fluctuations are supported by the junctions and not by the series inductors).  \\

\subsection{Non-linearity induced by the superinductors}
The two superinductors in our circuit are made of chains of $N=25$ Josephson junctions, inducing  non-linearity in the modes they participate in. In detail, assuming that all junctions of the chain to be identical, that self-resonant modes of the structure are never excited~\cite{viola2015collective} and that phase-slips occur at a negligible rate~\cite{manucharyan2012superinductance}, internal degrees of freedom of the chain are frozen and the phase drop $\psi$ across the chain is equally distributed among the junctions. The chain potential then reads
\begin{equation}
V_{\mathrm{chain}}(\psi)=-N E^0_J \cos(\psi/N)\simeq -\frac{E^0_J}{2N} \psi^2 + \frac{E^0_J}{24 N^3} \psi^4
\end{equation}
where $E_J^0$ is the energy of each junction and we have truncated the potential at fourth order. This approximation is motivated by the large value of $N$ resulting in a small phase drop across each junction.\\

In order to quantitatively estimate the Kerr non-linearity induced on the modes, we adapt the energy participation ratio method~\cite{minev2021energy}. We perform finite-element electromagnetic simulations to estimate the phase fluctuations $\varphi^{ZPF}_{i,k}$ of each mode $i=a,b,r$ across each superinductor $k=1,2$. Letting $E_L=\frac{E^0_J}{N}$ the inductive energy of each chain, we then estimate that each mode $i$ inherits a self-Kerr non-linearity of amplitude  $\chi_{ii}=\sum_k  \frac{E_{L}}{2 N^2}{\varphi_{i,k}^{ZPF}}^4$ and any pair of modes $(i,i')$ are dispersively coupled through a cross-Kerr non-linearity of amplitude $\chi_{ii'}=\sum_k  \frac{E_{L}}{N^2}{\varphi_{i,k}^{ZPF}}^2{\varphi_{i',k}^{ZPF}}^2$. We find these terms to have an amplitude in the tens of kHz range, with a negligible impact on the system dynamics. We still include them in numerical simulations.

\subsection{Miscalibration of the ATS flux biases}

While the ATS should in principle be driven with symmetric fluxes to activate the desired four-to-one photon exchange (see Sec.~\ref{sec:introduction}), we actually drive the circuit with an unknown combination of common and differential flux drives (see Sec.~\ref{sec:fab}). Dynamical renormalization of the modes non-linearity may then arise. \\

To understand this phenomenon, we consider the ideal circuit Hamiltonian~\eqref{eq:simpleATS} (other sources of non-linearity neglected for clarity), at the ATS operating point and in presence of flux drives biasing the loops both symmetrically (reduced amplitude $\epsilon_c(t)$) and antisymmetrically (reduced amplitude $-\epsilon_d(t)$).  In the rotating frame, it reads
\begin{equation}
\HH_{cd}(t) =E_{\Sigma}\sin \big(\epsilon_c(t) \big) \sin \big( \varphiop(t) + \epsilon_d(t)\big)
\end{equation}
where 
$\varphiop(t)=\sum_{j=a,b,r}\varphi^{ZPF}_j (\aop_j e^{-i\omega_j t} + \aop_i^{\dagger} e^{+i\omega_j t})$. Note that we expressed the Hamiltonian in a gauge where the differential drive appears inside the \textit{sine} term, which may lead to the apparition of an effective charge drive in the Hamiltonian~\cite{you2019circuit}. The amplitude of this term depends on the exact magnetic field distribution with respect to the capacitive structures in the circuit and is hard to predict \textit{a priori}. Since it does not yield resonant terms in the RWA, we neglect it in our analysis for simplicity. However, we point out that this drive---along with an effective flux drive stemming from the first-order term in the \textit{sine} expansion---displaces the states of the modes, yielding new dynamical corrections to the modes non-linearity at the second order of the RWA.\\

 Assuming that the drive amplitudes are small, we expand to second order in $\epsilon_{c,d}$ and find
\begin{equation}
\HH_{cd}(t) =E_{\Sigma}~\epsilon_c(t)  \sin \big( \varphiop(t) \big) +E_{\Sigma}~\epsilon_c(t)\epsilon_d(t)  \cos \big( \varphiop(t) \big)
\end{equation}
When driving the four-to-one photon transition, both drives $\epsilon_c(t)=\xi_c e^{i\omega_p t}+ \mathrm{c.c}$ and $\epsilon_d(t)=\xi_d e^{i\omega_p t}+ \mathrm{c.c}$ are applied at the frequency  $\omega_p=4\omega_a-\omega_b$. In the RWA, the first term yields the desired four-to-one photon exchange $\HH_4$. As for the second term, it yields an effective static Hamiltonian
\begin{equation}
\HH_{cd, RWA} =\overline{\HH_{cd}(t)}= 2 \Re \big( \xi_c \xi^{\ast}_d \big) E_{\Sigma} ~\overline{\cos \big( \varphiop(t) \big) }
\label{eq:Heffcd}
\end{equation}
This effective Hamiltonian is analogous to the non-linear term in the Hamiltonian~\eqref{eq:NLdelta} deriving from imbalance of the SQUID. In practice, it slightly renormalizes the modes non-linearity, which we do not account for in our simulations (as detailed in Sec.~\ref{sec:fourtoonedynamics}, the system dynamics under pump does not depend on the exact value of the non-linearity).\\

We also note that this term is resonant irrespective of the pump frequency. Thus,  it should be possible to apply an off-resonant pump ($\omega_p \neq 4\omega_a-\omega_b$) to dynamically cancel the non-linearity stemming from imbalance of the SQUID. One simply adjusts the symmetric and antisymmetric drive amplitudes and phases such that $2 \Re \big( \xi_c \xi^{\ast}_d \big)=E_{\Delta}$, where  $E_{\Delta}$ is the difference of Josephson energy for the two junctions. However, given that the above analysis is only valid in the limit $|\epsilon_c|,|\epsilon_d|\ll 1$, this method may only prove useful in the limit $E_{\Delta}\ll E_{\Sigma}$. Finally, we note that the effective Hamiltonian \eqref{eq:Heffcd} also induces a shift of the modes frequency. This mechanism is expected to contribute to the controlled frequency shift (induced by an off-resonant pump) that we apply  to avoid hybridization of our system with  two-level systems (see Sec.~\ref{sec:TLS}).

\section{Details of the calibration and measurement sequences}

\subsection{Photon-number resolving measurement}
\label{sm:meas}

\begin{figure}[htbp]
		\centering
		\includegraphics[width=1\columnwidth]{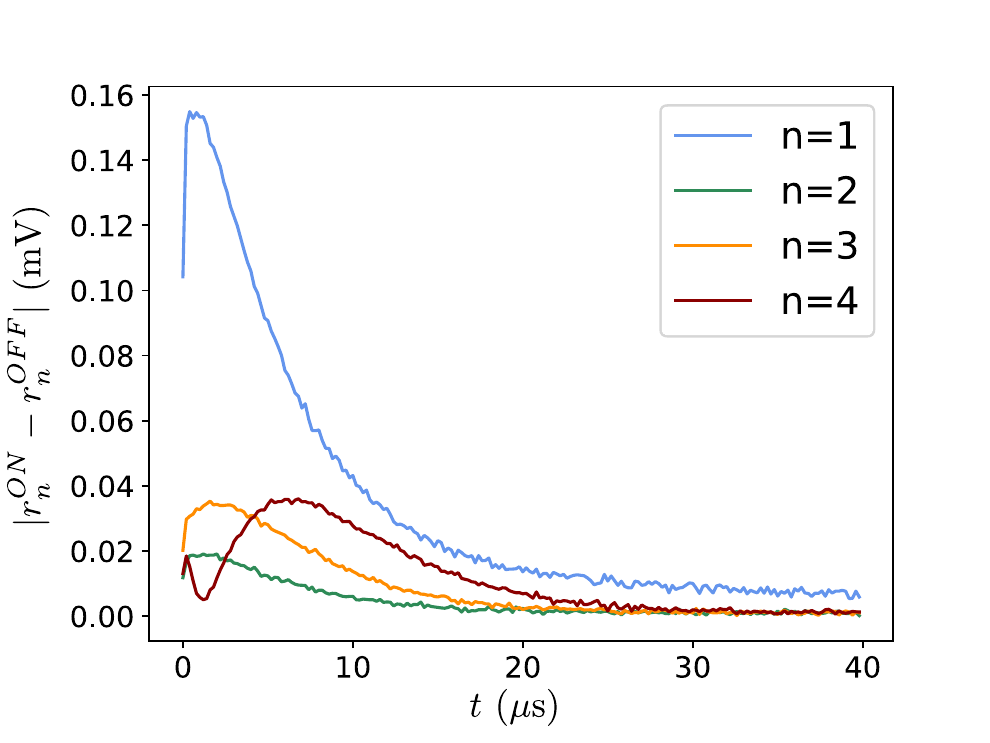}
		\caption{{\bf Reference signals for the photon-number resolving measurement}. In order to maximize SNR, we acquire reference signals $r_n^{ON}(t)$ and $r_n^{OFF}(t)$ after preparing the state $|n\rangle_a$ and applying (ON) or not (OFF) a pulse on the $|n\rangle_a \leftrightarrow |n+3\rangle_a$ transition. We use each signal  as an averaging envelope for subsequent measurement of the occupation of $|n\rangle_a$ for an unknown state (Eq.~\eqref{eq:waveform}). The non-zero signal at long time for $n=0$ is attributed to residual thermal occupation of state $|n=1\rangle_1$ at equilibrium. The maximum of each signal is expected to be delayed when $n$ increases, which is clearly visible for $n=4$ (the initial drop is  not understood). The pulses duration and amplitude (referenced by the voltage applied to the mixer generating the pulse) are 0.3~$\mu$s~/~0.4~V, 1~$\mu$s~/~0.15~V, 1~$\mu$s~/~0.15~V and 0.5~$\mu$s~/~0.1~V for $n=1,~2,~3$ and $4$ respectively. These characteristics are only optimized to maximize signal for $n=4$, which explains a higher signal than for $n=2,3$. Preparation of $|n=2\rangle_a$, which relies on two  successive $\pi$-pulses on single-photon transitions is less efficient than the preparation of $|n=3\rangle_a$ (single three-photon $\pi$-pulse), explaining a weaker signal.  }
		\label{fig:waveforms}
	\end{figure}

The principle of this measurement is described in the main text and depicted in Fig.~\ref{fig3}a. We recall here the motivation for the measurement sequence and give details on its calibration.\\

While the memory mode vacuum state $|0\rangle_a$ exerts a  dispersive pull onto the readout mode significantly different from the pull exerted by excited states, the differential pull exerted by one excited state and the next tails off rapidly. Indeed, the dispersive pulls stem from the $\cos(2\varphiop)$ contribution in the circuit static Hamiltonian (see Sec.~\ref{sec:smNL}) and follow the same trend as the memory mode anharmonicity (see Fig.~\ref{fig2}). This makes excited states hard to distinguish directly. We circumvent this shortcoming by applying (ON) or not (OFF) a  $\pi$-pulse on the $|n\rangle_a \leftrightarrow |n+3\rangle_a$ transition before probing the readout mode on resonance with a long (40~$\mu$s) pulse. The reflected field is demodulated and its quadratures averaged to obtain the complex signals $s_n^{ON}(t)$ and $s_n^{OFF}(t)$. Since the dynamics of the decay from $|n+3\rangle_a$ to the vacuum state is delayed compared to the decay from  $|n\rangle_a$,    $s_n^{ON}(t)$ detected after the application of the $\pi$-pulse is distorted and time-shifted  with respect to the signal $s_n^{OFF}(t)$ detected in absence of the pulse.\\

In order to maximize the measurement signal-to-noise ratio (SNR), we first prepare $|n\rangle_a$ with a sequence of $\pi$-pulses on single or three-photon transitions, and record an averaged reference signal $r_n^{ON}(t)$ and $r_n^{OFF}(t)$ (see Fig.~\ref{fig:waveforms}). Subsequently, when measuring an unknown state of the memory, we integrate the detected differential signal with an envelope provided by these reference signals:
\begin{widetext}
\begin{equation}
S_n=\frac{1}{T^{1/2}_{\mathrm{bin}}}\dfrac{\int_{t=0}^T\Re\Big(\big(s_n^{ON}(t) - s_n^{OFF}(t)\big)\big(r_n^{ON}(t) - r_n^{OFF}(t)\big)^{\ast}\Big)\mathrm{d}t}{\Big(\int_{t=0}^T|r_n^{ON}(t) - r_n^{OFF}(t)|^2~\mathrm{d}t\Big)^{1/2}}
\label{eq:waveform}
\end{equation}
\end{widetext}
where $T_{\mathrm{bin}}=200$~ns is the integration time for the acquisition of each point of the raw signals $s_n^{ON}$, $s_n^{OFF}$, $r_n^{ON}$, $r_n^{OFF}$.\\

We use this measurement to record the time-decay of various Fock states under pump, and make several key assumptions. First, we assume that before the measurement, only Fock states $|k\rangle_a$  with $k\leq n$ are occupied. Moreover, we assume that this property is preserved during the measurement. In other words, we neglect heating mechanisms such as coupling to a hot bath or measurement-induced transitions to higher-excited states. Finally, we assume that the $\pi$-pulse does not impact the system if  states $|k\rangle_a$  with $k< n$ are occupied (see below). Therefore, even though the $\pi$-pulse duration is of the order or longer than the states lifetimes, it only modifies the memory state when $|n\rangle_a$ is occupied beforehand. Moreover, the signal $S_n$  is proportional to the occupation probability. \\

\begin{figure}[htbp]
		\centering
		\includegraphics[width=0.8\columnwidth]{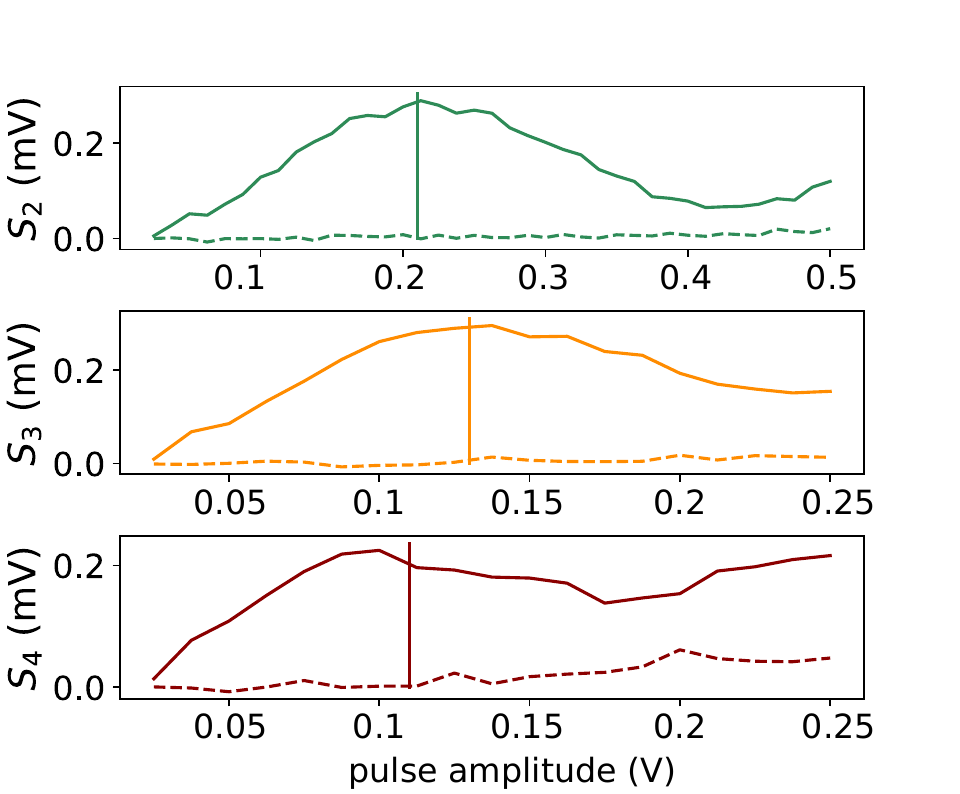}
		\caption{{\bf Optimization of the $\pi$-pulses parameters for the photon-number resolving measurement}. The measurement relies on the differential signal obtained when exciting or not the $|n\rangle_a \leftrightarrow |n+3\rangle_a$ transition. Holding the exciting pulse duration constant, we vary its amplitude and record the differential signal $S_n$ after preparing $|n\rangle_a$ (plain lines) or $|n-1\rangle_a$ (dashed lines). We select the amplitude yielding the strongest signal in the former case while yielding a negligible signal in the latter case (vertical lines). We repeat this calibration for various pulse durations and select the optimal amplitude combination of amplitude and duration. }
		\label{fig:rabionoff}
	\end{figure}

In order to maximize SNR while ensuring that the $\pi$-pulse does not impact the system if  only states $|k\rangle_a$  with $k< n$ are occupied, we need to adjust finely its amplitude and duration. Intuitively, we want to minimize the pulse duration while ensuring that it is sufficiently weak and well-resolved in frequency not to excite the $|n-1\rangle_a \leftrightarrow |n+2\rangle_a$ transition. To this end, we prepare in separate experiments the state $|n\rangle_a$ and the state $|n-1\rangle_a$ and vary the pulse amplitude and duration (square waveform with smoothed edges). We then select the characteristics yielding the highest signal when $|n\rangle_a$ is prepared while yielding a negligible signal when $|n-1\rangle_a$ is prepared (see Fig.~\ref{fig:rabionoff}). Note that after this calibration, we could again acquire reference signals for the integration of $S_n$ using the optimal $\pi$-pulse characteristics. This was done for $S_4$, but not for $S_n$ with $n\leq 3$ (we have verified that the reference signal shape does not depend significantly on the $\pi$-pulse characteristics as long as it does not excite unwanted transitions).

\subsection{Tuning the pump frequency  }
\label{sec:tuningPump}
\begin{figure}[htbp] 
		\centering
		\includegraphics[width=1\columnwidth]{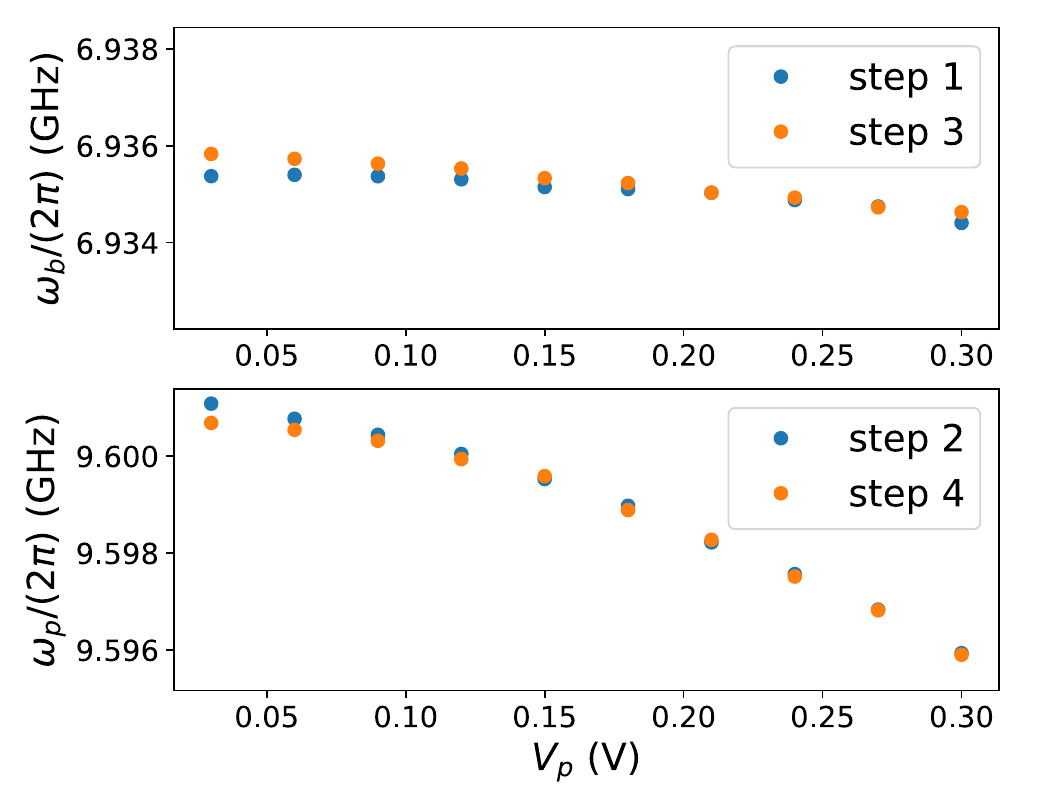}
		\caption{{\bf Pump frequency calibration}. The pump and buffer frequencies dressed by the pump itself are calibrated iteratively (see text). The  buffer frequency $\omega_{b_1}$ found at step 1 and $\omega_{b_3}$ found at step 3 are slightly mismatched at low pump amplitude. This  is attributed to  inaccuracy of the estimate made at step 3, in which the reflection was not properly fitted but only smoothed aggressively before locating the minimum   of the  signal magnitude. This mismatch then propagates to the calibrated pump frequency (step 2 vs step 4) as the pump resonance condition reads $\omega_{p_n}=4\omega_{a_n}-\omega_{b_{n-1}}$ where $\omega_{b_{n-1}}$ is the frequency at which the buffer is driven (found at the previous step) and $\omega_{a_n}$ is the memory frequency dressed by the pump at $\omega_{p_n}$. }
		\label{fig:tuningmemory}
\end{figure}
In Fig.~\ref{fig3}, we report the decay of population in state $|4\rangle_a$ and the buffer reflection spectrum as the  amplitude of the pump addressing the $|4\rangle_a|0\rangle_b \leftrightarrow |0\rangle_a|1\rangle_b$ transition is varied (encoded in color). For each pump amplitude, the pump frequency needs to be adjusted as the modes resonance frequency  gets shifted through by the pump itself (shifting mechanisms include---but are not limited to---miscalibration of the pump symmetry described in Sec.~\ref{sec:smNL} and direct Stark shift owing to the modes non-linearity). This shift depends non trivially on the pump frequency, mostly due to the uncalibrated dispersion of the cryostat input lines. For a given pump amplitude, the pump frequency is adjusted following four steps.

\begin{enumerate}
\item We apply the pump  at  frequency $\omega_{p_0}=4\omega_a -\omega_b$ corresponding to the four-to-one photon transition frequency in absence of pump. We then probe the buffer in reflection and find its resonance frequency $\omega_{b_1}$  dressed by the pump.
\item We  apply a weak drive at $\omega_{b_1}$ to the buffer  and sweep the  pump frequency   until we see the memory excite at $\omega_{p_2}$. This occurs when $\omega_{p_2}=4 \omega_{a_2}-\omega_{b_1}$, where  $\omega_{a_2}$ is the memory resonance frequency dressed by the pump at $\omega_{p_2}$. Note that $\omega_{b_1}$ is  slightly detuned from the frequency $\omega_{b_2}$ of the buffer  dressed by the pump. Photons from the buffer drive are still converted to quartets of excitations in the memory if $|\omega_{b_1} -\omega_{b_2} |  \lesssim \kappa_b$.
\item Again  as in step (1), we hold the pump at  $\omega_{p_2}$ and probe the buffer to find its resonance frequency $\omega_{b_3}$. Note that the buffer spectra acquired at this step are the ones reported in Fig.~\ref{fig3} and in Fig.~\ref{fig:bufferSpectra}. For the largest pump amplitudes, the buffer line splits  owing to the hybridization of $|4\rangle_a|0\rangle_b $ and $ |0\rangle_a|1\rangle_b$. Fitting to the full two-mode model presented in Sec.~\ref{sec:fourtoonedynamics} was not attempted at this step, so that the buffer resonance frequency was simply extracted by smoothing  each spectrum over a wide moving window  and locating the minimum   of the smoothed signal magnitude. 
\item Again as in step (2), we apply a weak drive at $\omega_{b_3}$ to the buffer and sweep the  pump frequency   until we see the memory excite at $\omega_{p_4}  = 4 \omega_{a_4}-\omega_{b_3}$, with  $\omega_{a_4}$ the memory frequency dressed by the pump at $\omega_{p_4}$
\end{enumerate}
One should iterate further the procedure until the resonance condition $\omega_{p_n}  = 4 \omega_{a_n}-\omega_{b_n}$ is met self-consistently. In practice, the variation of  the pump frequency  $|\omega_{p_2}-\omega_{p_4}|$ between step (2) and (4) was deemed small enough (less than 0.6~MHz at all considered pump amplitudes) that the buffer frequency $\omega_{b_4}$ dressed by the the pump at  $\omega_{p_4}$ would be near identical the frequency $\omega_{b_3}$ measured under the pump at $\omega_{p_2}$. This assumption, together with the rather coarse estimate of $\omega_{b_1}$ in step (3) explains a small detuning between the predictions from our model  (plain lines in Fig.~\ref{fig:bufferSpectra}) to the recorded spectra (dots). \\

For each pump amplitude, the time-decay of population in state $|4\rangle_a$ (presented in Fig.~\ref{fig3}b) was recorded after adjusting the pump frequency at $\omega_{p_2}$.

\subsection{Decay rate of low energy states under pump}
\label{sec:gammaslow}

\begin{figure}[htbp] 
		\centering
		\includegraphics[width=0.9\columnwidth]{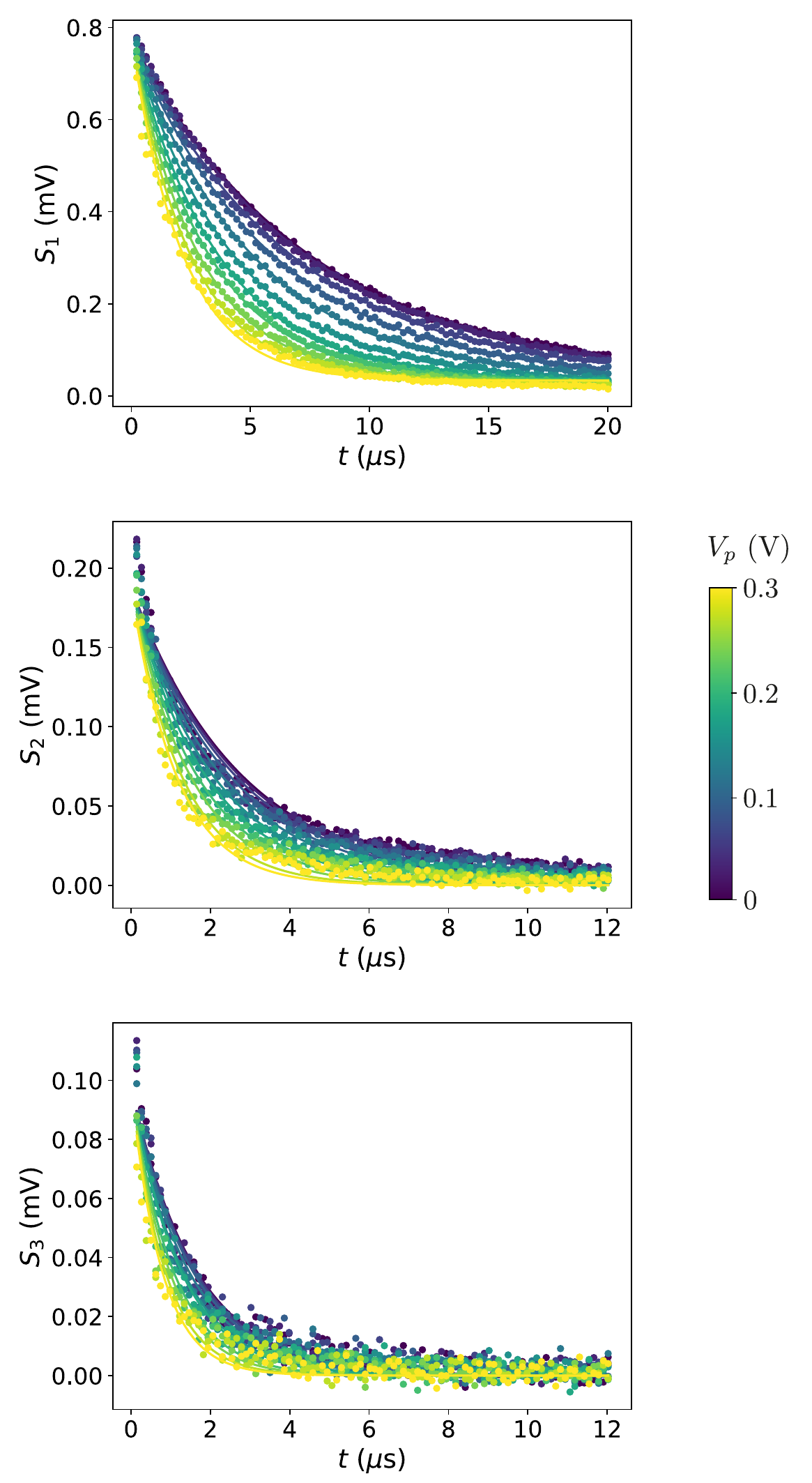}
		\caption{{\bf Decay rate of the low-energy states under pump}. The time decay under pump of population in states $|1\rangle_a$ (top), $|2\rangle_a$ (middle) and $|3\rangle_a$ (bottom) is recorded (filled circles) and fitted with an exponential law (lines). For each state $|n\rangle_a$, we fit a single scaling factor for the differential signal $S_n$, close to the value calibrated in Fig.~\ref{fig:rabionoff} (the slightly lower values recorded here, in particular for $S_3$, are attributed to drifts during the $\sim12$~hour acquisition time). The stationary value of the exponential law is fixed to zero, except for $S_1$ where we leave it as a free fit parameter to account for thermal excitations of the memory at equilibrium (see Fig.~\ref{fig:waveforms}). The initial drop of  $S_2$ and $S_3$  is not understood but does not seem directly related to the pump as it is visible at all pump amplitudes. }
		\label{fig:gammaslow}
\end{figure}

\begin{figure}[htbp] 
		\centering
		\includegraphics[width=0.9\columnwidth]{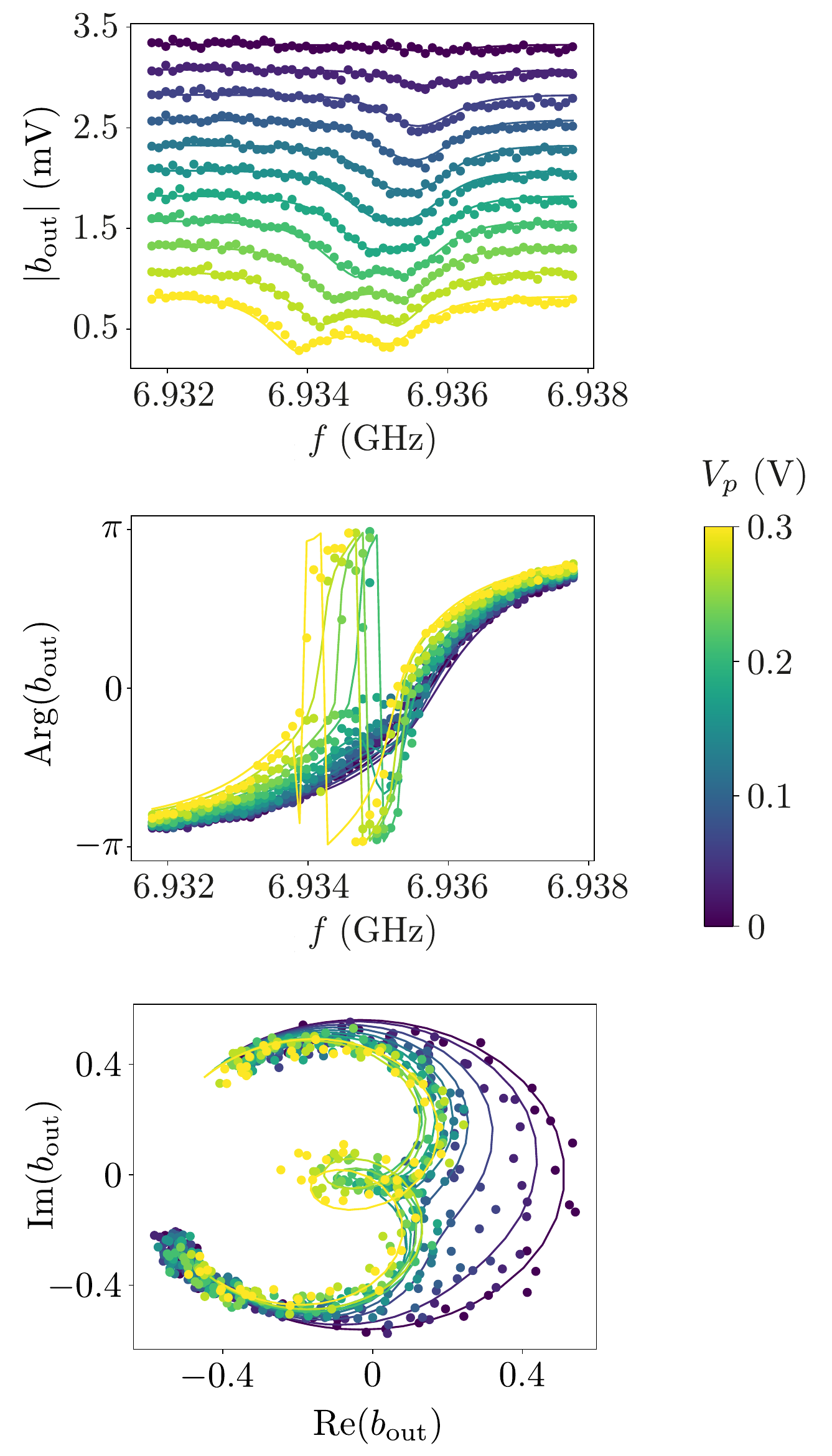}
		\caption{{\bf Buffer reflection spectrum under pump}. We represent the amplitude of the field reflected off the buffer as in Fig.~\ref{fig3}c (top, curves offset for readability), together with the phase of the reflected signal (middle) and a representation in the Fresnel plane (bottom). The spectra are well reproduced by simulations based on the effective Hamiltonian \eqref{eq:Hrwa} (lines). }
		\label{fig:bufferSpectra}
\end{figure}

As a sanity check of our experiment, we  record, for the same range of pump amplitudes activating the four-photon dissipation, the time-decay of populations prepared in states $|n\rangle_a$ for $n <4$ (see Fig.~\ref{fig:gammaslow}). For all states, we observe a slight increase of the decay rate induced by the pump, which is not captured by our model. For the states $|2\rangle_a$ and $|3\rangle_a$, we also record an initial sharp drop of the corresponding signals $S_2$ and $S_3$. This drop is not understood, but does not seem directly induced by the pump as it is also visible for a null pump amplitude (dark blue). We thus rather suspect an effect caused by the proximity of the memory preparation pulses and of the final pulse on the $|n\rangle_a \leftrightarrow |n+3\rangle_a$ transition (used in the measurement) for the shortest wait times. All signals are fitted with exponential laws and the fitted decay rates are reported in Fig.~\ref{fig4}.

\section{Numerical simulations and fitting procedures}
\label{sm:fits}
\subsection{Frequency maps}
\label{sec:maps}

In this section, we detail how the circuit parameters are estimated by analyzing the dependency of each mode resonance frequency against DC bias fluxes, in the neighborhood of the ATS operating points (see Fig.~\ref{fig2}a, operating points appearing as saddle points in the color map).  Our model   Hamiltonian  is derived from the three-mode effective circuit pictured in Fig.~\ref{fig2}a (obtained after Foster synthesis of the environment seen by the SQUID) and includes the non-idealities listed in Sec.~\ref{sec:smNL}, it  reads
\begin{equation}
\begin{split}
\HH=&\sum_i \hbar \omega_i \aop_i^{\dagger} \aop_i\\
& - E_{\Sigma}\cos(\varphi_{\Sigma}) \cos(\varphiop-\varphi_{\Delta}) -E_{\Delta}\sin(\varphi_{\Sigma}) \sin(\varphiop-\varphi_{\Delta}) \\
&+ e_{\Sigma}\cos(2\varphi_{\Sigma}) \cos(2\varphiop-2\varphi_{\Delta}) + \frac{E_L}{N^2} \big( \cos(\varphiop) + \frac{1}{2} \varphiop^2 \big) 
\label{eq:Htot}
\end{split}
\end{equation}
In this expression, $i \in \{a,b,r\}$ labels each mode, $\varphiop=\sum_i \varphi_i^{ZPF} (\aop_i + \aop_i^{\dagger}) $ is the reduced phase across the SQUID, $\varphi_i^{ZPF}=(\pi Z_i/R_q)^{1/2}$ represents the vacuum fluctuations of the phase of each mode across the SQUID. We have defined the reduced common and differential bias fluxes $(\varphi_{\Sigma},\varphi_{\Delta})$ in Eq.~\eqref{eq:cdfluxes}. As detailed in Sec.~\ref{sec:smNL},   we have allowed for the ATS junctions to be imbalanced with respective energies $E_{J_1}$ and $E_{J_2}$ and defined $E_{\Sigma}=E_{J_1}+E_{J_2}$ and $E_{\Delta}=E_{J_1}-E_{J_2}$. The last term on the first line stems from this imbalance (see Sec.~\ref{sec:smNL}). The first term on the second line stems from series inductances in the arms of the SQUID. We have assumed that series inductances are identical on both arms (inductive energy $\epsilon_L$) and defined $e_{\Sigma}=(E^2_{J_1}+E^2_{J_2})/(4\epsilon_L)$ (a last term in $(E^2_{J_1}-E^2_{J_2})/(4\epsilon_L)$ is neglected). Finally, the last term on the second line stems from the non-linear contribution of the superinductors potential (each a chain of $N=25$ for a total inductance $E_L$)~\cite{note10}.\\

In practice, we first record the frequency of the readout mode  over a full period of modulation for both fluxes. After locating three saddle points on the recorded map, we find the linear transform relating DC currents applied to the circuit and reduced fluxes $\varphi_{\Sigma}$ and $\varphi_{\Delta}$. We then measure all three modes resonance frequencies nearby the saddle points. More specifically,  the fluxes are varied along cuts intersecting $(\varphi_{\Sigma},\varphi_{\Delta})=(-\pi/2,\pm \pi/2)$, and we record the transition frequencies $|0\rangle_a \leftrightarrow |1\rangle_a $ and $|0\rangle_a \leftrightarrow |3\rangle_a $ of the memory and the lowest transition frequency of the buffer and readout modes. In order to facilitate the fit of these datasets with the model Hamiltonian~\eqref{eq:Htot}, we first fit the measured frequencies along each cut with a parabola, and report the extracted curvature against cut angle in Fig.~\ref{fig:curvatures}. We subsequently fit the curvature pattern with our model. We first detail a  simplified classical  analysis of the circuit around the frequency saddle points, which is the rationale for the fitting procedure to the full quantum model detailed in the following section.

\subsubsection*{Classical analysis of the frequency map curvature at the saddle points}

We here compute simple analytical formula for the frequency of each mode against externally applied fluxes in the neighborhood of the saddle points, using a classical model. We  neglect  all circuit non-idealities ($E_{\Delta}=e_{\Sigma}=E_L/N^2=0$ in~\eqref{eq:Htot}). Moreover, for simplicity, we consider two only modes $a$ and $b$ with phases $\varphi_a$ and $\varphi_b$. We will generalize the obtained formula to the three-mode case at the end of the calculation.\\

Letting $E_{L_a}$ and $E_{L_b}$ the modes inductive energies, and $E_{C_a}$ and $E_{C_b}$  their capacitive energies (related to the modes resonance frequencies and impedances by $\omega_{a,b}=(8E_{L_{a,b}}E_{C_{a,b}})^{1/2}$ and $Z_{a,b}=R_q (2E_{C_{a,b}}/E_{L_{a,b}})^{1/2}/\pi$), we express the circuit inductive potential in a coordinate system $(\epsilon_{\Sigma},\epsilon_{\Delta})=(\varphi_{\Sigma},\varphi_{\Delta})-(-\pi/2,- \pi/2)$ centered on the saddle point $\mathrm{SP}_2$ (see Fig.~\ref{fig2}b):
\begin{equation}
V(\varphi_a,\varphi_b)=\frac{1}{2}E_{L_a} \varphi_a^2+\frac{1}{2}E_{L_b} \varphi_b^2 +E_{\Sigma} \sin(\epsilon_{\Sigma}) \sin(\varphi_a+\varphi_b - \epsilon_{\Delta}) 
\label{eq:V0}
\end{equation}
We  consider cuts through the saddle point parametrized by an angle $\theta$
\begin{equation}
\begin{aligned}
\epsilon_{\Sigma}&=\epsilon \cos(\theta)\\
\epsilon_{\Delta}&=\epsilon \sin(\theta)
\label{eq:cutdef}
\end{aligned}
\end{equation}
with $\epsilon$ varying on a range much smaller than 1. We then expand the potential to second order in $\epsilon$. Noting  $c=\cos(\theta)$ and $s=\sin(\theta)$ for compactness, we obtain
\begin{equation}
\begin{aligned}
V(\varphi_a,\varphi_b)=&\frac{1}{2}E_{L_a} \varphi_a^2+\frac{1}{2}E_{L_b} \varphi_b^2 +E_{\Sigma} c \epsilon \Big( \\
 &\qquad -\epsilon s \big(\cos(\varphi_{a})\cos(\varphi_{b})-\sin(\varphi_{a})\sin(\varphi_{b}) \big) \\
 &\qquad + \sin(\varphi_{a})\cos(\varphi_{b})+\cos(\varphi_{a})\sin(\varphi_{b})\\
 &\qquad\Big)\\
 &+\mathcal{O}(\epsilon^3)
\end{aligned}
\label{eq:vsym}
\end{equation}

The first step is to look for the coordinates $(\varphi_a^m,\varphi_b^m)$ that minimize the inductive energy. We will then expand the potential to second order  to estimate the frequency of a classical particle oscillating around this minimum. The  coordinates of the minimum  are  approximated to second order in $\epsilon$. Since $(\varphi_a^m,\varphi_b^m)=(0,0)$ when $\epsilon=0$, they are parametrized as
\begin{equation}
\begin{aligned}
\varphi_a^m&= \alpha_a \epsilon +\beta_a \epsilon^2 +\mathcal{O}(\epsilon^3) \\
\varphi_b^m&= \alpha_b \epsilon +\beta_b \epsilon^2 +\mathcal{O}(\epsilon^3)
\end{aligned}
\end{equation}
Using $\frac{\partial V}{\partial \varphi_a}|_{(\varphi_a^m,\varphi_b^m)}=0$, we get 
\begin{equation}
\alpha_a \epsilon + \beta_a \epsilon^2 + \epsilon r_a c +\mathcal{O}(\epsilon^3) =0
\end{equation}
where we have let $r_a=\frac{E_{\Sigma}}{E_{L_a}}$.\\

\begin{figure}[htbp] 
		\centering
		\includegraphics[width=0.4\columnwidth]{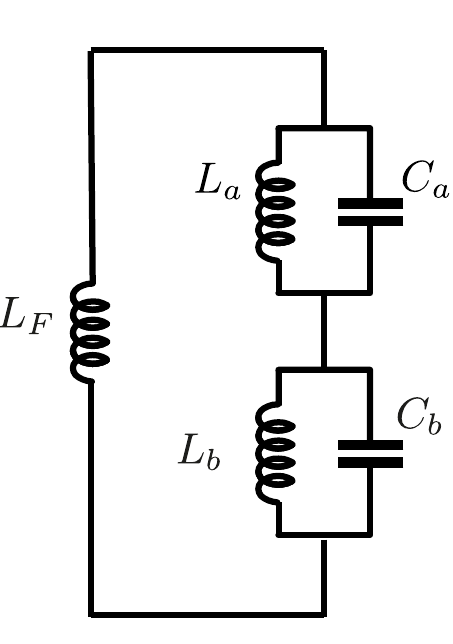}
		\caption{Effective circuit corresponding to the inductive potential encoded by the Hessian matrix \eqref{eq:hessian} and the charging energies found by Foster synthesis of the SQUID environment. }
		\label{fig:effcircuit}
\end{figure}

We can proceed similarly for the second variable. After identifying terms with same order in $\epsilon$ in the equations, we get
\begin{equation}
\begin{aligned}
\varphi_a^m&= -\epsilon r_a c +\mathcal{O}(\epsilon^3) \\
\varphi_b^m&= -\epsilon r_b c +\mathcal{O}(\epsilon^3)
\end{aligned}
\end{equation}
We can inject these expressions in the formulas giving the second derivatives of the potential around the minimum

\begin{equation}
\begin{aligned}
\frac{\partial^2 V}{\partial \varphi_a^2}|_{(\varphi_a^m,\varphi_b^m)}&=E_{L_a}\big (  1+\epsilon^2 r_a c (s + r_a c + r_b c)    \big) +\mathcal{O}(\epsilon^3)\\
\frac{\partial^2 V}{\partial \varphi_b^2}|_{(\varphi_a^m,\varphi_b^m)}&=E_{L_b}\big (  1+\epsilon^2 r_b c (s + r_a c + r_b c)    \big) +\mathcal{O}(\epsilon^3)\\
\frac{\partial^2 V}{\partial \varphi_a \partial \varphi_b}|_{(\varphi_a^m,\varphi_b^m)}&=E_{\Sigma}  \epsilon^2 c (s + r_a c + r_b c)     +\mathcal{O}(\epsilon^3)
\end{aligned}
\end{equation}
which, letting $F=\epsilon^2 E_{\Sigma} c (r_a c+ r_b c +s)$, is compactly encoded by the Hessian matrix  
\begin{equation}
\begin{pmatrix}
E_{L_a} + F& F  \\
F  & E_{L_b} + F  
\end{pmatrix}
\label{eq:hessian}
\end{equation}
at second order in $\epsilon$. Since the modes charging energies do not vary as a function of flux, this effective potential combined with the charging energies correspond to the circuit pictured in Fig.~\ref{fig:effcircuit}, where we have defined $L_F=\frac{\varphi_0^2}{F}$. We can write down the circuit Lagrangian and find the two dressed mode frequencies
\begin{equation}
\omega_{\pm}^2=\frac{\tilde{\omega}_a^2+\tilde{\omega}_b^2}{2}\pm \frac{\big( (\tilde{\omega}_a^2-\tilde{\omega}_b^2)^2+4 \omega_{F}^4  \big)^{1/2}}{2}
\label{eq:eigfreqs}
\end{equation}
where we use the notations $\tilde{\omega}_{a}=\frac{1}{\sqrt{(L_a \parallel L_F) C_a}}$, $\tilde{\omega}_{b}=\frac{1}{\sqrt{(L_b \parallel L_F) C_b}}$, ${\omega}^4_{F}=\frac{1}{{L_F^2 C_a C_b}}$.\\

At second order in $\epsilon$, $\omega_F^4=0$ and  we find the mode frequencies to be 
\begin{equation}
\begin{aligned}
\tilde{\omega}_{a}&=\omega_a \big( 1 + \frac{\epsilon^2}{2} r_a c(r_a c +r_b c+s)\big)\\
\tilde{\omega}_{b}&=\omega_b \big( 1 + \frac{\epsilon^2}{2}r_b c(r_a c +r_b c+s)\big)
\label{eq:finalsym}
\end{aligned}
\end{equation}
Finally, we generalize this formula to our three-mode circuit to get each mode curvature $C_i$ ($i=a,b,r$) against the cut angle $\theta$ at the saddle-point
\begin{equation}
C_i(\theta)=\omega_i r_i \cos(\theta) \big( (r_a+r_b+r_c) \cos(\theta) +\sin(\theta) \big)
\label{eq:classmodel}
\end{equation}
where we recall that $r_i=\frac{E_{\Sigma}}{E_{L_i}}$ and $\omega_i$ is the mode resonance frequency at the saddle point. \\

A few remarks are here in order.
\begin{itemize}
\item One can obtain a formula for the curvature at the other saddle point  located at $(\varphi_{\Sigma},\varphi_{\Delta})=(-\pi/2,+\pi/2)$ (corresponding to $\mathrm{SP}_1$ in Fig.~\ref{fig2}, up to a $2\pi$ offset in $\varphi_{\Delta}$)  by substituting $r_i \rightarrow -r_i$
\item Once the modes resonance frequencies at the saddle points are known (directly measurable), the curvatures of the modes frequencies against cut angle fully characterize the frequency maps in the neighborhood of the saddle points. Therefore, measuring and fitting these curvatures is an efficient and straightforward way to extract circuit parameters. The method is not sufficient to unambiguously estimate all circuit parameters though, which requires one extra constraint (below, this constraint is given by the curvature of the $|0\rangle_a \rightarrow |3\rangle_a$ transition).
\item  Imbalance of the SQUID junctions modifies the inductive  Hessian matrix \eqref{eq:hessian} following
\begin{equation}
H=\begin{pmatrix}
E_{L_a} - E_{\Delta} + F &-E_{\Delta}+ F  \\
-E_{\Delta}+F  & E_{L_b} -E_{\Delta} + F 
\end{pmatrix}
\end{equation}
where $F$ can be expressed as a function of the ratios $E_{\Sigma}/E_{L_i}$ and $E_{\Delta}/E_{L_i}$, with a similar expression substituting $E_{\Sigma}  \rightarrow -E_{\Sigma} $ and $E_{\Delta}  \rightarrow -E_{\Delta} $ at the other saddle point. We find that the frequency splitting of each mode between the two saddle point reads $\delta \omega_i = \omega_i E_{\Delta} / E_{L_i} $ which allows us to estimate $E_{\Delta}$, of the order of $10^{-5} E_{\Sigma}$ in our experiment. This small imbalance is not expected to impact the curvatures at the saddle point derived above.
\end{itemize}

\subsubsection*{Fits to the full quantum model}

\begin{figure}[htbp] 
		\centering
		\includegraphics[width=0.8\columnwidth]{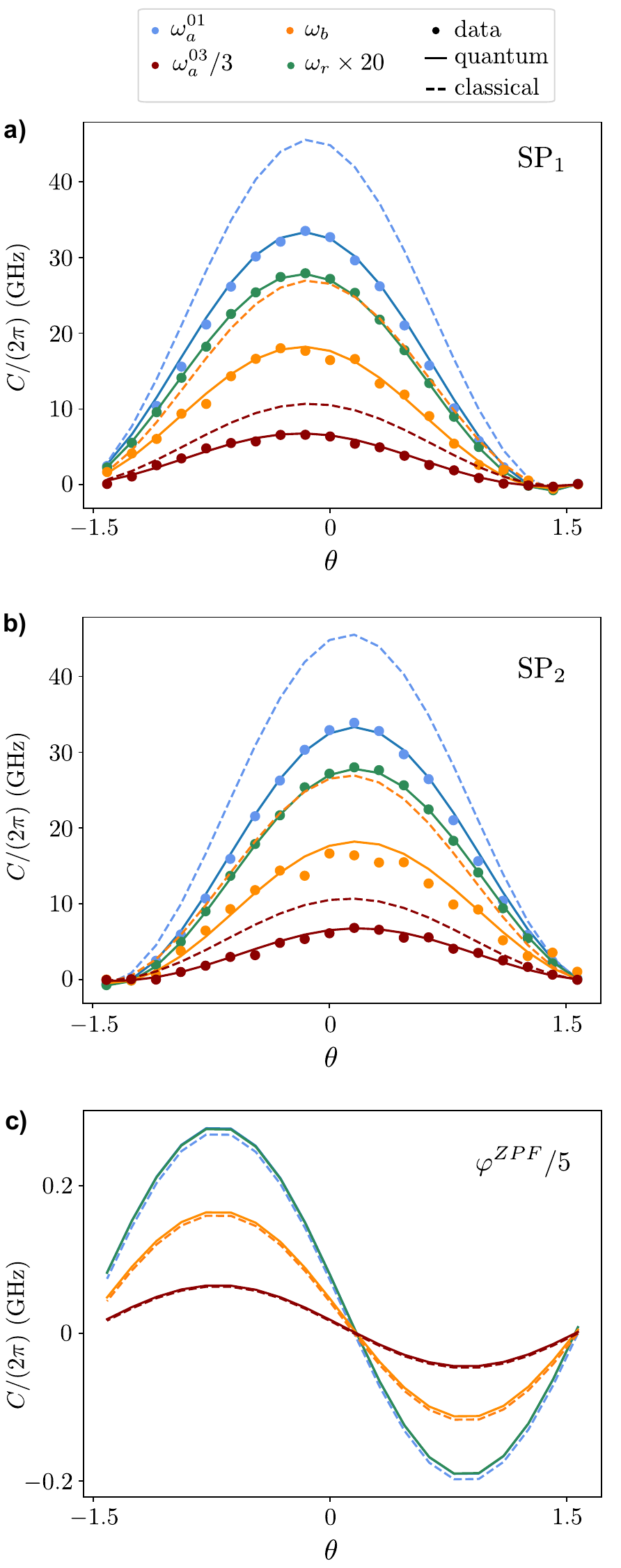}
		\caption{{\bf Curvature of the frequency maps  around the saddle points a-b)} Measured (filled circles) and simulated (plain lines) curvature of the memory and buffer low-energy transition frequency at the two non-equivalent saddle points, against cut angle $\theta$ (defined in Eq.~\eqref{eq:cutdef}). Simulated values use the model Hamiltonian \eqref{eq:Htot} fitted to the data. Dashed lines represent the predictions from the classical model \eqref{eq:classmodel} with the same parameters. These only  reproduce qualitatively the data. Note that the curvature of the readout mode has been enhanced by a factor 20 for visibility. {\bf c)} Predictions from the quantum and classical model for a hypothetical circuit  whose modes have  five times lower phase fluctuations. Here, the two models agree quantitatively. }
		\label{fig:curvatures}
\end{figure}

In the classical model detailed above, we have neglected the vacuum fluctuations of each mode. Therefore, we do not expect the predictions to be accurate when the modes vacuum fluctuations approach 1, which is the case in our experiment. After a first coarse estimate of the circuit parameters using our classical model, we refine our analysis by numerically diagonalizing the Hamiltonian \eqref{eq:Htot}. In Fig.~\ref{fig:curvatures}a-b, we represent the curvature $C_i$ of each mode resonance frequency around the saddle point against cut angle $\theta$ (see Eq.~\eqref{eq:cutdef}), together with the result of these simulations (for the estimated circuit parameters) and the predictions from the classical analysis. The latter only qualitatively reproduce the data. For comparison, we represent in Fig.~\ref{fig:curvatures}c the  predictions from the Hamiltonian diagonalization against classical predictions for a circuit whose modes have five times smaller vacuum fluctuations across the SQUID. Here, the agreement is quantitative.\\

As stated above, measuring and fitting a single transition frequency for each mode is not sufficient to  estimate independently the total Josephson energy of the SQUID $E_{\Sigma}$ and the value of the phase fluctuations $\varphi^{ZPF}_i$ of each across the SQUID (equivalently $E_{\Sigma}$ and each mode inductive energy $E_{L_i}$). Therefore, we additionally consider the dependency of the  $|0\rangle_a \rightarrow |3\rangle_a$ transition frequency against flux biases, which provides an additional constraint on the estimated parameters. Intuitively, the curvature of the anharmonicity $\omega_{03}-3\omega_{01}$ at the saddle point is set by $E_{\Sigma} {\varphi_a^{ZPF}}^4$ (at leading order of the non-linearity). On the other hand, the curvature of $\omega_{01}$ depends on $E_{\Sigma} {\varphi_a^{ZPF}}^2$ only, such that both $E_{\Sigma}$ and ${\varphi_a^{ZPF}}$  can be fitted independently.\\

We next extract the energy difference  between the two SQUID junctions $E_{\Delta}$ through the frequency splitting of the transition $|0\rangle_a \rightarrow |1\rangle_a$  at the two saddle points. As for  the parameter $e_{\Sigma}$---accounting for the non-linearity induced by series inductance in the arms of the SQUID---it is estimated via the the  value of $\omega_{03}-3\omega_{01}$ averaged over both saddle points. Given the small value of these two parameters, they only marginally impact the fits of the  curvatures considered above. Finally we note that the Kerr non-linearity inherited from the circuit superinductances  is also included in our numerical model  but has no significant impact on the numerical results ($E_L$ is estimated from the nominal value of the superinductors). \\

The fitted  values for all parameters are summarized in Table.~\ref{tableparam}.

\subsection{Dynamics under the four-to one photon exchange interaction}
\label{sec:fourtoonedynamics}

In this section, we detail our model for the memory and buffer dynamics under the pump microwave addressing the $|4\rangle_a|0\rangle_b \leftrightarrow |0\rangle_a|1\rangle_b$. We then detail the fitting procedure for the datasets presented in Fig.~\ref{fig3}b-c. \\

\subsubsection*{Model dynamics}

We start from the circuit Hamiltonian~\eqref{eq:Htot}, which is time-dependent for  flux biases $(\varphi_{\Sigma},\varphi_{\Delta})=(-\frac{\pi}{2}+\epsilon(t),-\frac{\pi}{2})$ that include a small AC component ($\epsilon(t)\ll 1$). Neglecting the readout mode, expanding at first order in $\epsilon$ and including a direct linear drive on the buffer (needed to reproduce the buffer reflection spectra in Fig.~\ref{fig3}c), we get  
\begin{equation}
\begin{split}
\HH(t)=& \hbar \omega_a \adag \aop + \hbar \omega_b \bdag \bop + \epsilon(t) E_{\Sigma} \sin(\varphiop)\\
&+ E_{\Delta} \cos(\varphiop) + e_{\Sigma}\cos(2\varphiop) + \frac{E_L}{N^2} \big( \cos(\varphiop) + \frac{1}{2} \varphiop^2 \big)\\
&+ \epsilon_b(t) (\bop + \bdag)
\end{split}
\label{eq:Hlab}
\end{equation}
where $\varphiop=\varphi_a^{ZPF}(\aop +\adag)+\varphi_b^{ZPF}(\bop +\bdag)$. Our goal is to derive an effective static Hamiltonian allowing for numerical simulation. To this end, we move to a rotating frame for both modes, and select resonant or near-resonant terms. \\

We assume the pump to be near-resonant with the four-to-one photon transition $\epsilon(t)=\xi e^{i(4 \tilde{\omega}_a - \tilde{\omega}_b+ \Delta)t} + \mathrm{c.c}.$ and  the linear drive  to be near resonant with the  buffer $\epsilon_b(t)=\xi_b e^{i( \tilde{\omega}_b + \delta) t} + \mathrm{c.c}.$ with $\Delta, \delta \ll \tilde{\omega}_a,\tilde{\omega}_b$. Here, $\tilde{\omega}_{a,b}$ denote the modes resonance frequencies dressed by the pump, which are assumed to be perfectly  calibrated by the four-step procedure detailed in Sec.~\ref{sec:tuningPump}.  When simulating the time-decay of the state $|4\rangle_a$, we thus set $\Delta=0$ and $\xi_b=0$ (pump on resonance, no buffer drive, see Fig.~\ref{fig3}b). When reproducing the buffer reflection spectra (Fig.~\ref{fig3}c), $\delta$ is the detuning between the applied buffer drive and the buffer resonance frequency dressed by the pump. We give a  non-zero value to  $\Delta=\omega_{b_3}-\omega_{b_1}$ to account for the slightly off-resonant pump (we use the convention of the four-step calibration procedure and recall that the spectra are recorded at step 3). In practice, this small detuning has no visible impact on the simulation results.\\

We now move  to frames rotating with both the pump and the linear drive
\begin{equation}
\begin{split}
a& \rightarrow a^{-i\big(\tilde{\omega}_a + \frac{\Delta+\delta}{4}\big)t}\\
b& \rightarrow b^{-i(\tilde{\omega}_b + \delta)t}.
\end{split}
\end{equation} \\
First considering the  terms on the second line of the Hamiltonian \eqref{eq:Hlab} stemming from non-idealities of the circuit, we discard contributions that are fast-oscillating in the rotating frame, i.e. which do not preserve photon number. This amounts to keeping only diagonal terms in the Fock basis and we end up with a diagonal Hamiltonian
\begin{equation}
\HH_{\chi} = \sum_{n,m} \chi_{nm}  |n\rangle_a |m\rangle_b \langle n|_a \langle m|_b
\label{eq:Hchi}
\end{equation}
where the dispersive shifts $\chi_{nm}$ are computed from the non-driven version of Hamiltonian~\eqref{eq:Hlab} (setting $\epsilon=\epsilon_b=0$). In detail, we first diagonalize the non-driven Hamiltonian and label $(n,m)$ each eigenstate by identifying it to the Fock state $|n\rangle_a |m\rangle_b$ it maximally overlaps with~\cite{note11}. Next, we label each state  energy $E_{n,m}$ after offsetting the energy scale such that $E_{0,0}=0$. Finally, we find the dispersive shift  $\chi_{nm}=E_{n,m} - n \frac{E_{4,0}}{4} - m E_{0,1}$. We note that with our definitions, the dispersive Hamiltonian $\HH_{\chi}$ does not contribute to the energy splitting between $|4\rangle_a |0\rangle_b$ and $|0\rangle_a |1\rangle_b$, which is expected given our choice of rotating frames. We also note that by computing $\HH_{\chi}$ from the non-driven Hamiltonian~\eqref{eq:Hlab}, we have neglected a small renormalization by the pump. In practice, the exact value of $\HH_{\chi}$ does not significantly impact numerical simulations in the regime we are in, as detailed below.\\

Now turning to the term $\epsilon(t) E_{\Sigma} \sin(\varphiop)$, we get in the RWA a resonant four-to-one photon exchange interaction at rate $g_4= \frac{E_{\Sigma}}{\hbar 4!} \xi e^{-(\varphi^{{ZPF}^2}_a+\varphi^{{ZPF}^2}_b)/2}\varphi^{{ZPF}^4}_a \varphi^{ZPF}_b $ (see Sec.~\ref{sec:targetparams}). Overall, we get the effective static Hamiltonian
\begin{equation}
\begin{split}
\HH_{RWA}=&-\frac{\Delta+\delta}{4} \aop^{\dagger} \aop - \delta \bop^{\dagger } \bop \\
&+ \HH_{\chi} + g_4  (\aop^{\dagger 4} \bop + \aop^4 \bdag) + \xi_b  \bop + \xi_b^{\ast} \bdag
\end{split}
\label{eq:Hrwa}
\end{equation} 
Finally, we model dissipation and dephasing in our system by considering the following Lindblad dissipators
\begin{itemize}
\item $\gamma_k\mathcal{D}[|k-1\rangle \langle k |]$ for $1 \leq i \leq 4 $. These  dissipators account for the  relaxation  of population in state $|k\rangle$ through single-photon emission.  The rates $\gamma_k$ for $k\leq 3$ are independently calibrated for each pump amplitude (see Sec.~\ref{sec:gammaslow}). Since we do not consider any other decay mechanism from these states, they  are identical to the rates $\Gamma_k$ reported in Fig.~\ref{fig4}. Note that their values  do not follow the expected linear scaling in $k$, preventing us from modeling single-photon dissipation with a single dissipator $\mathcal{D}[\aop]$. Given the memory anharmonicity, this behavior is expected if the coupling to the bath in which single-photons are emitted---or the bath density of states---is not constant in frequency.  As for $\gamma_4$, it cannot be unambiguously estimated for a non-zero pump amplitude since decay from state $|4\rangle_a$ under pump is dominated by the engineered four-photon dissipation channel. We therefore assume $\gamma_4$ to be constant over all considered pump amplitudes.
\item $\kappa_b\mathcal{D}[\bop]$. This dissipator accounts for single-photon dissipation in the buffer. Here, we assume $\kappa_b$ to be constant for all pump amplitudes as radiative decay through the charge  line is the dominant contribution to single-photon relaxation. Moreover the buffer is negligibly excited beyond the first excited state in our experiment, such that we do not need to adjust the relaxation rate for higher excited states in our model. $\kappa_b$ is independently calibrated by fitting the buffer reflection spectra when the pump is off.
\item $2\kappa_{\phi} \mathcal{D}[\adag \aop]$. This dissipator accounts for pure dephasing of the memory. The dephasing rate is fitted together with the datasets presented in Fig.~\ref{fig3} and is in good agreement with an independent measurement of the dephasing of the $|0\rangle_a \leftrightarrow |1\rangle_a$ transition. Note that this rate is enhanced by the application of the off-resonant pump employed to shift the memory transitions away from spurious TLS resonances (see Sec.~\ref{sec:TLS}) and was seen to fluctuate over the course of hours.
\end{itemize}

\subsubsection*{Fitting procedure}

Our goal here is to fit the datasets presented in Fig.~\ref{fig3}b-c with the model presented above. The fit parameters are  
\begin{itemize}
\item The value of $g_4$ for each pump amplitude. The fitted values are reported in \ref{fig3}d.
\item The displacement rate $\xi_b$ entailed by the linear drive on the buffer. The fitted value is $\xi_b=2\pi\times 175~$kHz.
\item The memory pure dephasing rate $\kappa_{\phi}$ (value reported in Table.~\ref{tableparam}).
\item A proportionality factor between the differential readout signal $S_4$ and the occupation of state $|4\rangle_a$ (determining the y-scale  in Fig.~\ref{fig3}b). 
\item A scaling factor accounting for the cryostat input and output lines attenuation (determining the amplitude of off-resonant waves reflected from the buffer in Fig.~\ref{fig3}c and Fig.~\ref{fig:bufferSpectra}).
\end{itemize}

In order to limit computational costs when fitting, we simplify our model based on the exact experiment that needs to be reproduced numerically. Before detailing and justifying these simplifications, we note that they are taken solely when fitting experimental data to extract unknown parameter values. Using the fitted values, we then run the model described above with no simplifying assumption to produce the simulated curves presented in Fig.~\ref{fig3}.  \\

\begin{figure}[htbp] 
		\centering
		\includegraphics[width=0.7\columnwidth]{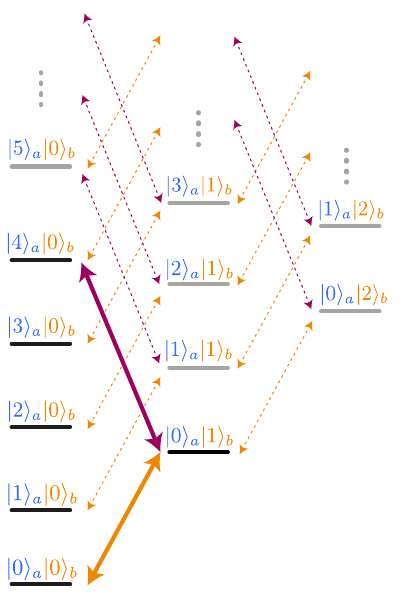}
		\caption{ {\bf Simplified six-level model} employed to fit the datasets presented in Fig.~\ref{fig4}. The pump (plain purple arrow) and the linear drive applied to the buffer (plain orange arrow) are each resonant with a single transition only (detuned transitions represented by dashed arrows). As a result, the system state is confined to a six-level manifold (levels pictured in black) and never visits other states (levels pictured in gray). }
		\label{fig:simplemodel}
\end{figure}
We first notice that for both datasets presented Fig.~\ref{fig3}b-c, the system state is constrained to a 6-dimension manifold spanned by $\{ |0\rangle_a |0\rangle_b, |1\rangle_a |0\rangle_b, |2\rangle_a |0\rangle_b, |3\rangle_a |0\rangle_b, |4\rangle_a |0\rangle_b, |0\rangle_a |1\rangle_b  \}$ (see Fig.~\ref{fig:simplemodel}). Indeed, the pump (purple arrow) is only resonant with the $|4\rangle_a |0\rangle_b \leftrightarrow |0\rangle_a |1\rangle_b$ transition and is weak enough that no other transition can be driven 
($g_4 \sqrt {\frac{(n+4)!}{n!}(m+1)}\ll |(\chi_{n+4,m}-\chi_{n,m+1})|$ for $n>0$ or $m>0$). Similarly, the linear drive on the buffer (only on when recording the reflection spectra in Fig.~\ref{fig3}c) is  resonant with the $|0\rangle_a |0\rangle_b \leftrightarrow |0\rangle_a |1\rangle_b$ transition only ($\xi_b \sqrt {m+1}\ll |(\chi_{n,m+1}-\chi_{n,m})|$ for $n>0$ or $m>0$).\\

For the data presented in Fig.~\ref{fig3}b, we can further simplify our model by noticing that when the system relaxes from state $|4\rangle_a |0\rangle_b$ (either though single-photon relaxation to $|3\rangle_a |0\rangle_b$ at rate $\gamma_4$ or through a  four-to-one photon exchange driven by the pump), it never returns to $|4\rangle_a |0\rangle_b$. Since we are only interested in computing the decay rate from $|4\rangle_a|0\rangle_b$, we can consider an effective three-level system $\{ |D\rangle, |A\rangle, |B\rangle \}$ with $|A\rangle =|4\rangle_a |0\rangle_b$, $|B\rangle =|0\rangle_a |1\rangle_b $ and $|D\rangle$  a dark state to which $|A\rangle$ relaxes at rate $\gamma_4$ and $|B\rangle$ relaxes at rate $\kappa_b$. Finally, we use the Lindblad-adjoint formalism~\cite{robin2024convergence} to encode the evolution of the expectation value of the operators $|A\rangle \langle A|$, $|B\rangle \langle B|$, $|A\rangle \langle B|$, and $|B\rangle \langle A|$ in a $4\times 4$ evolution matrix. Note that beside allowing fast simulations of the evolution of the expectation value of $|A\rangle \langle A|$ (used to fit the data in Fig.~\ref{fig3}b), we also perform spectral analysis of this matrix to extract the  relaxation rate $\Gamma_4$  corresponding to the  matrix eigenvalue closest to zero (reported in Fig.~\ref{fig4}). \\

For the data presented in Fig.~\ref{fig3}c, reduction of the system Hilbert space is less obvious. Here, we are interested in the steady-state value of $\langle \bop \rangle$. We directly apply the Lindblad adjoint formalism to encode the evolution of $\langle \bop \rangle$ along with the expectation value of ten other orthogonal operators (which we compute numerically and do not detail here) in a $11\times 11$ matrix $M$. The steady-state is directly found from the single vector in the kernel of $M$.\\

With these model simplifications, we  perform a global fit of the two datasets in Fig.~\ref{fig3}b-c with a simple least-square method. As a sanity check of our simplifying assumptions, we finally perform a full Lindblad master equation  featuring the effective Hamiltonian $\HH_{RWA}$ and the dissipators $\kappa_b\mathcal{D}[\bop]$, $2\kappa_{\phi} \mathcal{D}[\adag \aop]$ and $\gamma_k\mathcal{D}[|k-1\rangle \langle k |]$ for $1 \leq i \leq 4 $, with  parameter values extracted from the above fit. The results of these simulations are plotted in Fig.~\ref{fig3}b-c and Fig.~\ref{fig:bufferSpectra} and quantitatively reproduce our data.

\section{Tuning the system away from TLS resonances}
\label{sec:TLS}

\begin{figure}[htbp] 
		\centering
		\includegraphics[width=0.9\columnwidth]{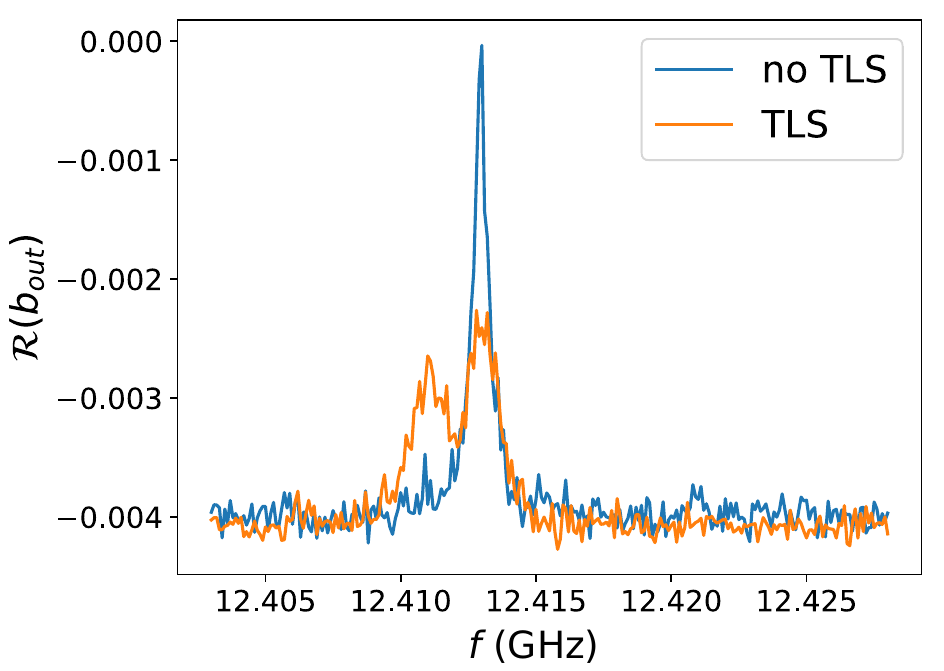}
		\caption{ {\bf Signature of hybridization to a Two-Level System.} Response of the system when probed in two-tone spectroscopy around the resonance frequency of the $|0\rangle_a \leftrightarrow |3\rangle_a$ transition (driven through the flux lines). The two traces are taken 8 minutes apart. On the later one (orange), the line is split in two peaks of approximately equal height.  }
		\label{fig:TLS}
\end{figure}

In our experiment,  some of the memory transitions---either  single-photon or  three-photon transitions---were  transiently observed  to shift or split. Typically, a given transition that would yield  a clean Lorentzian response in two-tone spectroscopy would suddenly yield a two-peak response. This behavior was either be permanent (until the system was cycled around 20~K before being cooled back down at base temperature) or appearing on and off on a minute timescale (see Fig.~\ref{fig:TLS}). The buffer lowest energy transition was also witnessed to split at one occasion.\\

We attribute this effect to  hybridization of the system to a bath of Two-Level-Systems (TLS). Such TLS baths, whose frequency landscape reconfigure on timescales of the order of a second, are frequently observed in superconducting circuits~\cite{oliver2013materials,muller2019towards}. While the exact coupling mechanism and density of TLS's was not characterized in our experiment,  the probability that at least one of the memory or buffer transitions be affected by the TLS bath was high enough that waiting for the bath to reconfigure in a more favorable configuration  was not a viable option. Instead, we use an off-resonant  flux drive of large amplitude  to shift the frequency of the circuit transitions away from any resonant coupling with TLS's.\\

Various mechanisms can contribute to a transition frequency shift proportional to the power of  an  off-resonant drive. In particular, a flux drive applied to the ATS loops with both a symmetric and an anti-symmetric component yields such a shift (see Sec.~\ref{sec:smNL}). The Stark effect~\cite{gambetta2006qubit} is another possible mechanism owing to the modes non-linearity and given that their states are linearly displaced by the drive. Such a displacement is expected   both for a symmetric flux drive (through the leading term in the ATS \textit{sine} Hamiltonian) and for an anti-symmetric flux drive (through induction in the circuit loops~\cite{you2019circuit}). This latter mechanism is also expected to degrade the system coherence through measurement-induced dephasing~\cite{gambetta2006qubit}. Measurement-induced transitions are also a concern for some specific drive frequencies~\cite{petrescu2020lifetime,venkatraman2024nonlinear,putterman2025preserving,carde2025flux}. While we did not quantitatively investigate these mechanisms in our experiment, the fact that a frequency shift of the circuit transitions is expected at any flux drive frequency lets us select the one which minimizes decoherence.\\

\begin{figure}[htbp] 
		\centering
		\includegraphics[width=0.9\columnwidth]{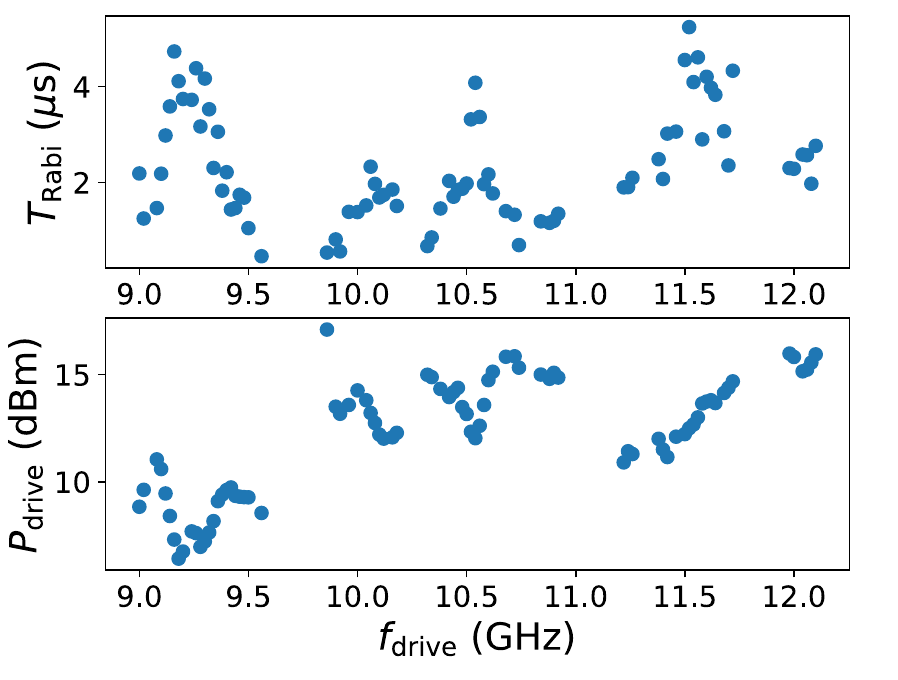}
		\caption{ {\bf Shifting the system transitions out of resonance from TLS's}. A controlled shift of all transitions is realized by the application of a strong off-resonant drive (labeled \textit{Stark shifting pump} in Fig.~\ref{fig:wiring}). In order to select the drive frequency and power minimizing decoherence, we sweep the drive frequency and adjust its power to reach a Stark shift of 17~MHz for the memory $|0\rangle_a \leftrightarrow |1\rangle_a$ transition (corresponding power referenced at the microwave source output port reported in the bottom panel). We then record Rabi oscillations while this off-resonant drive is on (top panel), and select the drive frequency for which the oscillations are longest-lived (at 9.25~GHz). }
		\label{fig:pumpstark}
\end{figure}

In detail, we use the characteristic decay time of Rabi oscillations driven on the memory $|0\rangle_a \leftrightarrow |1\rangle_a$ transition as a proxy for the system coherence. As illustrated in Fig.~\ref{fig:pumpstark}, we sweep the frequency of an off-resonant drive over a few GHz range. For each frequency, we adjust the drive power to get a 17~MHz shift of the  $|0\rangle_a \leftrightarrow |1\rangle_a$ transition (which is approximately the value used in our experiment) and record Rabi oscillations in these conditions. We then select the off-resonant drive frequency yielding the longest-lived Rabi oscillations (at 9.25~GHz).\\

When acquiring the datasets presented in Fig.~\ref{fig3}b-c, we apply this off-resonant drive at a power of 5.3~dBm, referenced as in Fig.~\ref{fig:pumpstark} (at the output of the microwave source labeled \textit{Stark shifting pump} in Fig.~\ref{fig:wiring}). The drive is turned on during the system preparation, application of the four-photon dissipation channel and the measurement. In these conditions, the $T_1$ of the $|0\rangle_a \leftrightarrow |1\rangle_a$ transition is measured to be of 7.5~$\mu$s (see Fig.~\ref{fig:gammaslow} for a null pump amplitude) and in a later measurement, we observed the Ramsey free-induction decay time  to be of 3.5~$\mu$s, corresponding to a pure dephasing rate of $2\pi\times35$~kHz. This is in good agreement with the value fitted from the datasets presented in Fig.~\ref{fig3}b-c. These lifetimes are slightly degraded when compared to those measured in absence of the off-resonant drive ($T_1\lesssim 11~\mu$s, $T_2\lesssim  6~\mu$s), even though these were seen to fluctuate over the course of days.

\bibliography{biblio.bib}

@article{rojkov2024stabilization,
  title={Stabilization of cat-state manifolds using nonlinear reservoir engineering},
  author={Rojkov, Ivan and Simoni, Matteo and Zapusek, Elias and Reiter, Florentin and Home, Jonathan},
  journal={arXiv preprint arXiv:2407.18087},
  year={2024}
}

@article{touzard2019gated,
  title={Gated conditional displacement readout of superconducting qubits},
  author={Touzard, S and Kou, A and Frattini, NE and Sivak, VV and Puri, S and Grimm, A and Frunzio, L and Shankar, S and Devoret, MH},
  journal={Physical Review Letters},
  volume={122},
  number={8},
  pages={080502},
  year={2019},
  publisher={APS}
}

@article{braumuller2020characterizing,
  title={Characterizing and optimizing qubit coherence based on {SQUID} geometry},
  author={Braum{\"u}ller, Jochen and Ding, Leon and Veps{\"a}l{\"a}inen, Antti P and Sung, Youngkyu and Kjaergaard, Morten and Menke, Tim and Winik, Roni and Kim, David and Niedzielski, Bethany M and Melville, Alexander and others},
  journal={Physical Review Applied},
  volume={13},
  number={5},
  pages={054079},
  year={2020},
  publisher={APS}
}

@article{putterman2025preserving,
  title={Preserving phase coherence and linearity in cat qubits with exponential bit-flip suppression},
  author={Putterman, Harald and Noh, Kyungjoo and Patel, Rishi N and Peairs, Gregory A and MacCabe, Gregory S and Lee, Menyoung and Aghaeimeibodi, Shahriar and Hann, Connor T and Jarrige, Ignace and Marcaud, Guillaume and others},
  journal={Physical Review X},
  volume={15},
  number={1},
  pages={011070},
  year={2025},
  publisher={APS}
}

@article{smith2023spectral,
  title={Spectral signature of high-order photon processes enhanced by Cooper-pair pairing},
  author={Smith, WC and Borgognoni, A and Villiers, M and Roverc’h, E and Palomo, J and Delbecq, MR and Kontos, T and Campagne-Ibarcq, P and Dou{\c{c}}ot, B and Leghtas, Z},
  journal={Nature Communications},
  volume={16},
  number={1},
  pages={8359},
  year={2025},
  publisher={Nature Publishing Group UK London}
}

@article{lescanne2020exponential,
  title={Exponential suppression of bit-flips in a qubit encoded in an oscillator},
  author={Lescanne, Rapha{\"e}l and Villiers, Marius and Peronnin, Th{\'e}au and Sarlette, Alain and Delbecq, Matthieu and Huard, Benjamin and Kontos, Takis and Mirrahimi, Mazyar and Leghtas, Zaki},
  journal={Nature Physics},
  volume={16},
  number={5},
  pages={509--513},
  year={2020},
  publisher={Nature Publishing Group}
}

@article{berdou2023one,
  title={One hundred second bit-flip time in a two-photon dissipative oscillator},
  author={Berdou, Camille and Murani, Anil and Reglade, Ulysse and Smith, William C and Villiers, Marius and Palomo, Jos{\'e} and Rosticher, Michael and Denis, A and Morfin, Pascal and Delbecq, Matthieu and others},
  journal={PRX Quantum},
  volume={4},
  number={2},
  pages={020350},
  year={2023},
  publisher={APS}
}

@article{de2022error,
  title={Error correction of a logical grid state qubit by dissipative pumping},
  author={De Neeve, Brennan and Nguyen, Thanh-Long and Behrle, Tanja and Home, Jonathan P},
  journal={Nature Physics},
  volume={18},
  number={3},
  pages={296--300},
  year={2022},
  publisher={Nature Publishing Group UK London}
}

@article{royer2020stabilization,
  title={Stabilization of finite-energy {G}ottesman-{K}itaev-{P}reskill states},
  author={Royer, Baptiste and Singh, Shraddha and Girvin, SM},
  journal={Physical Review Letters},
  volume={125},
  number={26},
  pages={260509},
  year={2020},
  publisher={APS}
}

@article{touzard2018coherent,
  title={Coherent oscillations inside a quantum manifold stabilized by dissipation},
  author={Touzard, Steven and Grimm, Alexander and Leghtas, Zaki and Mundhada, Shantanu O and Reinhold, Philip and Axline, Christopher and Reagor, Matt and Chou, Kevin and Blumoff, Jacob and Sliwa, Katrina M and others},
  journal={Physical Review X},
  volume={8},
  number={2},
  pages={021005},
  year={2018},
  publisher={APS}
}

@article{mundhada2019experimental,
  title={Experimental implementation of a {R}aman-assisted eight-wave mixing process},
  author={Mundhada, Shantanu O and Grimm, Alexander and Venkatraman, Jayameenakshi and Minev, Zlatko K and Touzard, Steven and Frattini, Nicholas E and Sivak, Volodymyr V and Sliwa, Katrina and Reinhold, Philip and Shankar, Shyam and others},
  journal={Physical Review Applied},
  volume={12},
  number={5},
  pages={054051},
  year={2019},
  publisher={APS}
}

@article{grimsmo2020quantum,
  title={Quantum computing with rotation-symmetric bosonic codes},
  author={Grimsmo, Arne L and Combes, Joshua and Baragiola, Ben Q},
  journal={Physical Review X},
  volume={10},
  number={1},
  pages={011058},
  year={2020},
  publisher={APS}
}

@article{marquet2024autoparametric,
  title={Autoparametric resonance extending the bit-flip time of a cat qubit up to 0.3 s},
  author={Marquet, Antoine and Essig, Antoine and Cohen, Joachim and Cottet, Nathana{\"e}l and Murani, Anil and Albertinale, E and Dupouy, Simon and Bienfait, Audrey and Peronnin, Th{\'e}au and Jezouin, S{\'e}bastien and others},
  journal={Physical Review X},
  volume={14},
  number={2},
  pages={021019},
  year={2024},
  publisher={APS}
}

@article{putterman2024hardware,
  title={Hardware-efficient quantum error correction via concatenated bosonic qubits},
  author={Putterman, Harald and Noh, Kyungjoo and Hann, Connor T and MacCabe, Gregory S and Aghaeimeibodi, Shahriar and Patel, Rishi N and Lee, Menyoung and Jones, William M and Moradinejad, Hesam and Rodriguez, Roberto and others},
  journal={Nature},
  volume={638},
  number={8052},
  pages={927--934},
  year={2025},
  publisher={Nature Publishing Group UK London}
}

@article{reglade2024quantum,
  title={Quantum control of a cat qubit with bit-flip times exceeding ten seconds},
  author={R{\'e}glade, Ulysse and Bocquet, Adrien and Gautier, Ronan and Cohen, J and Marquet, A and Albertinale, E and Pankratova, N and Hall{\'e}n, M and Rautschke, F and Sellem, L-A and others},
  journal={Nature},
  pages={1--6},
  year={2024},
  publisher={Nature Publishing Group UK London}
}

@article{mirrahimi2014dynamically,
  title={Dynamically protected cat-qubits: a new paradigm for universal quantum computation},
  author={Mirrahimi, Mazyar and Leghtas, Zaki and Albert, Victor V and Touzard, Steven and Schoelkopf, Robert J and Jiang, Liang and Devoret, Michel H},
  journal={New J. Phys.},
  volume={16},
  number={4},
  pages={045014},
  year={2014},
  publisher={IOP Publishing}
}

@article{leghtas2015confining,
  title={Confining the state of light to a quantum manifold by engineered two-photon loss},
  author={Leghtas, Zaki and Touzard, Steven and Pop, Ioan M and Kou, Angela and Vlastakis, Brian and Petrenko, Andrei and Sliwa, Katrina M and Narla, Anirudh and Shankar, Shyam and Hatridge, Michael J and others},
  journal={Science},
  volume={347},
  number={6224},
  pages={853--857},
  year={2015},
  publisher={American Association for the Advancement of Science}
}

@book{gottesman1997stabilizer,
  title={Stabilizer codes and quantum error correction},
  author={Gottesman, Daniel},
  year={1997},
  publisher={California Institute of Technology}
}

@article{ma2020path,
  title={Path-independent quantum gates with noisy ancilla},
  author={Ma, Wen-Long and Zhang, Mengzhen and Wong, Yat and Noh, Kyungjoo and Rosenblum, Serge and Reinhold, Philip and Schoelkopf, Robert J and Jiang, Liang},
  journal={Physical Review Letters},
  volume={125},
  number={11},
  pages={110503},
  year={2020},
  publisher={APS}
}

@article{vy2013error,
  title={Error-transparent evolution: the ability of multi-body interactions to bypass decoherence},
  author={Vy, Os and Wang, Xiaoting and Jacobs, Kurt},
  journal={New Journal of Physics},
  volume={15},
  number={5},
  pages={053002},
  year={2013},
  publisher={IOP Publishing}
}

@article{reinhold2020error,
  title={Error-corrected gates on an encoded qubit},
  author={Reinhold, Philip and Rosenblum, Serge and Ma, Wen-Long and Frunzio, Luigi and Jiang, Liang and Schoelkopf, Robert J},
  journal={Nature Physics},
  volume={16},
  number={8},
  pages={822--826},
  year={2020},
  publisher={Nature Publishing Group UK London}
}

@article{ding2025quantum,
  title={Quantum control of an oscillator with a {K}err-cat qubit},
  author={Ding, Andy Z and Brock, Benjamin L and Eickbusch, Alec and Koottandavida, Akshay and Frattini, Nicholas E and Corti{\~n}as, Rodrigo G and Joshi, Vidul R and de Graaf, Stijn J and Chapman, Benjamin J and Ganjam, Suhas and others},
  journal={Nature Communications},
  volume={16},
  number={1},
  pages={1--7},
  year={2025},
  publisher={Nature Publishing Group}
}

@article{puri2018stabilized,
  title = {Stabilized Cat in a Driven Nonlinear Cavity: A Fault-Tolerant Error Syndrome Detector},
  author = {Puri, Shruti and Grimm, Alexander and Campagne-Ibarcq, Philippe and Eickbusch, Alec and Noh, Kyungjoo and Roberts, Gabrielle and Jiang, Liang and Mirrahimi, Mazyar and Devoret, Michel H. and Girvin, S. M.},
  journal = {Phys. Rev. X},
  volume = {9},
  issue = {4},
  pages = {041009},
  numpages = {29},
  year = {2019},
  month = {Oct},
  publisher = {American Physical Society},
  doi = {10.1103/PhysRevX.9.041009},
  url = {https://link.aps.org/doi/10.1103/PhysRevX.9.041009}
}

@article{petrescu2020lifetime,
  title={Lifetime renormalization of driven weakly anharmonic superconducting qubits. {I}{I}. The readout problem},
  author={Petrescu, Alexandru and Malekakhlagh, Moein and T{\"u}reci, Hakan E},
  journal={Physical Review B},
  volume={101},
  number={13},
  pages={134510},
  year={2020},
  publisher={APS}
}

@article{venkatraman2024nonlinear,
  title={Nonlinear dissipation in a driven superconducting circuit},
  author={Venkatraman, Jayameenakshi and Xiao, Xu and Corti{\~n}as, Rodrigo G and Devoret, Michel H},
  journal={Physical Review A},
  volume={110},
  number={4},
  pages={042411},
  year={2024},
  publisher={APS}
}

@article{rosenblum2018fault,
  title={Fault-tolerant detection of a quantum error},
  author={Rosenblum, Serge and Reinhold, Philip and Mirrahimi, Mazyar and Jiang, Liang and Frunzio, Luigi and Schoelkopf, Robert J},
  journal={Science},
  volume={361},
  number={6399},
  pages={266--270},
  year={2018},
  publisher={American Association for the Advancement of Science}
}

@article{nigg2012black,
  title={Black-box superconducting circuit quantization},
  author={Nigg, Simon E and Paik, Hanhee and Vlastakis, Brian and Kirchmair, Gerhard and Shankar, Shyam and Frunzio, Luigi and Devoret, MH and Schoelkopf, RJ and Girvin, SM},
  journal={Phys. Rev. Lett.},
  volume={108},
  number={24},
  pages={240502},
  year={2012},
  publisher={APS}
}

@article{smith2016quantization,
  title={Quantization of inductively shunted superconducting circuits},
  author={Smith, WC and Kou, A and Vool, U and Pop, IM and Frunzio, L and Schoelkopf, RJ and Devoret, MH},
  journal={Physical Review B},
  volume={94},
  number={14},
  pages={144507},
  year={2016},
  publisher={APS}
}

@book{nielsen2002quantum,
address = {Cambridge},
author = {Nielsen, Michael A. and Chuang, Isaac L.},
doi = {10.1017/CBO9780511976667},
isbn = {9780511976667},
publisher = {Cambridge University Press},
title = {{Quantum Computation and Quantum Information}},
year = {2010}
}

@article{gertler2021protecting,
  title={Protecting a bosonic qubit with autonomous quantum error correction},
  author={Gertler, Jeffrey M and Baker, Brian and Li, Juliang and Shirol, Shruti and Koch, Jens and Wang, Chen},
  journal={Nature},
  volume={590},
  number={7845},
  pages={243--248},
  year={2021},
  publisher={Nature Publishing Group}
}

@phdthesis{manucharyan2012superinductance,
  title={Superinductance},
  author={Manucharyan, VE},
  year={2012},
}

@article{minev2021energy,
  title={Energy-participation quantization of {Josephson} circuits},
  author={Minev, Zlatko K and Leghtas, Zaki and Mundhada, Shantanu O and Christakis, Lysander and Pop, Ioan M and Devoret, Michel H},
  journal={npj Quantum Information},
  volume={7},
  number={1},
  pages={131},
  year={2021},
  publisher={Nature Publishing Group UK London}
}

@article{putterman2022stabilizing,
  title={Stabilizing a Bosonic Qubit Using Colored Dissipation},
  author={Putterman, Harald and Iverson, Joseph and Xu, Qian and Jiang, Liang and Painter, Oskar and Brand{\~a}o, Fernando GSL and Noh, Kyungjoo},
  journal={Physical Review Letters},
  volume={128},
  number={11},
  pages={110502},
  year={2022},
  publisher={APS}
}

@article{foster1924reactance,
  title={A reactance theorem},
  author={Foster, Ronald M},
  journal={Bell System technical journal},
  volume={3},
  number={2},
  pages={259--267},
  year={1924},
  publisher={Wiley Online Library}
}

@misc{lecturenotesmirrahimirouchon,
	author = {Mirrahimi, Mazyar and Rouchon, Pierre},
	howpublished = {Lecture Notes},
	year = {2015},
	title = {Dynamics and Control of Open Quantum Systems},
	url = {https://cas.mines-paristech.fr/~rouchon/MasterUPMC/LectureNotes-03-18.pdf}
}

@article{siegele2023robust,
  title = {Robust suppression of noise propagation in {{Gottesman-Kitaev-Preskill}} error correction},
  author={Siegele, Christian and Campagne-Ibarcq, Philippe},
  journal = {Phys. Rev. A},
  volume = {108},
  issue = {4},
  pages = {042427},
  numpages = {18},
  year = {2023},
  month = {Oct},
  publisher = {American Physical Society},
  doi = {10.1103/PhysRevA.108.042427},
  url = {https://link.aps.org/doi/10.1103/PhysRevA.108.042427}
}

@article{sellem2025dissipative,
  title={Dissipative Protection of a {GKP} Qubit in a High-Impedance Superconducting Circuit Driven by a Microwave Frequency Comb},
  author={Sellem, L-A and Sarlette, A and Leghtas, Z and Mirrahimi, M and Rouchon, P and Campagne-Ibarcq, P},
  journal={Physical Review X},
  volume={15},
  number={1},
  pages={011011},
  year={2025},
  publisher={APS}
}

@inproceedings{robin2024convergence,
  title={Convergence of bipartite open quantum systems stabilized by reservoir engineering},
  author={Robin, R{\'e}mi and Rouchon, Pierre and Sellem, Lev-Arcady},
  booktitle={Annales Henri Poincar{\'e}},
  pages={1--51},
  year={2024},
  organization={Springer}
}

@incollection{albert2025bosonic,
  title={Bosonic coding: introduction and use cases},
  author={Albert, Victor V},
  booktitle={Quantum Fluids of Light and Matter},
  pages={79--107},
  year={2025},
  publisher={IOS Press}
}

@article{gottesman2001encoding,
  title={Encoding a qubit in an oscillator},
  author={Gottesman, Daniel and Kitaev, Alexei and Preskill, John},
  journal={Physical Review A},
  volume={64},
  number={1},
  pages={012310},
  year={2001},
  publisher={APS}
}

@article{muller2019towards,
  title={Towards understanding two-level-systems in amorphous solids: insights from quantum circuits},
  author={M{\"u}ller, Clemens and Cole, Jared H and Lisenfeld, J{\"u}rgen},
  journal={Reports on Progress in Physics},
  volume={82},
  number={12},
  pages={124501},
  year={2019},
  publisher={IOP Publishing}
}

@article{oliver2013materials,
  title={Materials in superconducting quantum bits},
  author={Oliver, William D and Welander, Paul B},
  journal={MRS bulletin},
  volume={38},
  number={10},
  pages={816--825},
  year={2013},
  publisher={Cambridge University Press}
}

@article{carde2025flux,
  title={Flux-pump-induced degradation of T 1 for dissipative cat qubits},
  author={Carde, L{\'e}on and Rouchon, Pierre and Cohen, Joachim and Petrescu, Alexandru},
  journal={Physical Review Applied},
  volume={23},
  number={2},
  pages={024073},
  year={2025},
  publisher={APS}
}

@article{viola2015collective,
  title={Collective modes in the fluxonium qubit},
  author={Viola, Giovanni and Catelani, Gianluigi},
  journal={Physical Review B},
  volume={92},
  number={22},
  pages={224511},
  year={2015},
  publisher={APS}
}

@article{lachance2024autonomous,
  title={Autonomous quantum error correction of {G}ottesman-{K}itaev-{P}reskill states},
  author={Lachance-Quirion, Dany and Lemonde, Marc-Antoine and Simoneau, Jean Olivier and St-Jean, Lucas and Lemieux, Pascal and Turcotte, Sara and Wright, Wyatt and Lacroix, Am{\'e}lie and Fr{\'e}chette-Viens, Jo{\"e}lle and Shillito, Ross and others},
  journal={Physical Review Letters},
  volume={132},
  number={15},
  pages={150607},
  year={2024},
  publisher={APS}
}

@article{you2019circuit,
  title={Circuit quantization in the presence of time-dependent external flux},
  author={You, Xinyuan and Sauls, James A and Koch, Jens},
  journal={Physical Review B},
  volume={99},
  number={17},
  pages={174512},
  year={2019},
  publisher={APS}
}

@article{willsch2024observation,
  title={Observation of {J}osephson harmonics in tunnel junctions},
  author={Willsch, Dennis and Rieger, Dennis and Winkel, Patrick and Willsch, Madita and Dickel, Christian and Krause, Jonas and Ando, Yoichi and Lescanne, Rapha{\"e}l and Leghtas, Zaki and Bronn, Nicholas T and others},
  journal={Nature Physics},
  pages={1--7},
  year={2024},
  publisher={Nature Publishing Group UK London}
}

@article{kolesnikow2024gottesman,
  title={{G}ottesman-{K}itaev-{P}reskill state preparation using periodic driving},
  author={Kolesnikow, Xanda C and Bomantara, Raditya W and Doherty, Andrew C and Grimsmo, Arne L},
  journal={Physical Review Letters},
  volume={132},
  number={13},
  pages={130605},
  year={2024},
  publisher={APS}
}

@article{nathan2024self,
  title={Self-Correcting Gottesman-Kitaev-Preskill Qubit and Gates in a Driven-Dissipative Circuit},
  author={Nathan, Frederik and O’Brien, Liam and Noh, Kyungjoo and Matheny, Matthew H and Grimsmo, Arne L and Jiang, Liang and Refael, Gil},
  journal={PRX Quantum},
  volume={6},
  number={3},
  pages={030352},
  year={2025},
  publisher={APS}
}

@article{sun2014tracking,
  title={Tracking photon jumps with repeated quantum non-demolition parity measurements},
  author={Sun, Luyan and Petrenko, Andrei and Leghtas, Zaki and Vlastakis, Brian and Kirchmair, Gerhard and Sliwa, KM and Narla, Aniruth and Hatridge, Michael and Shankar, Shyam and Blumoff, Jacob and others},
  journal={Nature},
  volume={511},
  number={7510},
  pages={444--448},
  year={2014},
  publisher={Nature Publishing Group UK London}
}

@article{ofek2016extending,
  title={Extending the lifetime of a quantum bit with error correction in superconducting circuits},
  author={Ofek, Nissim and Petrenko, Andrei and Heeres, Reinier and Reinhold, Philip and Leghtas, Zaki and Vlastakis, Brian and Liu, Yehan and Frunzio, Luigi and Girvin, Steven M and Jiang, Liang and others},
  journal={Nature},
  volume={536},
  number={7617},
  pages={441--445},
  year={2016},
  publisher={Nature Publishing Group UK London}
}

@article{cohen2017degeneracy,
  title={Degeneracy-preserving quantum nondemolition measurement of parity-type observables for cat qubits},
  author={Cohen, Joachim and Smith, W Clarke and Devoret, Michel H and Mirrahimi, Mazyar},
  journal={Physical Review Letters},
  volume={119},
  number={6},
  pages={060503},
  year={2017},
  publisher={APS}
}

@article{hu2019quantum,
  title={Quantum error correction and universal gate set operation on a binomial bosonic logical qubit},
  author={Hu, Ling and Ma, Yuwei and Cai, Weizhou and Mu, Xianghao and Xu, Yuan and Wang, Weiting and Wu, Yukai and Wang, Haiyan and Song, YP and Zou, C-L and others},
  journal={Nature Physics},
  volume={15},
  number={5},
  pages={503--508},
  year={2019},
  publisher={Nature Publishing Group UK London}
}

@article{ni2023beating,
  title={Beating the break-even point with a discrete-variable-encoded logical qubit},
  author={Ni, Zhongchu and Li, Sai and Deng, Xiaowei and Cai, Yanyan and Zhang, Libo and Wang, Weiting and Yang, Zhen-Biao and Yu, Haifeng and Yan, Fei and Liu, Song and others},
  journal={Nature},
  volume={616},
  number={7955},
  pages={56--60},
  year={2023},
  publisher={Nature Publishing Group UK London}
}

@article{wolinsky1988quantum,
  title={Quantum noise in the parametric oscillator: from squeezed states to coherent-state superpositions},
  author={Wolinsky, M and Carmichael, HJ},
  journal={Physical Review Letters},
  volume={60},
  number={18},
  pages={1836},
  year={1988},
  publisher={APS}
}

@article{zanardi1997noiseless,
  title={Noiseless quantum codes},
  author={Zanardi, Paolo and Rasetti, Mario},
  journal={Physical Review Letters},
  volume={79},
  number={17},
  pages={3306},
  year={1997},
  publisher={APS}
}

@article{lidar1998decoherence,
  title={Decoherence-free subspaces for quantum computation},
  author={Lidar, Daniel A and Chuang, Isaac L and Whaley, K Birgitta},
  journal={Physical Review Letters},
  volume={81},
  number={12},
  pages={2594},
  year={1998},
  publisher={APS}
}

@article{cohen2014dissipation,
  title={Dissipation-induced continuous quantum error correction for superconducting circuits},
  author={Cohen, Joachim and Mirrahimi, Mazyar},
  journal={Physical Review A},
  volume={90},
  number={6},
  pages={062344},
  year={2014},
  publisher={APS}
}

@article{kapit2016hardware,
  title={Hardware-efficient and fully autonomous quantum error correction in superconducting circuits},
  author={Kapit, Eliot},
  journal={Physical Review Letters},
  volume={116},
  number={15},
  pages={150501},
  year={2016},
  publisher={APS}
}

@article{kapit2017upside,
  title={The upside of noise: engineered dissipation as a resource in superconducting circuits},
  author={Kapit, Eliot},
  journal={Quantum Science and Technology},
  volume={2},
  number={3},
  pages={033002},
  year={2017},
  publisher={IOP Publishing}
}

@article{albert2016geometry,
  title={Geometry and response of {L}indbladians},
  author={Albert, Victor V and Bradlyn, Barry and Fraas, Martin and Jiang, Liang},
  journal={Physical Review X},
  volume={6},
  number={4},
  pages={041031},
  year={2016},
  publisher={APS}
}

@article{ticozzi2008quantum,
  title={Quantum Markovian subsystems: invariance, attractivity, and control},
  author={Ticozzi, Francesco and Viola, Lorenza},
  journal={IEEE Transactions on Automatic Control},
  volume={53},
  number={9},
  pages={2048--2063},
  year={2008},
  publisher={IEEE}
}

@article{barnes2000automatic,
  title={Automatic quantum error correction},
  author={Barnes, Jeff P and Warren, Warren S},
  journal={Physical Review Letters},
  volume={85},
  number={4},
  pages={856},
  year={2000},
  publisher={APS}
}

@article{taguchi2015mode,
  title={Mode engineering with a one-dimensional superconducting metamaterial},
  author={Taguchi, Masahiko and Basko, Denis M and Hekking, Frank WJ},
  journal={Physical Review B},
  volume={92},
  number={2},
  pages={024507},
  year={2015},
  publisher={APS}
}

@book{pozar2011microwave,
  title={Microwave engineering},
  author={Pozar, David M},
  year={2011},
  publisher={John Wiley \& sons}
}

@article{planat2020photonic,
  title={Photonic-crystal {J}osephson traveling-wave parametric amplifier},
  author={Planat, Luca and Ranadive, Arpit and Dassonneville, R{\'e}my and Puertas Mart{\'\i}nez, Javier and L{\'e}ger, S{\'e}bastien and Naud, C{\'e}cile and Buisson, Olivier and Hasch-Guichard, Wiebke and Basko, Denis M and Roch, Nicolas},
  journal={Physical Review X},
  volume={10},
  number={2},
  pages={021021},
  year={2020},
  publisher={APS}
}

@article{bronn2015broadband,
  title={Broadband filters for abatement of spontaneous emission in circuit quantum electrodynamics},
  author={Bronn, Nicholas T and Liu, Yanbing and Hertzberg, Jared B and C{\'o}rcoles, Antonio D and Houck, Andrew A and Gambetta, Jay M and Chow, Jerry M},
  journal={Applied Physics Letters},
  volume={107},
  number={17},
  year={2015},
  publisher={AIP Publishing}
}

@article{sarovar2005continuous,
  title={Continuous quantum error correction by cooling},
  author={Sarovar, Mohan and Milburn, Gerard J},
  journal={Physical Review A—Atomic, Molecular, and Optical Physics},
  volume={72},
  number={1},
  pages={012306},
  year={2005},
  publisher={APS}
}

@article{lihm2018implementation,
  title={Implementation-independent sufficient condition of the {Knill-Laflamme} type for the autonomous protection of logical qudits by strong engineered dissipation},
  author={Lihm, Jae-Mo and Noh, Kyungjoo and Fischer, Uwe R},
  journal={Physical Review A},
  volume={98},
  number={1},
  pages={012317},
  year={2018},
  publisher={APS}
}

@article{lebreuilly2021autonomous,
  title={Autonomous quantum error correction and quantum computation},
  author={Lebreuilly, Jos{\'e} and Noh, Kyungjoo and Wang, Chiao-Hsuan and Girvin, Steven M and Jiang, Liang},
  journal={arXiv preprint arXiv:2103.05007},
  year={2021}
}

@article{xu2023autonomous,
  title={Autonomous quantum error correction and fault-tolerant quantum computation with squeezed cat qubits},
  author={Xu, Qian and Zheng, Guo and Wang, Yu-Xin and Zoller, Peter and Clerk, Aashish A and Jiang, Liang},
  journal={npj Quantum Information},
  volume={9},
  number={1},
  pages={78},
  year={2023},
  publisher={Nature Publishing Group UK London}
}

@article{frattini2018optimizing,
  title={Optimizing the nonlinearity and dissipation of a {SNAIL} parametric amplifier for dynamic range},
  author={Frattini, NE and Sivak, VV and Lingenfelter, A and Shankar, S and Devoret, MH},
  journal={Physical Review Applied},
  volume={10},
  number={5},
  pages={054020},
  year={2018},
  publisher={APS}
}

@article{gambetta2006qubit,
  title={Qubit-photon interactions in a cavity: Measurement-induced dephasing and number splitting},
  author={Gambetta, Jay and Blais, Alexandre and Schuster, David I and Wallraff, Andreas and Frunzio, L and Majer, J and Devoret, Michel H and Girvin, Steven M and Schoelkopf, Robert J},
  journal={Physical Review A—Atomic, Molecular, and Optical Physics},
  volume={74},
  number={4},
  pages={042318},
  year={2006},
  publisher={APS}
}

@article{reiter2017dissipative,
  title={Dissipative quantum error correction and application to quantum sensing with trapped ions},
  author={Reiter, Florentin and S{\o}rensen, Anders S{\o}ndberg and Zoller, Peter and Muschik, CA},
  journal={Nature communications},
  volume={8},
  number={1},
  pages={1822},
  year={2017},
  publisher={Nature Publishing Group UK London}
}

@article{pastawski2011quantum,
  title={Quantum memories based on engineered dissipation},
  author={Pastawski, Fernando and Clemente, Lucas and Cirac, Juan Ignacio},
  journal={Physical Review A—Atomic, Molecular, and Optical Physics},
  volume={83},
  number={1},
  pages={012304},
  year={2011},
  publisher={APS}
}

@misc{note1, title = "This hierarchy is needed to adiabatically eliminate the mode $b$ from the dynamics, but does not need to be respected for the cat manifold to be stabilized~\cite{robin2024convergence}."}

@misc{note2, title = "The fluxes $\Phi_1$ and $\Phi_2$ that thread the physical circuit loops can be formally allocated to each arm of the SQUID, such that they are the same fluxes threading the effective circuit loops."}

@misc{note3, title = "The fluxes $\Phi_1$ and $\Phi_2$ are known up to an integer number of flux quanta and depend on applied currents through an affine transformation. The  map   is centered at non-zero applied currents, indicating significant flux offsets."}

@misc{note4, title = "The spurious non-linear term in the circuit  Hamiltonian is a small perturbation so that eigenstates have an overlap close to 1 with unperturbed Fock states."}

@misc{note5, title = "Note that in principle, the memory coherence should not be affected by this drive as the ATS remains at a dynamical sweet-spot, in the sense that its susceptibility to magnetic flux drifts averages out over a period of the drive."}

@misc{note6, title = "In that work, the auxiliary mode was not designed to dissipate excitations at a sufficiently large rate to turn the interaction into an effective four-photon dissipation channel."}

@misc{note7, title = "Large dissipation rates are desired for fast adiabatic control of the cat state~\cite{guillaud2019repetition} but are not needed to achieve macroscopic bit-flip times~\cite{berdou2023one}."}

@misc{note8, title = "We considered activating a longitudinal coupling between the memory and buffer by pumping the ATS at $\omega_b$, following the strategy presented in Ref.~\cite{reglade2024quantum}. In practice this measurement scheme proved hard to tune in our system with significant dispersive interaction between modes, and dispersive measurements were sufficient to achieve the same goal."}

@misc{note9, title = "This formula holds in the more general case where $v$ also varies, albeit after  replacing $v$ with its average value."}

@misc{note10, title = "This term is only valid in the RWA since the modes phase fluctuations  can have opposed signs accross both inductors. However, the only contributions that survives the RWA do not depend on this sign."}

@misc{note11, title = "The  maximum overlap (defined for two states $|\psi_1\rangle$ and $|\psi_2\rangle$ as $|\langle \psi_1 | \psi_2 \rangle |^2$) is  larger than 0.9995 for all $(n,m)$."}

\end{document}